\documentclass[10pt,twocolumn,pra,aps,showkeys,superscriptaddress]{revtex4-1}

\usepackage[T1]{fontenc}
\usepackage[utf8]{inputenc}
\usepackage{mathrsfs}
\usepackage{amstext}
\usepackage{amsmath}
\usepackage{amssymb}
\usepackage{graphicx}
\usepackage[unicode=true,
 bookmarks=false,
 breaklinks=false,pdfborder={0 0 1},colorlinks=true]
 {hyperref}

\usepackage[svgnames]{xcolor}
\usepackage{pdfcolmk}
\usepackage{bm}

\begin{document}

\title{Non-Abelian geometric potentials and spin-orbit coupling for periodically
driven systems}

\author{Povilas Ra\v{c}kauskas}
\email[]{racpovilas@gmail.com}
\affiliation{Institute of Theoretical Physics and Astronomy, Vilnius University,
Saul\.{e}tekio Ave.~3, LT-10257 Vilnius, Lithuania}

\author{Viktor Novi\v{c}enko}
\email[]{viktor.novicenko@tfai.vu.lt}
\homepage[]{http://www.itpa.lt/~novicenko/}
\affiliation{Institute of Theoretical Physics and Astronomy, Vilnius University,
Saul\.{e}tekio Ave.~3, LT-10257 Vilnius, Lithuania}

\author{Han Pu}
\email[]{hpu@rice.edu}
\homepage[]{http://www.owlnet.rice.edu/~hpu/}
\affiliation{Department of Physics and Astronomy, and Rice Center for Quantum Materials, Rice University, Houston, Texas 77251, USA}

\author{Gediminas Juzeli\={u}nas}
\email[]{gediminas.juzeliunas@tfai.vu.lt}
\homepage[]{http://www.itpa.lt/~gj/}
\affiliation{Institute of Theoretical Physics and Astronomy, Vilnius University,
Saul\.{e}tekio Ave.~3, LT-10257 Vilnius, Lithuania}

\date{\today}

\begin{abstract}
We demonstrate the emergence of the non-Abelian geometric potentials and thus the three-dimensional (3D) spin-orbit coupling (SOC) for ultracold atoms without using the laser beams. This is achieved by subjecting an atom to a periodic perturbation which is the product of a position-dependent Hermitian operator $\hat{V}\left(\mathbf{r}\right)$ and a fast oscillating periodic function $f\left(\omega t\right)$ with a zero average. To have a significant spin-orbit coupling (SOC), we analyze a situation where the characteristic energy of the periodic driving is not necessarily small compared to the driving energy $\hbar\omega$. Applying a unitary transformation to eliminate the original periodic perturbation, we arrive at a non-Abelian (non-commuting) vector potential term describing the 3D SOC. The general formalism is illustrated by analyzing the motion of an atom in a spatially inhomogeneous magnetic field oscillating in time. A cylindrically symmetric magnetic field provides the SOC involving the coupling between the spin $\mathbf{F}$ and all three components of the orbital angular momentum (OAM) $\mathbf{L}$. In particular, the spherically symmetric monopole-type synthetic magnetic field $\mathbf{B}\propto\mathbf{r}$ generates the 3D SOC of the $\mathbf{L}\cdot \mathbf{F}$ form, which resembles the fine-structure interaction of hydrodgen atom. However, the strength of the SOC here goes as $1/r^{2}$ for larger distances, instead of $1/r^3$ as in atomic fine structure. Such a longer-ranged SOC significantly affects not only the lower states of the trapped atom, but also the higher ones. Furthermore, by properly tailoring the external trapping potential, the ground state of the system can occur at finite OAM, while the ground state of hydrogen atom has zero OAM.
\end{abstract}

\maketitle

\section{Introduction\label{sec:Introduction}}

The periodic driving enriches topological \cite{sorensen05,Oka2009,Kitagawa2010,Galitski2011NP,Rudner2013,Nathan15NJP,Budich17PRL,Eckardt17RMP,Weinberg17PR,Weitenberb19NatPhys,Unal19PRL}
and many body \cite{eckardt05,zenesini09,Eckardt2010,neupert11,regnault11,Struck11NP,wu12,Lewenstein2012,Chin2013NP,Struck:2013,bergholtz13,parameswaran13,Greschner14,Anisimovas15PRB,Nagerl2016,Esslinger17PRA,Eckardt17RMP,Chin18PRL,Sacha18RMP}
properties of physical systems. This can be used to generate the artificial
gauge fields for ultracold atoms \cite{kolovsky11,struck12,Hauke:2012,Ketterle:2013,Aidelsburger:2013,Anderson2013,Xu2013,atala14,Goldman2014RPP,Flaschner16,Luo16Sci_Rep,Shteynas19PRL,Galitski19PT}
and photonic systems \cite{Haldane:2008cc,Rechtsman:2013fe,Mukherjee17Ncommun,Rechtsman18PRL,Cardano16NPhoton,Ozawa19RMP},
as well as to alter the topological properties of condensed matter
systems \cite{Oka2009,Galitski2011NP,Kitagawa2011,Galitski13PRB,tong13majorana,grushin14,usaj14,quelle15,Gavensky18PRB}.
In many cases the periodic driving changes in time, which applies {\it inter alia}
to experiments on ultracold atoms where the periodic driving is often slowly
ramped up \cite{Esslinger17PRA}. In such a situation the
evolution of the system can be described in terms of a slowly changing
effective Floquet Hamiltonian and a fast oscillating micromotion operator
\cite{Novicenko2017}. In particular, this is the case if the time
periodic Hamiltonian is a product of a slowly changing operator
$\hat{V}\left(\boldsymbol{\lambda}\left(t\right)\right)$ and a fast
oscillating function $f\left(\omega t\right)=f\left(\omega t+2\pi\right)$
with a zero average, where the vector $\boldsymbol{\lambda}\left(t\right)$
represents a set of slowly changing parameters \cite{Novicenko2017,Novicenko19PRA}.
If the operator $\hat{V}\left(\boldsymbol{\lambda}\left(t\right)\right)$
does not commute with itself at different times, 
the effective evolution of the periodically driven system can be accompanied
by non-Abelian (non-commuting) geometric phases after the vector $\boldsymbol{\lambda}\left(t\right)$
undergoes a cyclic change and returns to its original value \cite{Novicenko2017,Novicenko19PRA,Bigelow19Arxiv}. 

Here we study a way to generate non-Abelian geometric potentials when
the incident parameter $\boldsymbol{\lambda}\left(t\right)$ is replaced
by a radius vector of a particle $\mathbf{r}=x\mathbf{e}_x+y\mathbf{e}_y+z\mathbf{e}_z$ representing a dynamical
variable, and a kinetic energy operator is added. If the operator
$\hat{V}\left(\mathbf{r}\right)$ featured in the periodic coupling
term $\hat{V}\left(\mathbf{r}\right)f\left(\omega t\right)$ does
not commute with itself at different positions, $\left[\hat{V}\left(\mathbf{r}\right),\hat{V}\left(\mathbf{r}^{\prime}\right)\right]\ne0$,
the adiabatic evolution of the system within a Floquet band can be
accompanied by a non-Abelian (non-commuting) geometric vector potential
providing a three-dimensional (3D) spin-orbit coupling (SOC). The
2D and 3D SOC can be also generated optically by using degenerate
eigenstates of the atom-light coupling operator known as dressed states
\cite{Ruseckas2005,Stanescu2007,Jacob2007,Juzeliunas2008PRA,Stanescu2008,Campbell2011,Anderson2012PRL,Huang16NP,Meng16PRL,Spielman19_2D_SOC}.
This requires a considerable amount of efforts \cite{Campbell2011,Huang16NP,Meng16PRL,Spielman19_2D_SOC}. 
Furthermore the formation of the SOC is accompanied by unwanted heating due to the radiative decay of atoms in the dressed states.

The present approach does not rely on the degenerate atom-light dressed
states. Instead, employment of the time-periodic interaction of the form $\hat{V}\left(\mathbf{r}\right)f\left(\omega t\right)$
provides degenerate Floquet states \cite{Novicenko19PRA}.
The spatial and temporal dependence of these states yields the non-Abelian vector potential and thus the 3D SOC. 
The general formalism is illustrated by analyzing the motion of an atom
in a spatially inhomogeneous magnetic field oscillating in time.
We study a cylindrically symmetric magnetic field and analyze the
coupling between the atomic spin $\mathbf{F}$ and the orbital angular momentum (OAM) $\mathbf{L}$ for such
a system. In particular, the monopole-type magnetic field generates
the 3D SOC of the $\mathbf{L}\cdot\mathbf{F}$ form involving the
coupling between the atomic spin $\mathbf{F}$  and OAM. 

We shown that the strength
of the $\mathbf{L}\cdot\mathbf{F}$ SOC is long ranged and goes as
$1/r^{2}$ for larger distances, rather than as $1/r^3$, as experienced by an electron
in a Coulomb potential via the fine-structure interaction \cite{Landau:1987}. 
For such larger 
distances exceeding a characteristic SOC range $r_0$, the SOC contribution reduces to $-\mathbf{L}^{2}/2mr^{2}$ and thus it cancels the centrifugal term featured in the kinetic energy operator.
Therefore  the SOC significantly affects all atomic states. In the case of the (quasi)spin $1/2$ atom, the SOC makes
the atomic states nearly degenerate with respect to the orbital quantum number $l$ for a fixed total angular momentum 
quantum number $j=l\pm1/2$
and a fixed radial quantum number $n_{r}$. 
Furthermore,  for a harmonic trap the ground state with $j=1/2$ and $l=0$ has a slightly lower energy
than the one with $j=1/2$ and $l=1$. The situation can be changed
by adding an extra anti-trapping potential for small $r$. In that
case the ground state of the system acquires a non-zero orbital quantum
number $l=1$ and thus is affected by the SOC. This is a consequence
of the periodic driving; normally the ground state of a spherically
symmetric SOC system  corresponds to $l=0$. 

The paper is organized as follows. The general formalism is presented in the subsequent Sec.~\ref{sec:Formulation}.
We define a periodically driven system and apply a unitary transformation
eliminating the time-oscillating part of the original Hamiltonian.
The evolution of the transformed state vector is then governed by
a new Hamiltonian $W\left(\omega t\right)$ containing a vector potential type contribution $\mathbf{\hat{A}}\left(\mathbf{r},\omega t\right)$ which gives the spin-dependent 
momentum shift and provides the SOC in the effective Floquet Hamiltonian. We also
present a general analysis of the vector potential and discuss ramping
of the periodic diving. In Sec.~\ref{sec:Spin-in-oscillating}, as a specific example, we
study the spin in an oscillating magnetic field. We present an explicit
expression for the time-periodic vector potential, and consider the
coupling between the spin and OAM for a cylindrically symmetric
magnetic field. In Sec.~\ref{sec:Monopole-field} we analyze the
SOC for the spherically symmetric monopole magnetic field. Section~\ref{sec:Adiabatic-conditions-and-Implementation}
considers the adiabatic condition and discusses possible experimental
implementations. Section~\ref{sec:Concluding-remarks} presents concluding
remarks. 
Details of some technical calculations are contained in two Appendices~\ref{sec:Appendix-A} and ~\ref{sec:A^2-term}.
In particular, in Appendix~\ref{sec:Appendix-A} we present a way of producing the interaction of an atom with an effective magnetic field violating the Maxwell equations,
including the monopole magnetic field \cite{Pu18PRL}. 

\section{Periodically driven system with a modulated driving\label{sec:Formulation}}

\subsection{Hamiltonian and equations of motion}

Let us consider the center of mass motion of a quantum particle such
as an ultracold atom. The particle is subjected to a periodic driving
described by the operator $\hat{V}\left(\mathbf{r}\right)f\left(\omega t\right)$
which is a product of a position-dependent Hermitian operator $\hat{V}\left(\mathbf{r}\right)$
and a time-periodic function $f\left(\omega t+2\pi\right)=f\left(\omega t\right)$
with a zero average ${\intop_{-\pi}^{\pi}f\left(\omega t\right)dt=0}$.
A hat over the operator $\hat{V}\left(\mathbf{r}\right)$ indicates
that it depends on the internal degrees of freedom of the
particle. The state-dependent operator $\hat{V}\left(\mathbf{r}\right)$
generally does not commute with itself at different positions $\left[\hat{V}\left(\mathbf{r}\right),\hat{V}\left(\mathbf{r}^{\prime}\right)\right]\ne0$.
Including the kinetic energy, the system is described by the time-periodic
Hamiltonian $\hat{H}\left(\omega t\right)=\hat{H}\left(\omega t+2\pi\right)$
given by: 
\begin{equation}
\hat{H}\left(\omega t\right)=\frac{\mathbf{p}^{2}}{2m}+\hat{V}\left(\mathbf{r}\right)f\left(\omega t\right)+V_{\mathrm{ex}}\left(\mathbf{r}\right)\,,\label{eq:H_full}
\end{equation}
where $\mathbf{p}=-i\hbar\boldsymbol{\nabla}$
is the momentum operator, $m$ is the mass of the particle, and we
have also added an extra potential $V_{\mathrm{ex}}\left(\mathbf{r}\right)$
to confine the particle in a trap. The external potential is considered
to be state-independent, so that $\left[V_{\mathrm{ex}}\left(\mathbf{r}\right),V_{\mathrm{ex}}\left(\mathbf{r}^{\prime}\right)\right] = 0$.

The system is described by a state-vector $\left|\phi\left(t\right)\right\rangle $
obeying the time-dependent Schr\"{o}dinger equation (TDSE):
\begin{equation}
i\hbar\frac{\partial}{\partial t}\left|\phi\left(t\right)\right\rangle =\hat{H}\left(\omega t\right)\left|\phi\left(t\right)\right\rangle \,.\label{eq:Schroed-time-dep-H}
\end{equation}
An example of such a system is an atom in a spatially inhomogeneous
magnetic field $\mathbf{B}\left(\mathbf{r}\right)f\left(\omega t\right)$
with a fast oscillating amplitude ${\propto f\left(\omega t\right)}$
and a slowly changing magnitude or direction of the amplitude $\mathbf{B}\left(\mathbf{r}\right)$.
In that case the position-dependent part of the operator $\hat{V}\left(\mathbf{r}\right)f\left(\omega t\right)$
is given by 
\begin{equation}
\hat{V}\left(\mathbf{r}\right)=g_{F}\hat{\mathbf{F}}\cdot\mathbf{B}\left(\mathbf{r}\right)\,,\label{eq:V-spin}
\end{equation}
where $g_{F}$ is a gyromagnetic factor, and $\hat{\mathbf{F}}=\hat{F}_{1}\mathbf{e}_{x}+\hat{F}_{2}\mathbf{e}_{y}+\hat{F}_{3}\mathbf{e}_{z}$
is the spin operator with the Cartesian components obeying the commutation
relations $\left[\hat{F}_{s},\hat{F}_{q}\right]=i\hbar\epsilon_{squ}\hat{F}_{u}$.
Here $\epsilon_{squ}$ is a Levi-Civita symbol, and the summation
over a repeated Cartesian index $u=x,y,z$ is implied. The operator
$\hat{V}\left(\mathbf{r}\right)$ does not commute with itself at
different positions $\left[\hat{V}\left(\mathbf{r}\right),\hat{V}\left(\mathbf{r}^{\prime}\right)\right]\ne0$
if $\mathbf{B}\left(\mathbf{r}\right)$ and $\mathbf{B}\left(\mathbf{r}^{\prime}\right)$
are oriented along different axes. This leads to the SOC for the spin in the spatially non-uniform magnetic field oscillating in time, to be studied in the subsequent Secs.~\ref{sec:Spin-in-oscillating}--\ref{sec:Adiabatic-conditions-and-Implementation}.

\subsection{Transformed representation }

To have a significant SOC, we consider a situation where the matrix
elements of the periodic driving $\hat{V}\left(\mathbf{r}\right)f\left(\omega t\right)$
are not necessarily small compared to the driving energy $\hbar\omega$.
In that case one cannot apply the high frequency expansion \cite{Goldman2014,Eckardt2015,Novicenko2017}
of an effective Floquet Hamiltonian in the original representation.
To by-pass the problem, we go to a new representation via a unitary
transformation eliminating the operator $\hat{V}\left(\mathbf{r}\right)f\left(\omega t\right)$:
\begin{equation}
\hat{R}=\hat{R}\left(\mathbf{r},\omega t\right)=\exp\left[-i\frac{\mathcal{F}\left(\omega t\right)}{\hbar\omega}\hat{V}\left(\mathbf{r}\right)\right]\,,\label{eq:R-Definition}
\end{equation}
where $\mathcal{F}\left(\theta\right)$ is a primitive function of
$f\left(\theta\right)=d\mathcal{F}\left(\theta\right)/d\theta$
with a zero average ($\intop_{-\pi}^{\pi}\mathcal{F}\left(\theta^{\prime}\right)d\theta^{\prime}=0$)
and a calligraphy letter $\mathcal{F}$ is used to avoid a confusion
with the spin operator $\mathbf{F}$ featured in Eq.~(\ref{eq:V-spin}).
Since $\mathcal{F}\left(\omega t\right)=\mathcal{F}\left(\omega t+2\pi\right)$,
the transformation $\hat{R}\left(\mathbf{r},\omega t\right)=\hat{R}\left(\mathbf{r},\omega t+2\pi\right)$
has the same periodicity as the original Hamiltonian $\hat{H}\left(\omega t\right)=\hat{H}\left(\omega t+2\pi\right)$. 

The transformed state-vector 
\begin{equation}
\left|\psi\left(t\right)\right\rangle =\hat{R}^{\dagger}\left(\mathbf{r},\omega t\right)\left|\phi\left(t\right)\right\rangle \,\label{eq:|chi_theta>}
\end{equation}
obeys the TDSE 
\begin{equation}
i\hbar\frac{\partial}{\partial t}\left|\psi\left(t\right)\right\rangle =\hat{W}\left(\omega t\right)\left|\psi\left(t\right)\right\rangle \,\label{eq:Schroed-time-dep-K_R}
\end{equation}
governed by the Hamiltonian 
\begin{equation}
\hat{W}\left(\omega t\right)=\frac{1}{2m}\left[\mathbf{p}-\mathbf{\hat{A}}\left(\mathbf{r},\omega t\right)\right]^{2}+V_{\mathrm{ex}}\left(\mathbf{r}\right)\,,\label{eq:W-definition}
\end{equation}
where a time-periodic vector potential type operator
\begin{equation}
\mathbf{\hat{A}}\left(\mathbf{r},\omega t\right)=i\hbar\hat{R}^{\dagger}\left(\mathbf{r},\omega t\right)\boldsymbol{\nabla}\hat{R}\left(\mathbf{r},\omega t\right)\,\label{eq:A-definition}
\end{equation}
is added to the momentum operator $\mathbf{p}$ due to the position-dependence
of the unitary transformation $\hat{R}\left(\mathbf{r},\omega t\right)$.
On the other hand, the transformation $\hat{R}\left(\mathbf{r},\omega t\right)$
does not affect the state-independent trapping potential $V_{\mathrm{ex}}\left(\mathbf{r}\right)$.
The transformed Hamiltonian $\hat{W}\left(\omega t\right)$ no longer
contains the time-periodic term $\hat{V}\left(\mathbf{r}\right)f\left(\omega t\right)$.
The periodic driving is now represented by the operator $\mathbf{\hat{A}}\left(\mathbf{r},\omega t\right)=\mathbf{\hat{A}}\left(\mathbf{r},\omega t+2\pi\right)$
featured in the transformed Hamiltonian (\ref{eq:W-definition}).
This leads to the SOC to be studied in the next Subsection.

In this way the properly chosen transformation $\hat{R}\left(\mathbf{r},\omega t\right)$ eliminates the interaction operator $V(r)f(\omega t)$ in the original Hamiltonian (\ref{eq:H_full}). The position-dependence of $\hat{R}\left(\mathbf{r},\omega t\right)$ yields the spin-dependent momentum shift $\mathbf{\hat{A}}\left(\mathbf{r},\omega t\right)$ in Eq.~(\ref{eq:W-definition}), so the SOC appears directly from the unitary transformation.  

\subsection{Floquet adiabatic approach\label{subsec:Floquet-adiabatic-approach}}

The transformed Hamiltonian $\hat{W}\left(\omega t\right)=\hat{W}\left(\omega t+2\pi\right)$
can be expanded in the Fourier components:
\begin{equation}
\hat{W}\left(\omega t\right)=\sum_{n=-\infty}^{\infty}W^{\left(n\right)}e^{in\omega t},\label{eq:W-periodic-expansion}
\end{equation}
with
\begin{equation}
\hat{W}^{\left(n\right)}=\frac{1}{2\pi}\intop_{-\pi}^{\pi}\hat{W}\left(\theta\right)e^{-in\theta}d\theta.
\label{eq:fourier_comp}
\end{equation}
In what follows the driving energy $\hbar\omega$ is assumed to be
much larger than the matrix elements of the Fourier components of
the transformed Hamiltonian $\hat{W}^{\left(n\right)}$,
\begin{equation}
\hbar\omega\gg\left|\hat{W}_{\alpha\beta}^{\left(n\right)}\right|\,,\label{eq:adiabatic_condition}
\end{equation}
where the superscript $\left(n\right)$ refers to the
$n$-th Fourier component. The condition (\ref{eq:adiabatic_condition}) allows one to consider the adiabatic
evolution of the system in a selected Floquet band by neglecting the non-zeroth
(with $n \neq 0$) Fourier components $\hat{W}^{\left(n\right)}$ of the
transformed Hamiltonian $\hat{W}\left(\omega t\right)$. Thus one
replaces the exact evolution governed by the time-dependent transformed
Hamiltonian $\hat{W}\left(\omega t\right)$ by the approximate one
governed by the time-independent effective Floquet Hamiltonian $\hat{W}_{\mathrm{eff}(0)}=\hat{W}^{\left(0\right)}$
equal to the zeroth Fourier component: 
\begin{equation}
i\hbar\frac{\partial}{\partial t}\left|\psi^{\left(0\right)}\left(t\right)\right\rangle =\hat{W}^{\left(0\right)}\left|\psi^{\left(0\right)}\left(t\right)\right\rangle \,,\label{eq:Schroed-time-dep-K_R-adiabatic}
\end{equation}
where $\left|\psi^{\left(0\right)}\left(t\right)\right\rangle $ is
the corresponding approximate state vector representing the slowly
changing part of the exact state-vector $\left|\psi\left(t\right)\right\rangle $.
The slowly changing state-vector $\left|\psi^{\left(0\right)}\left(t\right)\right\rangle $
deviates little from the exact time-evolution of the state vector
$\left|\psi\left(t\right)\right\rangle $ if the adiabatic condition
(\ref{eq:adiabatic_condition}) holds. 

The effective Floquet Hamiltonian corresponding to the transformed 
Hamiltonian (\ref{eq:W-definition}) reads
\begin{equation}
\hat{W}^{\left(0\right)}=\frac{\mathbf{p}^{2}}{2m}+\hat{W}_{\mathrm{SOC}}+\hat{V}_{\mathrm{total}}\left(\mathbf{r}\right)\,,\label{eq:W^0-definition}
\end{equation}
where 
\begin{equation}
\hat{V}_{\mathrm{total}}\left(\mathbf{r}\right)=\frac{1}{2m}\left\langle \left[\mathbf{\hat{A}}\left(\mathbf{r},\omega t\right)\right]^{2}\right\rangle +V_{\mathrm{ex}}\left(\mathbf{r}\right)\label{eq:V_total}
\end{equation}
is the total scalar potential and 
\begin{equation}
\hat{W}_{\mathrm{SOC}}=-\frac{1}{2m}\left[\mathbf{\hat{A}}^{\left(0\right)}\left(\mathbf{r}\right)\cdot\mathbf{p}+\mathbf{p}\cdot\mathbf{\hat{A}}^{\left(0\right)}\left(\mathbf{r}\right)\right]\label{eq:W_SOC}
\end{equation}
describes the SOC emerging via the zeroth Fourier component of the
oscillating vector potential: $\mathbf{\hat{A}}^{\left(0\right)}\left(\mathbf{r}\right)=\left\langle \mathbf{\hat{A}}\left(\mathbf{r},\omega t\right)\right\rangle =\frac{1}{2\pi}\intop_{-\pi}^{\pi}\mathbf{\hat{A}}\left(\mathbf{r},\theta\right)d\theta$.
Here the brackets $\left\langle \ldots\right\rangle $ signify the
zero-frequency component (the time average) of an oscillating operator.
In this way, the effective Hamiltonian $\hat{W}^{\left(0\right)}$
is determined by the time averages of the oscillating vector potential
$\mathbf{\hat{A}}\left(\mathbf{r},\omega t\right)$ and its square.
The vector potential $\mathbf{\hat{A}}^{\left(0\right)}\left(\mathbf{r}\right)$ generally contains three non-commuting Cartesian components 
leading to the 3D SOC. 
Note that the Floquet adiabatic approach applied here corresponds to the zero
order of the high frequency expansion \cite{Goldman2014,Eckardt2015,Bukov2015,Novicenko19PRA}
of the effective Floquet Hamiltonian $W_{\mathrm{eff}}=\hat{W}^{\left(0\right)}+O\left(1/\omega\right)$. 

The present perturbation analysis relies on the condition (\ref{eq:adiabatic_condition})
involving the Fourier components of the time-periodic operator $\mathbf{\hat{A}}\left(\mathbf{r},\omega t\right)=\mathbf{\hat{A}}\left(\mathbf{r},\omega t+2\pi\right)$
given by Eq.~(\ref{eq:A-definition}). The operator $\mathbf{\hat{A}}\left(\mathbf{r},\omega t\right)$
emerges via the $\mathbf{r}$-dependence of the ratio $\frac{\mathcal{F}\left(\omega t\right)}{\hbar\omega}\hat{V}\left(\mathbf{r}\right)$
featured in the exponent of the unitary transformation $\hat{R}\left(\mathbf{r},\omega t\right)$.
Therefore the operator $\mathbf{\hat{A}}\left(\mathbf{r},\omega t\right)$
is determined by the the spatial
changes of the ratio $\hat{V}\left(\mathbf{r}\right)/\omega$, and
the Floquet adiabatic condition (\ref{eq:adiabatic_condition}) requires
the smallness of the spatial changes of the operator $\hat{V}\left(\mathbf{r}\right)$
rather than on the smallness of the operator $\hat{V}\left(\mathbf{r}\right)$
itself with respect to the driving frequency $\omega$. 

The condition
(\ref{eq:adiabatic_condition}) can hold even if the matrix elements
of the periodic driving $\hat{V}\left(\mathbf{r}\right)f\left(\omega t\right)$
are not small compared to the driving energy $\hbar\omega$. In such a situation the high frequency expansion of the effective Hamiltonian \cite{Goldman2014,Eckardt2015,Bukov2015,Novicenko19PRA} is not applicable
in the original representation where the evolution is given by Eq.~(\ref{eq:Schroed-time-dep-H}). Yet it is applicable in the transformed representation corresponding to the equation of motion~(\ref{eq:Schroed-time-dep-K_R}). Therefore the present approach allows
one to realize the SOC which is much larger than the one relying
on the perturbation treatment in the original representation, as it was done in a very recent study~\cite{Cheng19arXiv}. The adiabatic condition (\ref{eq:adiabatic_condition}) will be analyzed
in more details for a spin in an oscillating magnetic field in Sec.
\ref{subsec:Adiabatic-condition}.

Returning to the original representation, the adiabatic evolution
of the state vector is given by 
\begin{equation}
\left|\phi\left(t\right)\right\rangle \equiv\left|\phi\left(\omega t,t\right)\right\rangle \approx \hat{R}\left(\mathbf{r},\omega t\right)\left|\psi^{\left(0\right)}\left(t\right)\right\rangle \,.\label{eq:phi_inverse transformation}
\end{equation}
Since $\hat{R}\left(\mathbf{r},\omega t\right)=\hat{R}\left(\mathbf{r},\omega t+2\pi\right)$,
the original state-vector $\left|\phi\left(\omega t,t\right)\right\rangle =\left|\phi\left(\omega t+2\pi,t\right)\right\rangle $
is $2\pi$ periodic with respect to the first variable. Therefore
$\hat{R}\left(\mathbf{r},\omega t\right)$ describes the fast micromotion
of the original state vector $\left|\phi\left(\omega t,t\right)\right\rangle $.
Additionally the state-vector $\left|\phi\left(\omega t,t\right)\right\rangle $
changes slowly with respect to the second variable due to the slow
changes of the transformed state-vector $\left|\psi^{\left(0\right)}\left(t\right)\right\rangle $. 

\subsection{Equation for $\mathbf{\hat{A}}\left(\mathbf{r},\omega t\right)$
and its expansion\label{sec:Analysis-of operator A}}

To obtain an equation for the vector potential ${\mathbf{\hat{A}}=\mathbf{\hat{A}}\left(\mathbf{r},\omega t\right)}$,
let us treat it as a function of the coordinate $\mathbf{r}$ and
a parameter $c=c\left(\omega t\right) \equiv \mathcal{F}\left(\omega t\right)/\hbar\omega$.
Differentiating $\mathbf{\hat{A}}\left(\mathbf{r},\omega t\right)=\mathbf{\hat{A}}\left(\mathbf{r};c\right)$
with respect to $c$ for fixed $\mathbf{r}$, and using Eqs.~(\ref{eq:R-Definition})
and (\ref{eq:A-definition}), one arrives at the following differential equation
for the Cartesian components $\hat{A}_{u}$ of the vector potential

\begin{equation}
\frac{\partial\hat{A}_{u}}{\partial c}=\hbar\frac{\partial\hat{V}}{\partial u}+i\left[\hat{V},\hat{A}_{u}\right]\,\quad(u=x,y,z)\,,\label{eq:A-dif equation}
\end{equation}
subject to the initial condition 
\begin{equation}
\hat{A}_{u}=0\,\quad\mathrm{for}\quad c=0\,.\label{eq:A-initial condition}
\end{equation}
A solution to Eq.~(\ref{eq:A-dif equation}) can be expanded in the
powers of $c\propto1/\omega$, giving
\begin{align}
\hat{\mathbf{A}}\left(\mathbf{r},\omega t\right) & =\frac{\mathcal{F}\left(\omega t\right)}{\omega}\boldsymbol{\nabla}\hat{V}+i\frac{\mathcal{F}^{2}\left(\omega t\right)}{2!\hbar\omega^{2}}\left[\hat{V},\boldsymbol{\nabla}\hat{V}\right]\nonumber \\
 & +i^{2}\frac{\mathcal{F}^{3}\left(\omega t\right)}{3!\hbar^{2}\omega^{3}}\left[\hat{V},\left[\hat{V},\boldsymbol{\nabla}\hat{V}\right]\right]+\cdots\,.\label{eq:A-expansion-general}
\end{align}

If for any $\mathbf{r}$ and $\mathbf{r}^\prime$ the commutator $\left[\hat{V}\left(\mathbf{r}\right),\hat{V}\left(\mathbf{r}^{\prime}\right)\right]=0$, then 
only the first term remains in the expansion (\ref{eq:A-expansion-general}):
\begin{equation}
\hat{\mathbf{A}}\left(\mathbf{r},\omega t\right)=\frac{\mathcal{F}\left(\omega t\right)}{\omega}\boldsymbol{\nabla}\hat{V}\,,\label{eq:A-Abelian}
\end{equation}
giving $\left\langle \mathbf{\hat{A}}\left(\mathbf{r},\omega t\right)\right\rangle =\mathbf{\hat{A}}^{\left(0\right)}\left(\mathbf{r}\right)=0$.
In that case no SOC is generated, and the time average $\left\langle \left[\mathbf{\hat{A}}\left(\mathbf{r},\omega t\right)\right]^{2}\right\rangle $
provides an extra trapping potential in Eq.~(\ref{eq:V_total}).
In particular, this applies to a state-independent potential $\hat{V}\left(\mathbf{r}\right)=V\left(\mathbf{r}\right)$
for which the time-periodic Hamiltonian (\ref{eq:H_full}) describes
the Kapitza problem \cite{Bukov2015}.

Here we go beyond the situation where $\left[\hat{V}\left(\mathbf{r}\right),\hat{V}\left(\mathbf{r}^{\prime}\right)\right]=0$,
so the commutators are non-zero in the expansion (\ref{eq:A-expansion-general}).
As a result, the vector potential $\mathbf{\hat{A}}\left(\mathbf{r},\omega t\right)$
has a non-zero average $\mathbf{\hat{A}}^{\left(0\right)}\left(\mathbf{r}\right)\ne0$
providing the SOC which acts in all three dimensions. Such a 3D SOC
can be realized for a spinful atom in an inhomogeneous magnetic field
oscillating in time, to be considered in Section \ref{sec:Spin-in-oscillating}.
In that case the vector potential $\mathbf{\hat{A}}\left(\mathbf{r},\omega t\right)$
is obtained exactly in Eq.~(\ref{eq:A-spin-solution-general}) which is valid for
an arbitrary driving frequency, not necessarily small compared to
the strength of the periodic driving. Thus the solution (\ref{eq:A-spin-solution-general})
effectively takes into account all the terms in the expansion of $\mathbf{\hat{A}}\left(\mathbf{r},\omega t\right)$ given by Eq.~(\ref{eq:A-expansion-general}). 

\subsection{Ramping of the periodic perturbation\label{subsec:Ramping-of-the}}

Up to now the operator $\hat{V}\left(\mathbf{r}\right)$ defining
the periodic driving in the Hamiltonian (\ref{eq:H_full}) was considered
to be time-independent, so the driving was strictly periodic in time. The analysis
can be extended to a situation where the operator $\hat{V}\left(\mathbf{r}\right)$
has an extra slow temporal dependence \cite{Novicenko2017,Novicenko19PRA,Bigelow19Arxiv}.
This can describe ramping of the periodic perturbation. It is quite
common to have no periodic driving at an initial time $t=t_{\mathrm{in}}$
and ramp up the driving slowly afterwards over the time much large
than the driving period $T=2\pi/\omega$. This can be described by
a slowly changing factor $\alpha\left(t\right)$ multiplying $\hat{V}\left(\mathbf{r}\right)$:
\begin{equation}
\hat{V}\left(\mathbf{r},\alpha\left(t\right)\right)=\alpha\left(t\right)\hat{V}\left(\mathbf{r}\right)\,,\label{eq:V(r,alpha)}
\end{equation}
where $\alpha\left(t\right)$ changes smoothly from $\alpha\left(t\right)=0$
at the initial time $t=t_{\mathrm{in}}$ to $\alpha\left(t\right)=1$
at the final stage of the ramping. In particular, Eq.~(\ref{eq:V(r,alpha)})
with $\hat{V}\left(\mathbf{r}\right)$ given by Eq.~(\ref{eq:V-spin})
describes a spin in an oscillating magnetic field with a
slowly ramped amplitude $\mathbf{B}(\mathbf{r},\alpha\left(t\right))=\alpha\left(t\right)\mathbf{B}(\mathbf{r})$. 

The slow temporal dependence of $\hat{V}\left(\mathbf{r},\alpha\left(t\right)\right)$
featured in the unitary transformation $\hat{R}$ provides an additional
term $\hat{W}_{\mathrm{add}}\left(\mathbf{r},\omega t,t\right)$ to the
transformed Hamiltonian $\hat{W}\left(\omega t,t\right)$ \cite{Novicenko19PRA}:
\begin{equation}
\hat{W}_{\mathrm{add}}\left(\mathbf{r},\omega t,t\right)=-i\hbar\dot{\alpha}\hat{R}^{\dagger}\left(\mathbf{r},\omega t,\alpha\right)\frac{\partial\hat{R}\left(\mathbf{r},\omega t,\alpha\right)}{\partial\alpha}.
\label{eq:W_add}
\end{equation}
The operator $\hat{V}\left(\mathbf{r},\alpha\left(t\right)\right)$
commutes with itself at different times, $\left[\hat{V}\left(\mathbf{r},\alpha\left(t\right)\right),\hat{V}\left(\mathbf{r},\alpha\left(t^{\prime}\right)\right)\right]=0$,
giving
\begin{equation}
\hat{W}_{\mathrm{add}}\left(\mathbf{r},\omega t,t\right)=-\frac{\mathcal{F}\left(\omega t\right)\dot{\alpha}}{\omega}\frac{\partial \hat{V}\left(\mathbf{r},\alpha\right)}{\partial\alpha}.
\label{eq:Wadd}
\end{equation}
Since $\left\langle \mathcal{F}\left(\omega t\right)\right\rangle =0$,
the extra term $\hat{W}_{\mathrm{add}}\left(\mathbf{r},\omega t,t\right)$
averages to zero and thus has no zero Fourier component $\hat{W}_{\mathrm{add}}^{\left(0\right)}\left(\mathbf{r},\omega t,t\right)=0$.
In this way, the ramping of the periodic driving described by Eq.~(\ref{eq:V(r,alpha)})
does not provide an extra contribution to the effective Hamiltonian
and thus does not affect the effective dynamics of the system. 

\section{Spin in time-oscillating magnetic field\label{sec:Spin-in-oscillating}}

\subsection{Vector potential}

The general formalism is illustrated by considering motion of a spinful
atom in a time-oscillating magnetic field with the interaction operator
$\hat{V}\left(\mathbf{r}\right)$ given by Eq.~(\ref{eq:V-spin}).
In that case the operator $\mathbf{\hat{A}}=\mathbf{\hat{A}}\left(\mathbf{r},\omega t\right)$
can be derived exactly for an arbitrary strength of the magnetic field.
Specifically, by solving Eq.~(\ref{eq:A-dif equation}) one arrives
at the following Cartesian components of the vector potential $\mathbf{\hat{A}}\left(\mathbf{r},\omega t\right)$:
\begin{align}
\hat{A}_{u}\left(\mathbf{r},\omega t\right) & =a\mathcal{F}\frac{\left(\mathbf{B}\cdot\partial\mathbf{B}/\partial u\right)\left(\mathbf{B}\cdot\hat{\mathbf{F}}\right)}{B^{3}}\nonumber \\
+ & \sin\left(a\mathcal{F}\right)\frac{\left[\left(\mathbf{B}\times\partial\mathbf{B}/\partial u\right)\times\mathbf{B}\right]\cdot\hat{\mathbf{F}}}{B^{3}}\nonumber \\
+ & \left[\cos\left(a\mathcal{F}\right)-1\right]\frac{\left(\mathbf{B}\times\partial\mathbf{B}/\partial u\right)\cdot\hat{\mathbf{F}}}{B^{2}}\,,\label{eq:A-spin-solution-general}
\end{align}
where
\begin{equation}
a=\frac{Bg_{F}}{\omega}\,,\label{eq:a}
\end{equation}
defines the frequency of the magnetic interaction in the units of
the driving frequency. Here we keep implicit the time dependence of
the oscillating function $\mathcal{F}=\mathcal{F}\left(\omega t\right)$,
as well as the $\mathbf{r}$-dependence of $\mathbf{B}=\mathbf{B}(\mathbf{r})$
and $a=a(\mathbf{r})$. The derivation of Eq.~(\ref{eq:A-spin-solution-general})
is analogous to the one presented in the Appendix of Ref.~\cite{Novicenko19PRA}
subject to replacement of the time-derivatives $\partial V/\partial t$
and $\partial\mathbf{B}/\partial t$ by the space-derivatives $-\partial V/\partial u$
and $-\partial\mathbf{B}/\partial u$, respectively.

\subsection{Time averaged vector potential and SOC term\label{subsec:Time-averaged-vector-potential}}

To simplify the subsequent analysis, we assume the original Hamiltonian
(\ref{eq:H_full}) to have a time reversal symmetry. This is the case
if the function $f\left(\omega t\right)$ describing the periodic
driving is even: $f\left(\omega t\right)=f\left(-\omega t\right)$
(subject to a proper choice of the origin of time). Consequently the
function $\mathcal{F}\left(\omega t\right)=\omega\int_{0}^{t}f\left(\omega t^{\prime}\right)dt^{\prime}$
featured in the vector potential $\mathbf{\hat{A}}\left(\mathbf{r},\omega t\right)$
is an odd function: $\mathcal{F}\left(\omega t\right)=-\mathcal{F}\left(-\omega t\right)$.
In particular, this holds for a harmonic driving with
\begin{equation}
f\left(\omega t\right)=\cos\left(\omega t\right)\,\quad\,\mathrm{and}\quad\mathcal{F}\left(\omega t\right)=\sin\left(\omega t\right).\label{eq:f-harmonic}
\end{equation}
 For $\mathcal{F}\left(\omega t\right)=-\mathcal{F}\left(-\omega t\right)$
the first two lines of Eq.~(\ref{eq:A-spin-solution-general}) are
odd functions of time and thus average to zero, giving
\begin{equation}
\hat{A}_{u}^{\left(0\right)}\left(\mathbf{r}\right)=\frac{\left\langle \cos\left(a\mathcal{F}\right)\right\rangle-1}{B^{2}}\left(\mathbf{B}\times\partial\mathbf{B}/\partial u\right)\cdot\hat{\mathbf{F}}\,.\label{eq:A^(0)-spin-solution-general}
\end{equation}
Note that for the harmonic driving (\ref{eq:f-harmonic}), the time
average $\left\langle \cos\left(a\mathcal{F}\right)\right\rangle $
is given by the Bessel functions of the first kind: 
\begin{equation}
\left\langle \cos\left(a\mathcal{F}\right)\right\rangle =\mathcal{J}_{0}\left(a\right)\,,\quad\mathrm{with}\quad\mathcal{J}_{0}\left(a\right)=\frac{1}{2\pi}\intop_{-\pi}^{\pi}e^{ia\sin\theta}d\theta\,.\label{eq:average-Bessel}
\end{equation}

Substituting Eq.~(\ref{eq:A^(0)-spin-solution-general}) into (\ref{eq:W_SOC}),
the SOC term takes the form
\begin{align}
&\hat{W}_{\mathrm{SOC}} =   \nonumber \\
&\frac{1}{2m} \left[\frac{1-\left\langle \cos\left(a\mathcal{F}\right)\right\rangle }{B^{2}}\mathbf{Z}\cdot\hat{\mathbf{F}}+\mathbf{Z}^\dagger\cdot\hat{\mathbf{F}}\frac{1-\left\langle \cos\left(a\mathcal{F}\right)\right\rangle }{B^{2}}\right]\label{eq:W_SOC-spin} \,,
\end{align}
where
\begin{equation}
\mathbf{Z}=\mathbf{B}\times\sum_{u=x,y,z}\frac{\partial \mathbf{B}}{\partial u}p_{u}\label{eq:Z-general} \,,
\end{equation}
is an orbital operator, which makes it clear that Eq.~(\ref{eq:W_SOC-spin}) represents SOC. 

In addition to the SOC term $\hat{W}_{\mathrm{SOC}}$, the effective Hamiltonian
(\ref{eq:W^0-definition}) contains also the $\left\langle \left[\mathbf{\hat{A}}\left(\mathbf{r},\omega t\right)\right]^{2}\right\rangle $
term which is featured in scalar potential $V_{\mathrm{total}}\left(r\right)$ given by Eq.~(\ref{eq:V_total}).
This contribution is analyzed in Appendix \ref{sec:A^2-term} for
specific configurations of the magnetic field. 

The form of the SOC term in Eq.~(\ref{eq:W_SOC-spin}) is valid for an arbitrary magnetic configuration. However, to give a concrete example and to make the physics more clear, in the next Subsection we consider a cylindrically symmetric magnetic field given by Eq.~(\ref{eq:B-linear}). Subsequently in Sec.~\ref{sec:Monopole-field} we analyze an important  particular situation where the magnetic field takes the form of a monopole, in which case we can take advantage of the spherical symmetry of the system and simplify the calculations. In the Appendix~\ref{sec:Appendix-A}  we present details on how to generate various effective magnetic fields with a non-zero divergence field including the cylindrically symmetric magnetic field (\ref{eq:B-linear}) and the effective monopole field (\ref{eq:B-monopole}).

\subsection{Cylindrical magnetic field\label{subsec:Linear-magnetic-field}}

\subsubsection{Magnetic field\label{subsec:Magnetic-field}}

Let us consider the magnetic field $\mathbf{B}=\mathbf{B}(\mathbf{r})$
which changes linearly in space and has a cylindric symmetry:
\begin{equation}
\mathbf{B}(\mathbf{r})=\alpha_{\bot}\left(x\mathbf{e}_{x}+y\mathbf{e}_{y}\right)+\alpha_{z}z\mathbf{e}_{z}\,,\label{eq:B-linear}
\end{equation}
where the ratio between $\alpha_{z}$ and $\alpha_{\bot}$ is considered
to be arbitrary. By taking $\alpha_{z}=-2\alpha_{\bot}$, Eq.~(\ref{eq:B-linear})
describes a quadrupole magnetic field \cite{Suchet_2016}. On the
other hand, for $\alpha_{z}\ne-2\alpha_{\bot}$ the cylindrically
symmetric magnetic field has a non-zero divergence and thus does not
obey the Maxwell equation $\boldsymbol{\nabla} \cdot \mathbf{B}(\mathbf{r})=0$.
In particular, this is the case for a spherically symmetric monopole
field corresponding to $\alpha_{z}=\alpha_{\bot}=\alpha$ and considered in Sec.~\ref{sec:Monopole-field}.

\subsubsection{SOC operator  \label{subsec:SOC-operator}}

The operator $\mathbf{Z}\cdot\hat{\mathbf{F}}$ entering 
Eq.~(\ref{eq:W_SOC-spin}) for the SOC operator 
$\hat{W}_{\mathrm{SOC}}$ reads for the cylindrically symmetric magnetic field (\ref{eq:B-linear})
\begin{equation}
\mathbf{Z}\cdot\hat{\mathbf{F}}=\mathbf{Z}^\dagger\cdot\hat{\mathbf{F}}=\alpha_{\bot}^{2}L_{z}\hat{F}_{z}+\alpha_{\bot}\alpha_{z}\left(L_{x}\hat{F}_{x}+L_{y}\hat{F}_{y}\right)\,,\label{eq:Z.F}
\end{equation}
where $L_{x}$, $L_{y}$ and $L_{z}$ are the Cartesian components
of the OAM operator $\mathbf{L}=\mathbf{r}\times\mathbf{p}$. Note that the function $\left(1-\left\langle \cos\left(a\mathcal{F}\right)\right\rangle \right)/B^{2}$ featured in Eq.~(\ref{eq:W_SOC-spin}) is cylindrically symmetric  and thus preserves the $z$-projection of the OAM for the magnetic field (\ref{eq:B-linear}).

 The term $L_{z}\hat{F}_{z}$ in Eq.~(\ref{eq:Z.F}) provides the
spin-dependent shift to eigenenergies of the OAM operator $L_{z}$.
On the other hand, the term $L_{x}\hat{F}_{x}+L_{y}\hat{F}_{y}=L_{+}\hat{F}_{-}+L_{-}\hat{F}_{+}$
represents transitions between different spin and OAM projection states
described by the raising / lowering operators $L_{\pm}=L_{x}\pm i L_{y}$
and $\hat{F}_{\pm}=\hat{F}_{x}\pm i \hat{F}_{y}$. Therefore
the present SOC has some similarities to the coupling between the
spin and OAM induced by Raman laser beams carrying optical vortices
\cite{Marzlin97PRL,Ruostekoski04PRL,Nandi04PRA,Juzeliunas2004,Juzeliunas2005,Bigelow08PRA,Bigelow09PRL,Pu15PRA,Qu15PRA,Lin18PRL-a,Lin18PRL-b,Zhang19PRL}.
Yet, unlike the Raman case now the coupling between the spin and OAM
is described by all three OAM projections $L_{x}$, $L_{y}$ and $L_{z}$
as long as $\alpha_{\bot}\ne0$ and $\alpha_{z}\ne0$, so the coupling
is truly three dimensional. In particular for a monopole magnetic
field where $\alpha_{\bot}=\alpha_{z}=\alpha$, Eqs.~(\ref{eq:W_SOC-spin})
and (\ref{eq:Z.F}) yield
a spherically symmetric coupling between the spin and OAM $\propto  \mathbf{L}\cdot\hat{\mathbf{F}}$ presented by Eq.~(\ref{eq:W_SOC-Monopole}) below.

\section{Monopole field\label{sec:Monopole-field}}

\subsection{Effective Hamiltonian for monopole field\label{subsec:Effective-Hamiltonian-for}}

For $\alpha_{z}=\alpha_{\bot}=\alpha$, Eq.~(\ref{eq:B-linear}) reduces
to the centrally symmetric monopole-like magnetic field 
\begin{equation}
\mathbf{B}=\alpha\mathbf{r}=\frac{2\omega}{r_{0}g_{F}}\mathbf{r}\,,\label{eq:B-monopole}
\end{equation}
where $r_{0}=2\omega/\alpha g_{F}$ defines the radius $r=r_{0}$
at which a characteristic frequency of the magnetic interaction $g_{F}B/2=\omega r/r_{0}$
becomes equal to the driving frequency $\omega$. In such a situation, the operator $\mathbf{Z}\cdot\hat{\mathbf{F}}= \alpha^2 \mathbf{L}\cdot\hat{\mathbf{F}}$ 
commutes with the spherically
symmetric magnetic field $B=\alpha r$, so ordering of operators is not important in the SOC term (\ref{eq:W_SOC-spin}), giving
\begin{equation}
\hat{W}_{\mathrm{SOC}}=\hbar\omega_{\mathrm{SOC}}\left(r\right)\frac{\mathbf{L}\cdot\hat{\mathbf{F}}}{\hbar^{2}}\,,\label{eq:W_SOC-Monopole}
\end{equation}
where the frequency
\begin{equation}
\omega_{\mathrm{SOC}}\left(r\right)=\frac{\hbar}{m}\frac{1-\left\langle \cos\left(2r\mathcal{F}/r_{0}\right)\right\rangle }{r^{2}}\,\label{eq:omega_SOC-Explicit}
\end{equation}
characterizes the SOC strength. On the other hand, the term $\left\langle \left[\mathbf{\hat{A}}\left(\mathbf{r},\omega t\right)\right]^{2}\right\rangle $
featured in the total scalar potential $V_{\mathrm{total}}\left(r\right)$ is
given by Eq.~(\ref{eq:A^2-spin-solution-Monopole-vector-result-averaged})
in Appendix \ref{sec:A^2-term}. Combining Eqs.~(\ref{eq:W^0-definition}),
(\ref{eq:V_total}), (\ref{eq:W_SOC-Monopole}), (\ref{eq:omega_SOC-Explicit})
and (\ref{eq:A^2-spin-solution-Monopole-vector-result-averaged}),
the effective Hamiltonian takes the form 
\begin{align}
\hat{W}^{\left(0\right)}=& \frac{p^{2}}{2m}+\hbar\omega_{\mathrm{SOC}}\left(r\right)\left[\frac{\mathbf{L}\cdot\hat{\mathbf{F}}}{\hbar^{2}}+\frac{r^{2}\mathbf{F}^{2}-\left(\mathbf{r}\cdot\hat{\mathbf{F}}\right)^{2}}{\hbar^{2}r^{2}}\right] \nonumber \\
&+\frac{2}{mr_{0}^{2}}\left\langle \mathcal{F}^{2}\right\rangle \frac{\left(\mathbf{r}\cdot\hat{\mathbf{F}}\right)^{2}}{r^{2}}+V_{\mathrm{ex}}\left(\mathbf{r}\right)\,,
\label{eq:W^0-result-general}
\end{align}
where the spin-dependent operator $\left(\mathbf{r}\cdot\hat{\mathbf{F}}\right)^{2}$ emerges from the $\left\langle \left[\mathbf{\hat{A}}\left(\mathbf{r},\omega t\right)\right]^{2}\right\rangle $ term entering the total scalar potential (\ref{eq:V_total}). Generally the operator  $\left(\mathbf{r}\cdot\hat{\mathbf{F}}\right)^{2}$
does not commute with $\mathbf{L}^{2}$, and thus mixes the states
with different orbital quantum numbers $l$. Specifically, the term $\left(\mathbf{r}\cdot\hat{\mathbf{F}}\right)^{2}$ can provide coupling between orbit and spin or even spin tensor involving the radius vector $\mathbf{r}$ rather than the momentum operator, as in ref. \cite{Zhang17PRL}. However, no such extra SOC appears for the spin-$1/2$ atom to be considered next.

\subsection{Spin-$1/2$\label{subsec:Spin 1/2}}

\subsubsection{Effective Hamiltonian}

Let us now consider the effective Hamiltonian (\ref{eq:W^0-result-general})
for a spin-$1/2$ atom for which 
\begin{equation}
\hat{\mathbf{F}}=\frac{\hbar}{2}\hat{\boldsymbol{\sigma}}\,,\label{eq:F spin 1/2}
\end{equation}
where $\hat{\boldsymbol{\sigma}}=\hat{\sigma}_{x}\mathbf{e}_{x}+\hat{\sigma}_{y}\mathbf{e}_{y}+\hat{\sigma}_{z}\mathbf{e}_{z}$
and $\hat{\sigma}_{x,y,z}$ are the Pauli matrices. In that case the
operator $\left(\mathbf{r}\cdot\hat{\mathbf{F}}\right)^{2}=\hbar^{2}r^{2}/4$
is spin-independent and spherically symmetric, making $\mathbf{L}^{2}$
a conserving quantity. Using $2\mathbf{L}\cdot\hat{\mathbf{F}}=\hat{\mathbf{J}}^{2}-\mathbf{L}^{2}-\mathbf{F}^{2}$,
the effective Hamiltonian (\ref{eq:W^0-result-general}) takes the
form
\begin{equation}
\hat{W}^{\left(0\right)}=\frac{p^{2}}{2m}+\hbar\omega_{\mathrm{SOC}}\left(r\right)\left(\frac{\hat{\mathbf{J}}^{2}-\mathbf{L}^{2}}{2\hbar^{2}}+\frac{1}{8}\right)+V_{\mathrm{ex}}\left(\mathbf{r}\right)\,,\label{eq:W^0-result_spin_1/2}
\end{equation}
where 
\begin{equation}
\hat{\mathbf{J}}=\mathbf{L}+\hat{\mathbf{F}}\,\label{eq:J-definition}
\end{equation}
is the total angular momentum, and a uniform energy shift $\hbar^{2}\left\langle \mathcal{F}^{2}\right\rangle /2mr_{0}^{2}$
has been omitted in Eq.~(\ref{eq:W^0-result_spin_1/2}). 

\begin{figure}[h]
\centering
\includegraphics[width=0.95\columnwidth]{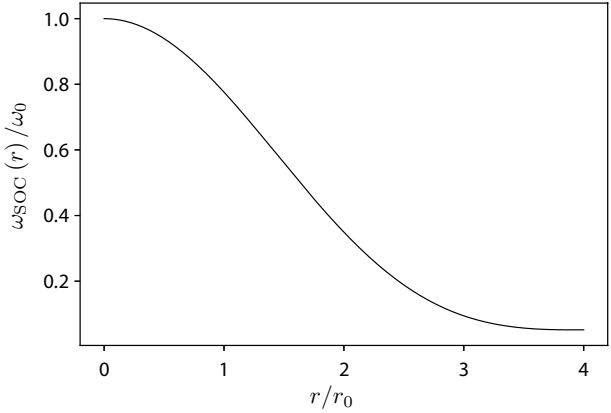}
\caption{\label{fig:SOC_plot} The radial dependence of the SOC energy $\omega_{\mathrm{SOC}}\left(r\right)$ given by Eq.~(\ref{eq:omega_SOC-Explicit}) for the sinusoidal driving
(\ref{eq:f-harmonic}) for which $\left\langle \cos\left(2r\mathcal{F}/r_{0}\right)\right\rangle =\mathcal{J}_{0}\left(2r/r_{0}\right)$.
The distance is measured in the units of $r_{0}$ and the frequency
is measured in the units of the SOC frequency $\omega_{0}=\omega_{\mathrm{SOC}}(0)$ given
by Eq.~(\ref{eq:omega_0}). }
\end{figure}

As one can see in Fig.~\ref{fig:SOC_plot}, the SOC frequency $\omega_{\mathrm{SOC}}\left(r\right)$
decreases with the radius $r$ and goes as $r^{-2}$ for distances
exceeding the SOC radius $r_{0}$:
\begin{equation}
\omega_{\mathrm{SOC}}\left(r\right)\approx\frac{\hbar}{mr^{2}}\,\quad\,\mathrm{for}\quad r\gg r_{0}\,.\label{eq:omega_SOC-Explicit-asymptotic}
\end{equation}
Such an asymptotic behavior of $\omega_{\mathrm{SOC}}\left(r\right)$
does not depend on the magnetic field strength, and is determined
exclusively by the ratio $\hbar/m$. The asymptotic Hamiltonian
\begin{equation}
\hat{W}^{\left(0\right)}=\frac{p^{2}}{2m}+\frac{1}{2mr^{2}}\left(\hat{\mathbf{J}}^{2}-\mathbf{L}^{2}+\frac{\hbar^{2}}{4}\right)+V_{\mathrm{ex}}\left(\mathbf{r}\right)\,, \quad\left(r\gg r_{0}\right)\label{eq:W^0-result_spin_1/2-3}
\end{equation}
contains a contribution $\propto-\mathbf{L}^{2}$ which cancels the
centrifugal term featured in the kinetic energy operator 
\begin{equation}
\frac{p^{2}}{2m}=-\frac{\hbar^{2}}{2m}\left[\frac{1}{r^{2}}\frac{\partial}{\partial r}\left(r^{2}\frac{\partial}{\partial r}\right)\right]+\frac{\mathbf{L}^{2}}{2mr^{2}}\,.\label{eq:E_kin}
\end{equation}
Thus, if the atomic eigenfunctions extend over distances exceeding
the radius $r_{0}$, the corresponding eigen-energies are determined
predominantly by the total angular momentum $\hat{\mathbf{J}}^{2}$ and
are nearly degenerate with respect to the orbital quantum number $l=j\mp1/2$
for fixed $j$ and fixed radial quantum number $n_{r}$, showing a
peculiar manifestation of the SOC. This is confirmed by numerical
calculations presented in Sec.\ref{subsec:Eigenstates-spin1/2} and
displayed in Fig.~\ref{fig:E_j,l-harmonic-with-higher-levels}. The
long-range behavior of $\omega_{\mathrm{SOC}}\left(r\right)$ makes
the SOC effects significant not only for the lower atomic states, but also also for higher ones situated
further away from the center. Note that for an electron in a Coulomb
potential $\propto-1/r$ the SOC strength is shorter ranged and goes
as $\propto r^{-3}$ \cite{Landau:1987}, affecting mostly the lower
electronic states situated closer to $r=0$. 

The SOC alone does not trap the atoms. Therefore an external trapping
potential $V_{\mathrm{ex}}\left(\mathbf{r}\right)$ is needed to have
bound states, like in the case of the light induced geometric potentials
\cite{Juzeliunas2005,Juzeliunas2006}. The external potential can
be chosen freely. For example, one can take $V_{\mathrm{ex}}\left(\mathbf{r}\right)$
to be a spherical harmonic trapping potential 
\begin{equation}
V_{\mathrm{ex}}\left(\mathbf{r}\right)=\frac{m}{8}\eta^{2}\omega_{0}^{2}r^{2},\label{eq:V-ex-harmonic}
\end{equation}
where $\eta$ defines the trapping frequency $\omega_{\mathrm{ex}}=\eta\omega_{0}/2$
and $\omega_{0}=\omega_{\mathrm{SOC}}\left(0\right)$ is the SOC frequency
at zero distance. For the sinusoidal driving (\ref{eq:f-harmonic})
one has 
\begin{equation}
\omega_{0}=\frac{\hbar}{mr_{0}^{2}}\,.\label{eq:omega_0}
\end{equation}

\subsubsection{Angular states\label{subsec:Eigenstates-spin1/2} }

In what follows we shall consider a spherically symmetric external potential $V_{\mathrm{ex}}\left(\mathbf{r}\right)=V_{\mathrm{ex}}\left(r\right)$. The Hamiltonian $\hat{W}^{\left(0\right)}$ given by Eq.~(\ref{eq:W^0-result_spin_1/2})
contains the commuting operators $\hat{\mathbf{J}}^{2}$ and $\mathbf{L}^{2}$
characterized by the eigenvalues $\hbar^{2}j\left(j+1\right)$ and
$\hbar^{2}l\left(l+1\right)$. The eigenstates $\left|j,l,f,m_{j}\right\rangle $
of $\hat{W}^{\left(0\right)}$ are thus described by the quantum numbers
$j$, $l$, $f$ and $m_{j}$, with $l-f\le j\le l+f$ and $f=1/2$.
The eigenstates $\left|j,l,f,m_{j}\right\rangle =\left|j,l,f,m_{j}\left(\theta,\phi\right)\right\rangle $
are degenerate with respect to projection of the total angular momentum
$-j\le m_{j}\le j$. They can be cast in terms of the angular momentum
states $Y_{l,m}\left(\theta,\phi\right)$ (the spherical harmonics)
and the spin states $\left|f,m_{f}\right\rangle $ with $f=1/2$:
\begin{align}
\left|j,l,f,m_{j}\left(\theta,\phi\right)\right\rangle =& \sum_{m_{l}=-l}^{l}\sum_{m_{f}=\pm\frac{1}{2}}Y_{l,m_{l}}\left(\theta,\phi\right)\left|f,m_{f}\right\rangle \nonumber \\
&\times \left\langle l,m_{l},f,m_{f}\right.\left|j,m_{j}\right\rangle \,,\label{eq:Angular states}
\end{align}
where $\theta$ and $\phi$ are the polar and azimuthal angles, $\left\langle l,m_{l},f,m_{f}\right.\left|j,m_{j}\right\rangle $
is the Clebsch--Gordan coefficient, and the summation is over the projections
of the spin and the orbital angular momentum with $m_{l}+m_{f}=m_{j}$.

\subsubsection{Radial eigen-equation}

The full eigenstate of the effective Hamiltonian (\ref{eq:W^0-result_spin_1/2})
contains also the radial part 
\begin{equation}
\left|\psi_{n_{r},j,l,f}\left(r,\theta,\phi\right)\right\rangle =\left|j,l,f,m_{j}\left(\theta,\phi\right)\right\rangle \psi_{n_{r},j,l,f}\left(r\right)\,,\label{eq:Full-state-vector}
\end{equation}
where $n_{r}$ is a radial quantum number. Substituting
\begin{equation}
\psi_{n_{r},j,l,f,m_{j}}\left(r\right) \equiv \frac{\phi_{n_{r},j,l,f,m_{j}}\left(r\right)}{r}\,,\label{eq:Radial function substitution}
\end{equation}
one arrives at the eigenvalue equation for the scaled radial function
$\phi_{n_{r},j,l,f,m_{j}}\left(r\right)$ 
\begin{align}
&\left[-\frac{\hbar^{2}}{2m}\frac{\partial^{2}}{\partial r^{2}}+V_{j,l}\left(r\right)\right]\phi_{n_{r},j,l,f,m_{j}}\left(r\right)\nonumber \\
&=E_{n_{r},j,l}\phi_{n_{r},j,l,f,m_{j}}\left(r\right),
\label{eq:Eigenvalue equation radial-Substituted}
\end{align}
subject to the condition $\phi_{n_{r},j,l,f,m_{j}}\left(0\right)=0$,
where 
\begin{align}
V_{j,l}\left(r\right)=&\frac{1}{2}\left[j\left(j+1\right)-l\left(l+1\right)+\frac{1}{4}\right]\hbar\omega_{\mathrm{SOC}}\left(r\right)\nonumber \\
&+\frac{\hbar^{2}l\left(l+1\right)}{2mr^{2}}+V_{\mathrm{ex}}\left(r\right),\label{eq:V-radial-explicit}
\end{align}
is the radial potential and $E_{n_{r},j,l}$ is an eigen-energy.

\begin{figure}[h]
\centering
\includegraphics[width=0.99\columnwidth]{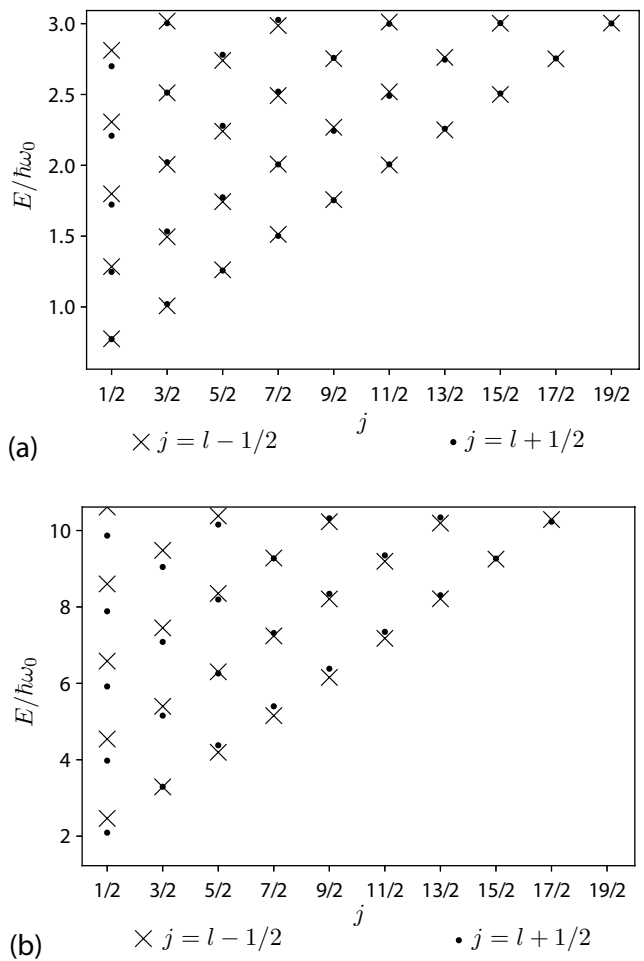}
\caption{\label{fig:E_j,l-harmonic-with-higher-levels} Dependence of eigen-energies $E_{n_{r},j,l}$ on $j$ for $l=j\mp1/2$ and up to five lowest radial
quantum numbers $n_{r}$. The harmonic trapping potential (\ref{eq:V-ex-harmonic})
is added. Panel~(a) corresponds to a softer trap with $\eta=0.5$; panel~(b) corresponds to a tighter trap with $\eta=2$.
The spectrum is calculated for the sinusoidal (\ref{eq:f-harmonic})
driving for which $\left\langle \cos\left(2r\mathcal{F}/r_{0}\right)\right\rangle =\mathcal{J}_{0}\left(2r/r_{0}\right)$.}
\end{figure}

\subsubsection{Analysis of eigen-energies\label{subsec:Analysis-of-eigenenergies}}

Figure \ref{fig:E_j,l-harmonic-with-higher-levels} displays the dependence
of the eigen-energies $E_{n_{r},j,l}$ on $j$ for $l=j\mp1/2$ and up
to five lowest radial quantum numbers $n_{r}$. The calculations are
carried out for the sinusoidal driving, Eq.~(\ref{eq:f-harmonic}),
and the harmonic trapping potential given by Eq.~(\ref{eq:V-ex-harmonic})
with $\eta=0.5$ and $\eta=2$. For a softer trap ($\eta=0.5$) and
$j\ge3/2$, there is an almost perfect degeneracy of the eigen-energies
$E_{n_{r},j,l}$ with the same $j$ and $n_{r}$ but different $l=j\mp$1/2.
For such a softer trap the atomic wave-functions extend to distances 
$r\gg r_{0}$ in which $\omega_{\mathrm{SOC}}\left(r\right)\approx\hbar/mr^{2}$, corresponding to the strong driving regime ($g_{F}B/2\gg \omega$).
Consequently the $l$-dependent part of the SOC term cancels the centrifugal
term in Eq.~(\ref{eq:V-radial-explicit}), and the eigenstates depend
weakly on $l$. 
For a tighter trap ($\eta=2$) the atom is localized
closer to the center leading to a larger difference in the eigen-energies
$E_{n_{r},j,l}$ with different $l=j\mp1/2$ but the same $j$ and
$n_{r}$. 

It is noteworthy that an infinite set of degenerate eigenstates (3D
Landau levels \cite{Congjun-Wu13PRL}) is formed if a particle is
subjected to the SOC term $\pm\omega_{\mathrm{C}}\mathbf{L}\cdot\hat{\boldsymbol{\sigma}}$
with a constant frequency $\omega_{\mathrm{C}}$, and a 3D isotropic
harmonic trap is added with the frequency $\omega_{\mathrm{C}}$ \cite{Ui-Takeda84PTP,Bagchi2001,Congjun-Wu13PRL}.
The eigenstates with $j=l\mp1/2$ then have eigen-energies $\hbar\omega_{\mathrm{C}}\left(2n_{r}+1\mp1/2\right)$
which depend only on the principal quantum number $n_{r}$ and thus
are degenerate with respect to the $l$ and $j$. In the present study
the situation is different. The SOC frequency $\omega_{\mathrm{SOC}}\left(r\right)$
decreases with the radius and has a special asymptotic behavior at
large distances, $\omega_{\mathrm{SOC}}\left(r\right)\approx\hbar/mr^{2}$,
leading to pairs of close energy levels with $l=j\pm1/2$ for fixed
$j$, as discussed above.

\begin{figure}[h]
\centering
\includegraphics[width=0.99\columnwidth]{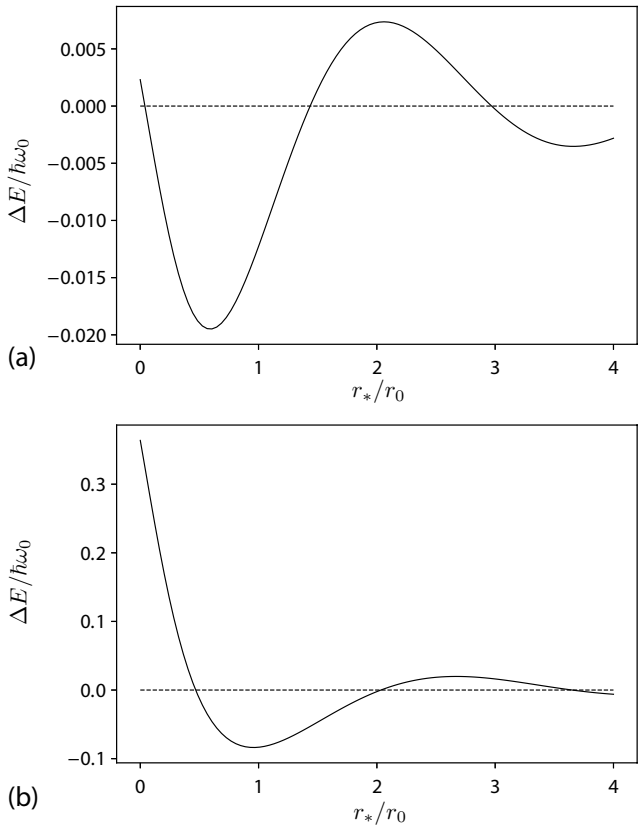}
\caption{\label{fig:Energy levels with r_*neq 0} The energy difference $\Delta E=E_{j=\frac{1}{2},l=1}-E_{j=\frac{1}{2},l=0}$
vs. the radius $r_{*}$ of the additional hard core potential for
an atom in a harmonic trapping potential (\ref{eq:V-ex-harmonic})
with $\eta=0.5$ (panel~(a)) and $\eta=2$ (panel~(b)). The calculations
are done for the sinusoidal driving (\ref{eq:f-harmonic}) for which
$\left\langle \cos\left(2r\mathcal{F}/r_{0}\right)\right\rangle =\mathcal{J}_{0}\left(2r/r_{0}\right)$.}
\end{figure}

For the sinusoidal driving, Eq.~(\ref{eq:f-harmonic}), the difference
in the radial potentials with $l=j\pm1/2$ for fixed $j$
\begin{align}
\Delta V_{j}\left(r\right)&=V_{j,l = j+1/2}\left(r\right)-V_{j,l=j-1/2}\left(r\right) \nonumber \\
&= \frac{\hbar\left(2j+1\right)\mathcal{J}_{0}(2r/r_{0})}{mr^{2}},
\label{eq:V-radial-difference}
\end{align}
is determined by the Bessel function $\mathcal{J}_{0}(2r/r_{0})$ which is positive
for distances $2r/r_{0}$ smaller than $2.4$. Consequently the ground
state with $j=1/2$ and $l=0$ has a slightly lower energy than the
one with $j=1/2$ and $l=1$, as one can see in Fig.~\ref{fig:E_j,l-harmonic-with-higher-levels}.
The situation can be reversed by adding an extra anti-trapping potential
for small $r$. Such a potential pushes the atomic probability distribution
to a region of larger distances, $2r/r_{0}>2.4$, where the Bessel function $\mathcal{J}_{0}(2r/r_{0})$
becomes negative and reaches the maximum negative value of $-0.36$
at $2r/r_{0}=3.83$. Figure \ref{fig:Energy levels with r_*neq 0}
shows the difference in the ground states energies $\Delta E=E_{j=\frac{1}{2},l=1}-E_{j=\frac{1}{2},l=0}$
for the external trapping potential $V_{\mathrm{ex}}\left(r\right)$
composed of a spherically symmetric harmonic potential with with $\eta=0.5$
(upper plot) or $\eta=2$ (lower plot), and an additional hard core
potential of a radius $r=r_{*}$ preventing the atom to be at distances
$r\le r_{*}$. For $\eta=0.5$ ($\eta=2$) the energy difference $\Delta E$
becomes negative at $r_{*}/r_{0}=0.14$ ($r_{*}/r_{0}=0.55$) and
reaches the maximum negative value of $\Delta E=-0.019\hbar\omega_{0}$
($\Delta E=-0.084\hbar\omega_{0}$) at $r_{*}/r_{0}=0.55$ ($r_{*}/r_{0}=0.97$). 

In this way, the ground state of the system can be the state with
$j=1/2$ and the orbital quantum number equal to $l=1$ rather than
$l=0$. For conventional spherically symmetric systems, such as the
hydrogen-like atoms, the ground state is always characterized by $l=0$
and thus is not affected by the SOC. The formation of the ground state
with $l=1$ is now facilitated by the time-periodic driving which
induces a longer-ranged SOC $\propto1/r^{2}$ and allows one to reverse
the sign of the potential difference $\Delta V_{j}\left(r\right)$
in Eq.~(\ref{eq:V-radial-difference}) due to the change of sign
in the Bessel function $\mathcal{J}_{0}\left(2r/r_{0}\right)$. Note
that the periodic driving can be also used to reverse the sign of the matrix
elements of tunneling of atoms in optical lattices \cite{Kenkre1986PRB,Arimondo07PRL,Arimondo2012,Struck11NP}.
By shaking the lattice sufficiently strongly, the Bessel function
renormalizes the tunneling matrix elements making them to change the
sign \cite{Arimondo07PRL,Arimondo2012,Struck11NP}. 

\section{Adiabatic conditions and implementation\label{sec:Adiabatic-conditions-and-Implementation}}

\subsection{Adiabatic condition\label{subsec:Adiabatic-condition}}

The general adiabatic condition (\ref{eq:adiabatic_condition}) was
discussed in Sec.\ref{subsec:Floquet-adiabatic-approach}. Now we
will consider in more details the adiabatic condition for an atom
in the magnetic field. Using Eqs.~(\ref{eq:W-definition}), (\ref{eq:adiabatic_condition}), (\ref{eq:A-spin-vector-general-alternative-App})
and (\ref{eq:d_123-monopole-App}), one arrives at the adiabatic condition
for the atom in a centrally symmetric magnetic field 

\begin{equation}
\omega\gg\omega_{0}\quad\mathrm{and}\quad\omega\gg \omega_{\mathrm{kin}} ,\label{eq:adiabatic_condition-Monopole-Specific}
\end{equation}
where $\omega_{0}=\hbar/mr_{0}^{2}$ is the SOC frequency and $\hbar\omega_{\mathrm{kin}}$
is the kinetic energy of the atomic motion. The first requirement
in Eq.~(\ref{eq:adiabatic_condition-Monopole-Specific}) is due to
the $\left[\mathbf{\hat{A}}\left(\mathbf{r},\omega t\right)\right]^{2}$ term in the transformed Hamiltonian $\hat{W}$, Eq.~(\ref{eq:W-definition}), the second one coming
from the mixed terms $\mathbf{\hat{A}}\left(\mathbf{r}\right)\cdot\mathbf{p}$ and $\mathbf{p} \cdot \mathbf{\hat{A}}\left(\mathbf{r}\right)$.
A condition similar to Eq.~(\ref{eq:adiabatic_condition-Monopole-Specific})
can be obtained also for a more general cylindrically symmetric magnetic
field given by Eq.~(\ref{eq:B-linear}). 

The adiabatic condition (\ref{eq:adiabatic_condition-Monopole-Specific})
does not rely on the smallness of the frequency of the magnetic interaction
$g_{F}B\left(r\right)/2=\omega r/r_{0}$ compared to the driving frequency
$\omega$, so we do not restrict ourselves to distances smaller than
the SOC radius $r_{0}$. We only require the SOC frequency $\omega_{0}$
and the atomic kinetic frequency $\omega_{\mathrm{kin}}$ to be much
smaller than the driving frequency $\omega$. Hence the present approach allows
one to study the SOC at distances exceeding the ones accessible via
the perturbative treatment in the original representation. Specifically, the SOC frequency  
$\omega_{0}=\hbar/mr_{0}^{2}$ can now be considerably larger than the one accessible by means of the 
original perturbation approach~\cite{Cheng19arXiv}
applicable if the radius of the atomic cloud is much smaller than the SOC radius $r_0$. 

The frequency
of magnetic interaction $g_{F}B\left(r\right)/2=\omega r/r_{0}$ reaches
the driving frequency at $r=r_{0}$. Consequently one can keep $r_{0}$
fixed by simultaneously increasing both the magnetic field strength
and the driving frequency until the latter $\omega$ becomes sufficiently
large compared to the SOC frequency $\omega_{0}$ to fulfill the adiabatic
condition (\ref{eq:adiabatic_condition-Monopole-Specific}). 

In many-body systems there can be additional losses due to atom-atom
interactions. The two-body losses have been studied recently for a
periodic driving of the form $\hbar k_{0}\sigma_{z}z\cos\left(\omega t\right)$
\cite{Ketterle19Floquet-heating}. For the driving energy $\hbar\omega$
exceeding both the kinetic energy and the the SOC energy $\hbar^{2}k_{0}^{2}/2m$,
the two-body heating rate was shown to increase as $\sqrt{\hbar\omega}$
due to an increase of the final density of states and the energy of
the quantum absorbed $\hbar\omega$ \cite{Ketterle19Floquet-heating}.
Yet the probability of such absorption events is proportional to $1/\sqrt{\hbar\omega}$
and thus goes to zero in the limit of an infinitely large driving
frequency. Thus the many-body heating can be minimized by increasing
the driving frequency and removing from the trap a handful of very
fast atoms which absorb the driving quantum $\hbar\omega$. This is
essentially the idea of evaporative cooling \cite{Metcalf1999}. If
$\hbar\omega$ exceeds the trap depth, then those few atoms absorbing
the quantum of $\hbar\omega$ are automatically ejected from the trap
which becomes shallow at large energies / large distances. Such a
trap can be produced optically by focusing a number of laser beams
within the atomic cloud. The many-body effects will be explored in
more details in a separate study. 

\subsection{Implementation \label{subsec:Implementation}}

In analyzing the 3D SOC induced by the oscillating magnetic field  
$\mathbf{B}\left(\mathbf{r}\right)f\left(\omega t\right)$ we used a cylindrically or spherically
symmetric magnetic fields with the amplitude  $\mathbf{B}(\mathbf{r})$ given by Eqs.~(\ref{eq:B-linear}) or (\ref{eq:B-monopole}).
Such a magnetic field generally has a non-zero divergence and thus 
does not obey the Maxwell equation.
Yet one can produce interaction between the atom and the magnetic field
characterized by an effective magnetic field with a non-zero divergence. In particular, one can generate 
the spherically symmetric monopole field given by Eq.~(\ref{eq:B-monopole}). As explained in details in Appendix~\ref{sec:Appendix-A}, this can be done by taking the actual (real) magnetic field $\mathbf{B}\mathrm{_{real}}$ entering the original interaction operator $\hat{V}_{\mathrm{real}}\left(\mathbf{r},t\right)=g_{F}\hat{\mathbf{F}}\cdot\mathbf{B}_{\mathrm{\mathrm{real}}}\left(\mathbf{r},t\right)$
to contain a time-independent bias magnetic field $B_{0}\mathbf{e}_{z}$
and an extra spatially inhomogeneous magnetic field proportional to the time-periodic function $f(\omega t)$:
\begin{equation}
\mathbf{B}\mathrm{_{real}}\left(\mathbf{r},t\right)=B_{0}\mathbf{e}_{z}+[\mathbf{B}_{1}\left(\mathbf{r}\right)+\mathbf{B}_{2}\left(\mathbf{r}\right)\cos(\omega_{\mathrm{B}}t)]f(\omega t)\,\label{eq:B-orig-main-text}
\end{equation}
where the frequency $\omega_{\mathrm{B}}$ is considered to be in an exact resonance with magnetic level splitting induced by the bias field: $\omega_{\mathrm{B}} = g_F B_{0}$. Furthermore $\omega_{\mathrm{B}}$ is taken to be much larger than the frequency of the periodic driving: $\omega_{\mathrm{B}} \gg \omega$. Transforming the spin to the frame rotating at the frequency $\omega_{\mathrm{B}}$ via the unitary transformation $U$ given by Eq.~(\ref{eq:U}), one can then apply the rotating wave approximation (RWA) and neglect the fast oscillating terms $\propto \exp(\pm \mathrm{i}\omega_{\mathrm{B}}t)$ and $\propto \exp(\pm 2\mathrm{i}\omega_{\mathrm{B}}t)$ 
in the transformed interaction operator. 
Consequently one arrives at the interaction of the spin with the time-periodic effective magnetic field $\mathbf{B}\left(\mathbf{r}\right)f(\omega t)$ characterized by the amplitude (see the Appendix~\ref{sec:Appendix-A} for more details) 
\begin{equation}
\mathbf{B}\left(\mathbf{r}\right)=\frac{1}{2}B_{2x}\left(\mathbf{r}\right)\mathbf{e}_{x}+\frac{1}{2}B_{2y}\left(\mathbf{r}\right)\mathbf{e}_{y}+B_{1z}\left(\mathbf{r}\right)\mathbf{e}_{z}\,.\label{eq:B_eff-main text}
\end{equation}

In this way, one makes use of two unitary transformations. The transformation $U$ given by Eq.~(\ref{eq:U}) eliminates the fast spin precession  around the bias magnetic field at the frequency $\omega_{\mathrm{B}}$. This provides an effective coupling of the atom with the effective magnetic field given by Eq.~(\ref{eq:B_eff-main text}). Subsequently one applies another unitary transformation $R$ given by  Eq.~(\ref{eq:R-Definition}) which eliminates the interaction operator $g_{F}\hat{\mathbf{F}}\cdot\mathbf{B}\left(\mathbf{r}\right)f(\omega t)$. The position-dependence of the unitary operator $R=R\left(\mathbf{r},t\right)$ yields then the SOC due to the spin-dependent momentum shift in the kinetic energy term of Eq.(\ref{eq:W-definition}).    

Using such an approach one can create various effective magnetic fields $\mathbf{B}\left(\mathbf{r}\right)$ which do not necessarily obey the Maxwell equations. In particular, by taking $\mathbf{B}_{1}\left(\mathbf{r}\right)$ and $\mathbf{B}_{2}\left(\mathbf{r}\right)$ to be the quadrupole magnetic fields, one generates the cylindrically symmetric field or the spherically symmetric the monopole field  given by Eqs.~(\ref{eq:B-linear}) and (\ref{eq:B-monopole}) respectively, see Appendix~\ref{sec:Appendix-A} and ref. \cite{Pu18PRL}. 



The method works if the frequency of the magnetic level splitting
$\omega_{B}$ is much larger than the driving frequency
$\omega$. In the experiment \cite{Lin2011} with $^{87}\mathrm{Rb}$
atoms, the magnetic splitting frequency $\omega_{B}=2\pi\times4.81$MHz
far exceeds the recoil frequency $\omega_{\mathrm{rec}}$ which equals to
$2\pi\times 3.77\mathrm{kHz}$ for the $780\mathrm{nm}$ $5^{2}S_{1/2}\rightarrow5^{2}P_{3/2}$
optical transition. Note that the bias magnetic field induces also
the quadratic Zeeman shift (QZS) with a frequency equal approximately
to $6\omega_{rec}$ in the experiment \cite{Lin2011}. The unwanted
QZS can be reduced by decreasing the bias magnetic field. For example,
by reducing $\omega_{B}$ to $2\pi\times50$kHz, the QZS
decreases to $0.001\omega_{rec}$, which is in the range of the few
Hz and thus can be completely neglected. 

On the other hand, QZS can be used to produce an effective quasi-spin
$1/2$ system for atoms characterized by larger spins \cite{Han15PRA,Shteynas19PRL}.
For this the oscillating magnetic field should be in resonance with
a selected pair of magnetic levels, and the QZS makes the coupling
with other spin states out of resonance. For example, the magnetic
field could resonantly couple the $m_{F}=-1$ and $m_{F}=0$ states
of the $F=1$ manifold of $^{87}\mathrm{Rb}$ or $^{23}\mathrm{Na}$
atoms \cite{Shteynas19PRL} representing the quasi-spin up and down
states, leaving the detuned $m_{F}=1$ state uncoupled, similar to
experiments on the Raman-induced SOC \cite{Lin2011}. To create the
quasi-spin $1/2$ system, the driving frequency $\omega$ should be
smaller than the frequency of the quadratic Zeeman shift $\omega_{q}$.
As mentioned above, $\omega_{q}$ equals to a few recoil frequencies
in the experiment \cite{Lin2011}. The QZS can be further increased
by increasing the bias magnetic field to reach the condition $\omega_{q}\gg\omega_{\mathrm{rec}}$,
so the driving frequency $\omega$ can be of the order of the recoil
frequency $\omega_{\mathrm{rec}}$ or a little above it. Therefore
the SOC frequency $\omega_{0}=\hbar/mr_{0}^{2}$ should then be smaller
than the recoil frequency to fulfill the adiabatic requirements (\ref{eq:adiabatic_condition-Monopole-Specific}).
For example, for $^{87}$Rb atoms, $\omega_{0}$ could be of the order
of a few tens to a few hundreds of Hz, which is comparable to
typical trapping frequencies. This provides the SOC radius $r_{0}$
of the order of a few optical wave-lengths. In this way, by taking
a width of an optical trap to be of the order of ten optical wave-lengths,
the SOC radius $r_{0}$ would be within the trapped atomic cloud,
and the atoms would experience a substantial SOC in the cloud.

\section{Concluding remarks\label{sec:Concluding-remarks}}

We have considered a method of creating the non-Abelian geometric
potential and thus the 3D SOC for the center-of-mass motion of the
particle subjected to periodic driving. The periodic perturbation
is a product of a position-dependent Hermitian operator $\hat{V}\left(\mathbf{r}\right)$
and a fast oscillating periodic function $f\left(\omega t\right)$
with a zero average. To have a significant SOC, we have analyzed a
situation where the matrix elements of the periodic operator $\hat{V}\left(\mathbf{r}\right)f\left(\omega t\right)$
are not necessarily small compared to the driving energy $\hbar\omega$,
so that one cannot apply the high frequency expansion of the effective
Floquet Hamiltonian \cite{Goldman2014,Eckardt2015,Bukov2015,Novicenko2017}
in the original representation.  To by-pass the problem, we have applied
a unitary transformation which eliminates the original periodic perturbation
and yields an oscillating vector potential term. The resulting periodic
perturbation is no longer proportional to the driving frequency $\omega$,
so the perturbation treatment is applicable to much stronger driving
(and thus over a larger range of distances) than in the original representation.
 We have considered a situation where $\hat{V}\left(\mathbf{r}\right)$
depends on internal (spin or quasi-spin) degrees of freedom of the
particle, and thus the operator $\hat{V}\left(\mathbf{r}\right)$
does not necessarily commute with itself at different positions: $\left[\hat{V}\left(\mathbf{r}\right),\hat{V}\left(\mathbf{r}^{\prime}\right)\right]\ne0$.
Consequently the adiabatic evolution of the system within a Floquet
band is accompanied by a non-Abelian (non-commuting) geometric vector
potential providing the 3D SOC. 

The periodic driving plays a vital role in our analysis. Without the periodic driving the interaction is given by a time-independent operator $V\left(\mathbf{r}\right)$,  like in ref. \cite{Pu18PRL}. In that case the spin adiabatically follows the magnetic field and is fully polarized along local field direction \cite{Pu18PRL}. In the present situation, the interaction operator $V\left(\mathbf{r}\right)f(\omega t)$ contains also the periodic function $f(\omega t)$ with the zero average. Therefore the spin no longer adiabatically follows the magnetic field and thus is no longer polarized along the local magnetic field. Specifically, the operator $V(r)f(\omega t)$ provides fully degenerate Floquet bands \cite{Novicenko19PRA}. The position-dependence of these Floquet eigenstates yields the spin-dependent momentum shift and thus the spin-orbit coupling. Therefore the current situation is very different from the case of the time-independent interaction $V\left(\mathbf{r}\right)$ where the eigen-energies of $V\left(\mathbf{r}\right)$  are non-degenerate and thus the spin is polarized \cite{Pu18PRL}. 


The general formalism has been illustrated by analyzing motion of a spinful atom in a magnetic field oscillating in time, subsequently concentrating on a spin-1/2 atom in a cylindrically symmetric magnetic field. This yields the SOC involving coupling between the spin and the orbital motion described by all three components of the OAM operator $\mathbf{L}$. In particular, the time-oscillating monopole-type magnetic
field $\mathbf{B}\propto\mathbf{r}$ generates the 3D SOC of the $\mathbf{L}\cdot\mathbf{F}$
form. The strength of this SOC goes as $1/r^{2}$ for larger distances,
rather than $1/r^{3}$, as for electrons in the Coulomb potential.
 Such a long-ranged SOC significantly affects not only the lower
states of the trapped atom, but also the higher ones. In particular,
the states with $l=j\pm1/2$ are nearly degenerate with fix $j$ and
$n_{r}$ for an atom characterized by the (quasi-)spin $1/2$. In the presence of a harmonic external trapping potential, the
ground state with $j=1/2$ and $l=0$ has a slightly lower energy
than the one with $j=1/2$ and $l=1$. The situation can be reversed
by adding an extra anti-trapping potential for small $r$, which makes
the ground state of the system to be characterized by the orbital
quantum number $l=1$. The $l=1$ ground state is affected by the
SOC, which can lead to interesting many-body phases to be explored
elsewhere. If the atom possesses higher spin, more complicated SOC terms can be generated. In this situation, the spin-dependent scalar potential featured in Eq.~(\ref{eq:W^0-result-general}) for the effective Hamiltonian $\hat{W}^{\left(0\right)}$, can lead to an additional coupling between orbit and spin or spin tensor. We plan to address this topic in a future study.  

Previously coupling between the spin and the linear momentum $\mathbf{p}$
was considered for ultracold atoms using time-periodic sequences of
magnetic pulses \cite{Anderson2013,Xu2013,Luo16Sci_Rep,Shteynas19PRL}.
To generate the 2D or 3D coupling between the spin and linear momentum
$\mathbf{p}$, the periodic driving involves rapid changes of the
magnetic field direction \cite{Anderson2013,Xu2013}. This would be
rather complicated to implement experimentally. It is much more straightforward
to generate a sizable coupling between the spin and the OAM using
the method considered here. For this one applies a simpler magnetic
field with properly designed spatial and temporal profiles rather
than the alternating magnetic pulses. Therefore the current scheme
is more realistic and can be implemented using experimental techniques
currently available. 

The 2D and 3D SOC can be also generated optically by using a degeneracy
of eigenstates of the atom-light coupling operator \cite{Ruseckas2005,Stanescu2007,Jacob2007,Juzeliunas2008PRA,Stanescu2008,Campbell2011,Anderson2012PRL,Huang16NP,Meng16PRL,Spielman19_2D_SOC},
which involves a considerable amount of efforts \cite{Campbell2011,Huang16NP,Meng16PRL,Spielman19_2D_SOC}.
The present approach does not require such a degeneracy. Instead,
the periodic driving yields degenerate Floquet states for the time-periodic
interaction operator $\hat{V}\left(\mathbf{r}\right)f\left(\omega t\right)$
in a straightforward way \cite{Novicenko19PRA}. 
The spatial and temporal
dependence of the operator $\hat{V}\left(\mathbf{r}\right)f\left(\omega t\right)$
provides the oscillating vector potential and hence the SOC. 
The present
SOC has also some similarities to the coupling between the spin and
OAM induced by Raman laser beams carying optical vortices \cite{Marzlin97PRL,Ruostekoski04PRL,Nandi04PRA,Juzeliunas2004,Juzeliunas2005,Bigelow08PRA,Bigelow09PRL,Pu15PRA,Qu15PRA,Lin18PRL-a,Lin18PRL-b,Zhang19PRL}. Yet, unlike the Raman case, now the coupling between the spin and OAM is described by all three OAM projections $L_{x}$, $L_{y}$ and $L_{z}$, so the SOC is truly three dimensional. Furthermore, our current scheme does not involve laser fields, and hence does not suffer from Raman-induced heating.

\section*{Acknowledgments}

We acknowledge helpful discussions with Arnoldas Deltuva, Julius Ruseckas
and Congjun Wu. This research has received funding from European Social
Fund (project No. 09.3.3--LMT--K--712--02--0065) under grant agreement with the Research Council of Lithuania (LMTLT). HP acknowledges support from the US NSF and the Welch Foundation (Grant No. C-1669).

\appendix

\section*{Appendix}

\section{Appendix A: Generation of effective magnetic field violating the Maxwell equations
\label{sec:Appendix-A}}

 In this Appendix we will explain a way of obtaining the coupling
of an atom with an effective magnetic field which does not necessarily
have a zero divergence and / or a zero curl \cite{Anderson2013,Pu18PRL,Shteynas19PRL}.
In particular, the effective magnetic field can describe a monopole-type
field \cite{Pu18PRL}.

\subsection{Original problem}

In the original problem, the atom is coupled with an actual (real) magnetic field
$\mathbf{B}\mathrm{_{real}}\left(\mathbf{r},t\right)$ obeying the
Maxwell equations. The corresponding interaction operator reads

\begin{equation}
\hat{V}_{\mathrm{real}}\left(\mathbf{r},t\right)=g_{F}\hat{\mathbf{F}}\cdot\mathbf{B}_{\mathrm{\mathrm{real}}}\left(\mathbf{r},t\right)\,.\label{eq:V_conv}
\end{equation}
Let us consider a magnetic field 
\begin{equation}
\mathbf{B}\mathrm{_{real}}\left(\mathbf{r},t\right)=B_{0}\mathbf{e}_{z}+[\mathbf{B}_{1}\left(\mathbf{r}\right)+\mathbf{B}_{2}\left(\mathbf{r}\right)\cos(\omega_{\mathrm{B}}t)]f(\omega t)\,\label{eq:B-orig}
\end{equation}
which is composed of a constant bias magnetic field $B_{0}\mathbf{e}_{z}$
pointing along a unit Cartesian vector $\mathbf{e}_{z}$, and a time
dependent term proportional to the periodic function $f(\omega t)=f(\omega t+2\pi)$
oscillating in time with the frequency $\omega$. The latter term
contains also a contribution $\propto \cos(\omega_{\mathrm{B}}t)$
oscillating with a frequency $\omega_{\mathrm{B}}$ much larger
than the driving frequency $\omega$. The frequency $\omega_{\mathrm{B}}$
is taken to be in an exact resonance with the frequency of the spitting
between be the spin states induced by the bias magnetic field, so
that 
\begin{equation}
\omega_{\mathrm{B}}=g_{F}B_{0}\quad\mathrm{with}\quad\omega_{\mathrm{B}}\gg\omega\,.\label{eq:omega_Bias}
\end{equation}
Replacing $\cos(\omega_{\mathrm{B}}t)$ by $\cos(\omega_{\mathrm{B}}t+\phi)$ one can also include the phase shift $\phi$ of the fast
oscillations in Eq.~(\ref{eq:B-orig}). This can be done within
the present framework by changing the origin of time $t\rightarrow t+\phi/\omega_{\mathrm{B}}$
and redefining the periodic function: $f(\omega t)\rightarrow f(\omega t-\phi\omega/\omega_{\mathrm{B}})$. 

\subsection{Transformed representation}

The bias magnetic field can be eliminated via a unitary transformation
which rotates the spin at the frequency $\omega_{\mathrm{B}}$
around the $z$ axis 
\begin{equation}
\hat{U}=\exp\left(-i\omega_{\mathrm{B}}t\hat{F}_{z}/\hbar\right)\,.\label{eq:U}
\end{equation}
The transformed Hamiltonian 
\begin{equation}
\hat{V}_{\mathrm{transf}}\left(\mathbf{r},t\right)=\hat{U}^{\dagger}\hat{V}_{\mathrm{real}}\left(\mathbf{r},t\right)\hat{U}-i\hbar\hat{U}^{\dagger}\partial_{t}\hat{U}\,\label{eq:V_transf-definition}
\end{equation}
no longer contains the bias term $g_{F}B_{0}\hat{F}_{z}$
which is canceled by the term $-i\hbar\hat{U}^{\dagger}\partial_{t}\hat{U}$,
giving  
\begin{equation}
\hat{V}_{\mathrm{transf}}\left(\mathbf{r},t\right)=g_{F}\hat{\tilde{\mathbf{F}}}\cdot[\mathbf{B}_{1}\left(\mathbf{r}\right)+\mathbf{B}_{2}\left(\mathbf{r}\right)\cos(\omega_{\mathrm{B}}t)]f(\omega t)\,.\label{eq:eq:V(r,t)}
\end{equation}
Here $\hat{\tilde{\mathbf{F}}}=\hat{U}^{\dagger}\hat{\mathbf{F}}\hat{U}$
is the transformed spin operator. Its $z$ component $\hat{\tilde{F}}_{z}=\hat{F}_{z}$
is not affected by the transformation. On the other hand, the $x$
and $y$ components of $\hat{\tilde{\mathbf{F}}}$ rotate at the
frequency $\omega_{\mathrm{B}}$: 
\begin{equation}
\hat{\tilde{F}}_{x}=\hat{F}_{x}\cos(\omega_{\mathrm{B}}t)-\hat{F}_{y}\sin(\omega_{\mathrm{B}}t)\,,\label{eq:F_x-tilde}
\end{equation}
\begin{equation}
\hat{\tilde{F}}_{y}=\hat{F}_{x}\sin(\omega_{\mathrm{B}}t)+\hat{F}_{y}\cos(\omega_{\mathrm{B}}t)\,.\label{eq:F_y-tilde}
\end{equation}

\subsection{Effective interaction operator and effective magnetic field}

Since $\omega_{\mathrm{B}} \gg\omega$, one can apply the rotating wave approximation (RWA) replacing the fast oscillating term $\hat{\tilde{\mathbf{F}}}\cdot\mathbf{B}_{1}\left(\mathbf{r}\right)f(\omega t)$
by its temporal average $\hat{F}_{z}B_{1z}\left(\mathbf{r}\right)f(\omega t)$
in Eq.~(\ref{eq:eq:V(r,t)}). In a similar manner, the fast oscillating
term $\hat{\tilde{\mathbf{F}}}\cdot\mathbf{B}_{2}\left(\mathbf{r}\right)\cos(\omega_{\mathrm{B}}t)$
averages to $\frac{1}{2}\hat{F}_{x}B_{2x}\left(\mathbf{r}\right)+\frac{1}{2}\hat{F}_{y}B_{2y}\left(\mathbf{r}\right)$
in Eq.~(\ref{eq:eq:V(r,t)}). Therefore by averaging $\hat{V}_{\mathrm{transf}}\left(\mathbf{r},t\right)$
over a small period $T_{\mathrm{B}}=2\pi/\omega_{\mathrm{B}}$,
one arrive at an effective interaction operator
\begin{equation}
\left\langle  \hat{V}_{\mathrm{transf}}\left(\mathbf{r},t\right)  \right\rangle = g_{F}\hat{\mathbf{F}}\cdot\mathbf{B}\left(\mathbf{r}\right)f(\omega t)\,,\label{eq:V_transf}
\end{equation}
where 
\begin{equation}
\mathbf{B}\left(\mathbf{r}\right)=\frac{1}{2}B_{2x}\left(\mathbf{r}\right)\mathbf{e}_{x}+\frac{1}{2}B_{2y}\left(\mathbf{r}\right)\mathbf{e}_{y}+B_{1z}\left(\mathbf{r}\right)\mathbf{e}_{z}\,\label{eq:B_eff}
\end{equation}
is a time-independent amplitude of the oscillating effective magnetic field.

Although the original
fields $\mathbf{B}_{1}\left(\mathbf{r}\right)$ and $\mathbf{B}_{2}\left(\mathbf{r}\right)$
entering Eq.~(\ref{eq:B-orig}) are divergence free,
the effective field $\mathbf{B}\left(\mathbf{r}\right)$ featured in  Eq.~(\ref{eq:V_transf}) 
does not necessarily obey this requirement. For example,
by taking $\mathbf{B}_{1}\left(\mathbf{r}\right)=0$ and $\mathbf{B}_{2}\left(\mathbf{r}\right)\propto x\mathbf{e}_{x}-z\mathbf{e}_{z}$,
one arrives at an effective magnetic field pointing along the $x$
axis $\mathbf{B}\left(\mathbf{r}\right)\propto x\mathbf{e}_{x}$ with
$\nabla \cdot \mathbf{B}\left(\mathbf{r}\right)\ne0$ \cite{Anderson2013}.

\subsection{Cylindrical and Monopole fields}

Suppose now that both $\mathbf{B}_{1}\left(\mathbf{r}\right)$ and $\mathbf{B}_{2}\left(\mathbf{r}\right)$
entering the original magnetic field (\ref{eq:B-orig}) are the quadrupole
fields: 
\begin{equation}
\mathbf{B}_{1}\left(\mathbf{r}\right)=-\frac{1}{2}\alpha_{z}\left(x\mathbf{e}_{x}+y\mathbf{e}_{y}-2z\mathbf{e}_{z}\right)\,,\label{eq:B_1}
\end{equation}
\begin{equation}
\mathbf{B}_{2}\left(\mathbf{r}\right)=2\alpha_{\bot}\left(x\mathbf{e}_{x}+y\mathbf{e}_{y}-2z\mathbf{e}_{z}\right)\,.\label{eq:B_2}
\end{equation}
where $\alpha_{z}$ and $\alpha_{\bot}$ characterize the strengths of the constituting
magnetic fields. In that case the effective magnetic field (\ref{eq:B-orig}) generally contains all three Cartesian components
and is given by: 
\begin{equation}
\mathbf{B}\left(\mathbf{r}\right)=\alpha_{\bot}\left(x\mathbf{e}_{x}+y\mathbf{e}_{y}\right)+\alpha_{z}z\mathbf{e}_{z}\,.\label{eq:B_eff-1}
\end{equation}
The field $\mathbf{B}\left(\mathbf{r}\right)$ has a non-zero divergence
$\nabla\cdot\mathbf{B}\left(\mathbf{r}\right)\ne0$ as long as $\alpha_{z}\ne-2\alpha_{\bot}$.
In particular, for $\alpha_{z}=\alpha_{\bot}=\alpha$ one arrives
at the monopole-type \cite{Pu18PRL} effective magnetic field for
which $\mathbf{B}\left(\mathbf{r}\right)=\alpha\mathbf{r}$.

\section{Calculation of $\left\langle \left[\mathbf{\hat{A}}\left(\mathbf{r},\omega t\right)\right]^{2}\right\rangle $ term\label{sec:A^2-term}}

Let us analyze the $\left\langle \left[\mathbf{\hat{A}}\left(\mathbf{r},\omega t\right)\right]^{2}\right\rangle $
term featured in Eq.~(\ref{eq:V_total}) for the scalar potential
$\hat{V}_{\mathrm{total}}\left(\mathbf{r}\right)$. For this let us
rewrite Eq.~(\ref{eq:A^(0)-spin-solution-general}) in vector notations:
\begin{equation}
\hat{\mathbf{A}}\left(\mathbf{r},\omega t\right)=a\mathcal{F}\hat{\mathbf{d}}_{1}-\sin\left(a\mathcal{F}\right)\hat{\mathbf{d}}_{2}-\left[\cos\left(a\mathcal{F}\right)-1\right]\hat{\mathbf{d}}_{3}\,,\label{eq:A-spin-vector-general-alternative-App}
\end{equation}
where $\hat{\mathbf{d}}_{1}$, $\hat{\mathbf{d}}_{2}$ and $\hat{\mathbf{d}}_{3}$ are
vectors with the Cartesian components: 
\begin{align}
\hat{d}_{1u}&=\frac{\left(\mathbf{B}\cdot\hat{\mathbf{F}}\right)\left(\mathbf{B}\cdot\partial\mathbf{B}/\partial u\right)}{B^{3}},\\
\hat{d}_{2u}&=\frac{\left[\mathbf{B}\times\left(\mathbf{B}\times\hat{\mathbf{F}}\right)\right]\cdot\partial\mathbf{B}/\partial u}{B^{3}}, \\
\hat{d}_{3u}&=\frac{\left(\mathbf{B}\times\hat{\mathbf{F}}\right)\cdot\partial\mathbf{B}/\partial u}{B^{2}}.
\label{eq:d_12-general-App}
\end{align}
Since $\mathcal{F}\left(\omega t\right)$ is an odd function of time
${\mathcal{F}\left(\omega t\right)=-\mathcal{F}\left(-\omega t\right)}$,
the terms containing $\hat{\mathbf{d}}_{1}\cdot\hat{\mathbf{d}}_{3}$ and $\hat{\mathbf{d}}_{2}\cdot\hat{\mathbf{d}}_{3}$
average to zero in $\left\langle \left[\mathbf{\hat{A}}\left(\mathbf{r},\omega t\right)\right]^{2}\right\rangle $.
Furthermore for the spherically symmetric monopole magnetic field
with $\mathbf{B}=\alpha\mathbf{r}$, one has
\begin{equation}
\hat{\mathbf{d}}_{1}=\frac{\left(\mathbf{r}\cdot\hat{\mathbf{F}}\right)\mathbf{r}}{r^{3}},\quad\hat{\mathbf{d}}_{2}=\frac{\mathbf{r}\times\left(\mathbf{r}\times\hat{\mathbf{F}}\right)}{r^{3}},\quad\hat{\mathbf{d}}_{3}=\frac{\left(\mathbf{r}\times\hat{\mathbf{F}}\right)}{r^{2}},
\label{eq:d_123-monopole-App}
\end{equation}
giving $\hat{\mathbf{d}}_{1}\cdot\hat{\mathbf{d}}_{2}=0$, as well as
\begin{equation}
\hat{d}_{1}^{2}=\frac{\left(\mathbf{r}\cdot\hat{\mathbf{F}}\right)^{2}}{r^{4}}\,\quad\mathrm{and}\quad \hat{d}_{2}^{2}=\hat{d}_{3}^{2}=\frac{F^{2}r^{2}-\left(\mathbf{r}\cdot\hat{\mathbf{F}}\right)^{2}}{r^{4}}\,.\label{eq:d_1^2-and-d_2^2-Monopole-result-App}
\end{equation}
Consequently one arrives at
\begin{equation}
\left\langle \left[\mathbf{\hat{A}}\left(\mathbf{r},\omega t\right)\right]^{2}\right\rangle =a^{2}\left\langle \mathcal{F}^{2}\right\rangle \hat{d}_{1}^{2}+2\hat{d}_{2}^{2}\left[1-\left\langle \cos\left(a\mathcal{F}\right)\right\rangle \right]\,.\label{eq:A^2-average-spin-general-result-App}
\end{equation}
Equations (\ref{eq:d_1^2-and-d_2^2-Monopole-result-App}) and (\ref{eq:A^2-average-spin-general-result-App})
provide the following explicit result for the monopole field
\begin{align}
\left\langle \left[\mathbf{\hat{A}}\left(\mathbf{r},\omega t\right)\right]^{2}\right\rangle =&\frac{4}{r_{0}^{2}}\left\langle \mathcal{F}^{2}\right\rangle \frac{\left(\mathbf{r}\cdot\hat{\mathbf{F}}\right)^{2}}{r^{2}} \nonumber \\
&+\frac{2-2\left\langle \cos\left(2r\mathcal{F}/r_{0}\right)\right\rangle }{r^{2}}\frac{F^{2}r^{2}-\left(\mathbf{r}\cdot\hat{\mathbf{F}}\right)^{2}}{r^{2}}.\label{eq:A^2-spin-solution-Monopole-vector-result-averaged}
\end{align}

The relation of the form (\ref{eq:A^2-average-spin-general-result-App})
holds also for the spin-$1/2$ atom ($\hat{\mathbf{F}}=\frac{\hbar}{2}\hat{\boldsymbol{\sigma}}$)
in a cylindrically symmetric magnetic field (\ref{eq:B-linear}) for
which $\hat{d}_{2}^{2}=\hat{d}_{3}^{2}$ and $\hat{\mathbf{d}}_{1}\cdot\hat{\mathbf{d}}_{2}+\hat{\mathbf{d}}_{2}\cdot\hat{\mathbf{d}}_{1}=0$.
In such a situation
\begin{equation}
d_{1}^{2}=\frac{\hbar^{2}}{4}\frac{\left(\alpha_{\bot}^{4}\rho^{2}+\alpha_{z}^{4}z^{2}\right)}{\left(\alpha_{z}^{2}z^{2}+\alpha_{\bot}^{2}\rho^{2}\right)^{2}}\,\label{eq:d_1^2_result-spin1/2-App}
\end{equation}
and

\begin{equation}
d_{2}^{2}=\frac{\hbar^{2}\alpha_{\bot}^{2}}{4}\frac{2\alpha_{z}^{2}z^{2}+\left(\alpha_{z}^{2}+\alpha_{\bot}^{2}\right)\rho^{2}}{\left(\alpha_{z}^{2}z^{2}+\alpha_{\bot}^{2}\rho^{2}\right)^{2}}\,.\label{eq:d_2^2_result-spin1/2-App}
\end{equation}
This is consistent with the spherically symmetric result (\ref{eq:d_1^2-and-d_2^2-Monopole-result-App})
for the spin-$1/2$ atom\textbf{.}

\bibliographystyle{apsrev4-1}
\bibliography{Floquet-extra-space2019}

\begin{thebibliography}{99}%
\makeatletter
\providecommand \@ifxundefined [1]{%
 \@ifx{#1\undefined}
}%
\providecommand \@ifnum [1]{%
 \ifnum #1\expandafter \@firstoftwo
 \else \expandafter \@secondoftwo
 \fi
}%
\providecommand \@ifx [1]{%
 \ifx #1\expandafter \@firstoftwo
 \else \expandafter \@secondoftwo
 \fi
}%
\providecommand \natexlab [1]{#1}%
\providecommand \enquote  [1]{``#1''}%
\providecommand \bibnamefont  [1]{#1}%
\providecommand \bibfnamefont [1]{#1}%
\providecommand \citenamefont [1]{#1}%
\providecommand \href@noop [0]{\@secondoftwo}%
\providecommand \href [0]{\begingroup \@sanitize@url \@href}%
\providecommand \@href[1]{\@@startlink{#1}\@@href}%
\providecommand \@@href[1]{\endgroup#1\@@endlink}%
\providecommand \@sanitize@url [0]{\catcode `\\12\catcode `\$12\catcode
  `\&12\catcode `\#12\catcode `\^12\catcode `\_12\catcode `\%12\relax}%
\providecommand \@@startlink[1]{}%
\providecommand \@@endlink[0]{}%
\providecommand \url  [0]{\begingroup\@sanitize@url \@url }%
\providecommand \@url [1]{\endgroup\@href {#1}{\urlprefix }}%
\providecommand \urlprefix  [0]{URL }%
\providecommand \Eprint [0]{\href }%
\providecommand \doibase [0]{http://dx.doi.org/}%
\providecommand \selectlanguage [0]{\@gobble}%
\providecommand \bibinfo  [0]{\@secondoftwo}%
\providecommand \bibfield  [0]{\@secondoftwo}%
\providecommand \translation [1]{[#1]}%
\providecommand \BibitemOpen [0]{}%
\providecommand \bibitemStop [0]{}%
\providecommand \bibitemNoStop [0]{.\EOS\space}%
\providecommand \EOS [0]{\spacefactor3000\relax}%
\providecommand \BibitemShut  [1]{\csname bibitem#1\endcsname}%
\let\auto@bib@innerbib\@empty
\bibitem [{\citenamefont {S{\o}rensen}\ \emph {et~al.}(2005)\citenamefont
  {S{\o}rensen}, \citenamefont {Demler},\ and\ \citenamefont
  {Lukin}}]{sorensen05}%
  \BibitemOpen
  \bibfield  {author} {\bibinfo {author} {\bibfnamefont {A.~S.}\ \bibnamefont
  {S{\o}rensen}}, \bibinfo {author} {\bibfnamefont {E.}~\bibnamefont {Demler}},
  \ and\ \bibinfo {author} {\bibfnamefont {M.~D.}\ \bibnamefont {Lukin}},\
  }\href {\doibase 10.1103/PhysRevLett.94.086803} {\bibfield  {journal}
  {\bibinfo  {journal} {Phys. Rev. Lett.}\ }\textbf {\bibinfo {volume} {94}},\
  \bibinfo {pages} {086803} (\bibinfo {year} {2005})}\BibitemShut {NoStop}%
\bibitem [{\citenamefont {Oka}\ and\ \citenamefont {Aoki}(2009)}]{Oka2009}%
  \BibitemOpen
  \bibfield  {author} {\bibinfo {author} {\bibfnamefont {T.}~\bibnamefont
  {Oka}}\ and\ \bibinfo {author} {\bibfnamefont {H.}~\bibnamefont {Aoki}},\
  }\href {\doibase 10.1103/PhysRevB.79.081406} {\bibfield  {journal} {\bibinfo
  {journal} {Phys. Rev. B}\ }\textbf {\bibinfo {volume} {79}},\ \bibinfo
  {pages} {081406(R)} (\bibinfo {year} {2009})}\BibitemShut {NoStop}%
\bibitem [{\citenamefont {Kitagawa}\ \emph {et~al.}(2010)\citenamefont
  {Kitagawa}, \citenamefont {Berg}, \citenamefont {Rudner},\ and\ \citenamefont
  {Demler}}]{Kitagawa2010}%
  \BibitemOpen
  \bibfield  {author} {\bibinfo {author} {\bibfnamefont {T.}~\bibnamefont
  {Kitagawa}}, \bibinfo {author} {\bibfnamefont {E.}~\bibnamefont {Berg}},
  \bibinfo {author} {\bibfnamefont {M.}~\bibnamefont {Rudner}}, \ and\ \bibinfo
  {author} {\bibfnamefont {E.}~\bibnamefont {Demler}},\ }\href {\doibase
  10.1103/PhysRevB.82.235114} {\bibfield  {journal} {\bibinfo  {journal} {Phys.
  Rev. B}\ }\textbf {\bibinfo {volume} {82}},\ \bibinfo {pages} {235114}
  (\bibinfo {year} {2010})}\BibitemShut {NoStop}%
\bibitem [{\citenamefont {Lindner}\ \emph {et~al.}(2011)\citenamefont
  {Lindner}, \citenamefont {Refael},\ and\ \citenamefont
  {Galitski}}]{Galitski2011NP}%
  \BibitemOpen
  \bibfield  {author} {\bibinfo {author} {\bibfnamefont {N.~H.}\ \bibnamefont
  {Lindner}}, \bibinfo {author} {\bibfnamefont {G.}~\bibnamefont {Refael}}, \
  and\ \bibinfo {author} {\bibfnamefont {V.}~\bibnamefont {Galitski}},\ }\href
  {\doibase 10.1038/nphys1926} {\bibfield  {journal} {\bibinfo  {journal} {Nat.
  Phys.}\ }\textbf {\bibinfo {volume} {7}},\ \bibinfo {pages} {490} (\bibinfo
  {year} {2011})}\BibitemShut {NoStop}%
\bibitem [{\citenamefont {Rudner}\ \emph {et~al.}(2013)\citenamefont {Rudner},
  \citenamefont {Lindner}, \citenamefont {Berg},\ and\ \citenamefont
  {Levin}}]{Rudner2013}%
  \BibitemOpen
  \bibfield  {author} {\bibinfo {author} {\bibfnamefont {M.~S.}\ \bibnamefont
  {Rudner}}, \bibinfo {author} {\bibfnamefont {N.~H.}\ \bibnamefont {Lindner}},
  \bibinfo {author} {\bibfnamefont {E.}~\bibnamefont {Berg}}, \ and\ \bibinfo
  {author} {\bibfnamefont {M.}~\bibnamefont {Levin}},\ }\href {\doibase
  10.1103/PhysRevX.3.031005} {\bibfield  {journal} {\bibinfo  {journal} {Phys.
  Rev. X}\ }\textbf {\bibinfo {volume} {3}},\ \bibinfo {pages} {031005}
  (\bibinfo {year} {2013})}\BibitemShut {NoStop}%
\bibitem [{\citenamefont {Nathan}\ and\ \citenamefont
  {Rudner}(2015)}]{Nathan15NJP}%
  \BibitemOpen
  \bibfield  {author} {\bibinfo {author} {\bibfnamefont {F.}~\bibnamefont
  {Nathan}}\ and\ \bibinfo {author} {\bibfnamefont {M.~S.}\ \bibnamefont
  {Rudner}},\ }\href@noop {} {\bibfield  {journal} {\bibinfo  {journal} {New J.
  Phys.}\ }\textbf {\bibinfo {volume} {17}},\ \bibinfo {pages} {125014}
  (\bibinfo {year} {2015})}\BibitemShut {NoStop}%
\bibitem [{\citenamefont {Budich}\ \emph {et~al.}(2017)\citenamefont {Budich},
  \citenamefont {Hu},\ and\ \citenamefont {Zoller}}]{Budich17PRL}%
  \BibitemOpen
  \bibfield  {author} {\bibinfo {author} {\bibfnamefont {J.~C.}\ \bibnamefont
  {Budich}}, \bibinfo {author} {\bibfnamefont {Y.}~\bibnamefont {Hu}}, \ and\
  \bibinfo {author} {\bibfnamefont {P.}~\bibnamefont {Zoller}},\ }\href@noop {}
  {\bibfield  {journal} {\bibinfo  {journal} {Phys. Phys. Lett.}\ }\textbf
  {\bibinfo {volume} {118}},\ \bibinfo {pages} {105302} (\bibinfo {year}
  {2017})},\ \Eprint {http://arxiv.org/abs/1608.05096} {arXiv:1608.05096}
  \BibitemShut {NoStop}%
\bibitem [{\citenamefont {Eckardt}(2017)}]{Eckardt17RMP}%
  \BibitemOpen
  \bibfield  {author} {\bibinfo {author} {\bibfnamefont {A.}~\bibnamefont
  {Eckardt}},\ }\href@noop {} {\bibfield  {journal} {\bibinfo  {journal} {Rev.
  Mod. Phys.}\ }\textbf {\bibinfo {volume} {89}},\ \bibinfo {pages} {011004}
  (\bibinfo {year} {2017})}\BibitemShut {NoStop}%
\bibitem [{\citenamefont {Weinberg}\ \emph {et~al.}(2017)\citenamefont
  {Weinberg}, \citenamefont {Bukov}, \citenamefont {D'Alessio}, \citenamefont
  {Polkovnikov}, \citenamefont {Vajna},\ and\ \citenamefont
  {Kolodrubetz}}]{Weinberg17PR}%
  \BibitemOpen
  \bibfield  {author} {\bibinfo {author} {\bibfnamefont {P.}~\bibnamefont
  {Weinberg}}, \bibinfo {author} {\bibfnamefont {M.}~\bibnamefont {Bukov}},
  \bibinfo {author} {\bibfnamefont {L.}~\bibnamefont {D'Alessio}}, \bibinfo
  {author} {\bibfnamefont {A.}~\bibnamefont {Polkovnikov}}, \bibinfo {author}
  {\bibfnamefont {S.}~\bibnamefont {Vajna}}, \ and\ \bibinfo {author}
  {\bibfnamefont {M.}~\bibnamefont {Kolodrubetz}},\ }\href@noop {} {\bibfield
  {journal} {\bibinfo  {journal} {Phys. Rep.}\ }\textbf {\bibinfo {volume}
  {688}},\ \bibinfo {pages} {1} (\bibinfo {year} {2017})},\ \Eprint
  {http://arxiv.org/abs/1606.02229} {arXiv:1606.02229} \BibitemShut {NoStop}%
\bibitem [{\citenamefont {Asteria}\ \emph {et~al.}(2019)\citenamefont
  {Asteria}, \citenamefont {Tran}, \citenamefont {Ozawa}, \citenamefont
  {Tarnowski}, \citenamefont {Rem}, \citenamefont {Fl{\"a}schner},
  \citenamefont {Sengstock}, \citenamefont {Goldman},\ and\ \citenamefont
  {Weitenberg}}]{Weitenberb19NatPhys}%
  \BibitemOpen
  \bibfield  {author} {\bibinfo {author} {\bibfnamefont {L.}~\bibnamefont
  {Asteria}}, \bibinfo {author} {\bibfnamefont {D.~T.}\ \bibnamefont {Tran}},
  \bibinfo {author} {\bibfnamefont {T.}~\bibnamefont {Ozawa}}, \bibinfo
  {author} {\bibfnamefont {M.}~\bibnamefont {Tarnowski}}, \bibinfo {author}
  {\bibfnamefont {B.~S.}\ \bibnamefont {Rem}}, \bibinfo {author} {\bibfnamefont
  {N.}~\bibnamefont {Fl{\"a}schner}}, \bibinfo {author} {\bibfnamefont
  {K.}~\bibnamefont {Sengstock}}, \bibinfo {author} {\bibfnamefont
  {N.}~\bibnamefont {Goldman}}, \ and\ \bibinfo {author} {\bibfnamefont
  {C.}~\bibnamefont {Weitenberg}},\ }\href {\doibase 10.1038/s41567-019-0417-8}
  {\bibfield  {journal} {\bibinfo  {journal} {Nature Physics}\ }\textbf
  {\bibinfo {volume} {15}},\ \bibinfo {pages} {449} (\bibinfo {year}
  {2019})}\BibitemShut {NoStop}%
\bibitem [{\citenamefont {\"Unal}\ \emph {et~al.}(2019)\citenamefont {\"Unal},
  \citenamefont {Seradjeh},\ and\ \citenamefont {Eckardt}}]{Unal19PRL}%
  \BibitemOpen
  \bibfield  {author} {\bibinfo {author} {\bibfnamefont {F.~N.}\ \bibnamefont
  {\"Unal}}, \bibinfo {author} {\bibfnamefont {B.}~\bibnamefont {Seradjeh}}, \
  and\ \bibinfo {author} {\bibfnamefont {A.}~\bibnamefont {Eckardt}},\ }\href
  {\doibase 10.1103/PhysRevLett.122.253601} {\bibfield  {journal} {\bibinfo
  {journal} {Phys. Rev. Lett.}\ }\textbf {\bibinfo {volume} {122}},\ \bibinfo
  {pages} {253601} (\bibinfo {year} {2019})}\BibitemShut {NoStop}%
\bibitem [{\citenamefont {Eckardt}\ \emph {et~al.}(2005)\citenamefont
  {Eckardt}, \citenamefont {Weiss},\ and\ \citenamefont
  {Holthaus}}]{eckardt05}%
  \BibitemOpen
  \bibfield  {author} {\bibinfo {author} {\bibfnamefont {A.}~\bibnamefont
  {Eckardt}}, \bibinfo {author} {\bibfnamefont {C.}~\bibnamefont {Weiss}}, \
  and\ \bibinfo {author} {\bibfnamefont {M.}~\bibnamefont {Holthaus}},\ }\href
  {\doibase 10.1103/PhysRevLett.95.260404} {\bibfield  {journal} {\bibinfo
  {journal} {Phys. Rev. Lett.}\ }\textbf {\bibinfo {volume} {95}},\ \bibinfo
  {pages} {260404} (\bibinfo {year} {2005})}\BibitemShut {NoStop}%
\bibitem [{\citenamefont {Zenesini}\ \emph {et~al.}(2009)\citenamefont
  {Zenesini}, \citenamefont {Lignier}, \citenamefont {Ciampini}, \citenamefont
  {Morsch},\ and\ \citenamefont {Arimondo}}]{zenesini09}%
  \BibitemOpen
  \bibfield  {author} {\bibinfo {author} {\bibfnamefont {A.}~\bibnamefont
  {Zenesini}}, \bibinfo {author} {\bibfnamefont {H.}~\bibnamefont {Lignier}},
  \bibinfo {author} {\bibfnamefont {D.}~\bibnamefont {Ciampini}}, \bibinfo
  {author} {\bibfnamefont {O.}~\bibnamefont {Morsch}}, \ and\ \bibinfo {author}
  {\bibfnamefont {E.}~\bibnamefont {Arimondo}},\ }\href {\doibase
  10.1103/PhysRevLett.102.100403} {\bibfield  {journal} {\bibinfo  {journal}
  {Phys. Rev. Lett.}\ }\textbf {\bibinfo {volume} {102}},\ \bibinfo {pages}
  {100403} (\bibinfo {year} {2009})}\BibitemShut {NoStop}%
\bibitem [{\citenamefont {Eckardt}\ \emph {et~al.}(2010)\citenamefont
  {Eckardt}, \citenamefont {Hauke}, \citenamefont {Soltan-Panahi},
  \citenamefont {Becker}, \citenamefont {Sengstock},\ and\ \citenamefont
  {Lewenstein}}]{Eckardt2010}%
  \BibitemOpen
  \bibfield  {author} {\bibinfo {author} {\bibfnamefont {A.}~\bibnamefont
  {Eckardt}}, \bibinfo {author} {\bibfnamefont {P.}~\bibnamefont {Hauke}},
  \bibinfo {author} {\bibfnamefont {P.}~\bibnamefont {Soltan-Panahi}}, \bibinfo
  {author} {\bibfnamefont {C.}~\bibnamefont {Becker}}, \bibinfo {author}
  {\bibfnamefont {K.}~\bibnamefont {Sengstock}}, \ and\ \bibinfo {author}
  {\bibfnamefont {M.}~\bibnamefont {Lewenstein}},\ }\href {\doibase
  10.1209/0295-5075/89/10010} {\bibfield  {journal} {\bibinfo  {journal}
  {Europhys. Lett.}\ }\textbf {\bibinfo {volume} {89}},\ \bibinfo {pages}
  {10010} (\bibinfo {year} {2010})}\BibitemShut {NoStop}%
\bibitem [{\citenamefont {Neupert}\ \emph {et~al.}(2011)\citenamefont
  {Neupert}, \citenamefont {Santos}, \citenamefont {Chamon},\ and\
  \citenamefont {Mudry}}]{neupert11}%
  \BibitemOpen
  \bibfield  {author} {\bibinfo {author} {\bibfnamefont {T.}~\bibnamefont
  {Neupert}}, \bibinfo {author} {\bibfnamefont {L.}~\bibnamefont {Santos}},
  \bibinfo {author} {\bibfnamefont {C.}~\bibnamefont {Chamon}}, \ and\ \bibinfo
  {author} {\bibfnamefont {C.}~\bibnamefont {Mudry}},\ }\href {\doibase
  10.1103/PhysRevLett.106.236804} {\bibfield  {journal} {\bibinfo  {journal}
  {Phys. Rev. Lett.}\ }\textbf {\bibinfo {volume} {106}},\ \bibinfo {pages}
  {236804} (\bibinfo {year} {2011})}\BibitemShut {NoStop}%
\bibitem [{\citenamefont {Regnault}\ and\ \citenamefont
  {Bernevig}(2011)}]{regnault11}%
  \BibitemOpen
  \bibfield  {author} {\bibinfo {author} {\bibfnamefont {N.}~\bibnamefont
  {Regnault}}\ and\ \bibinfo {author} {\bibfnamefont {B.~A.}\ \bibnamefont
  {Bernevig}},\ }\href {\doibase 10.1103/PhysRevX.1.021014} {\bibfield
  {journal} {\bibinfo  {journal} {Phys. Rev. X}\ }\textbf {\bibinfo {volume}
  {1}},\ \bibinfo {pages} {021014} (\bibinfo {year} {2011})}\BibitemShut
  {NoStop}%
\bibitem [{\citenamefont {Struck}\ \emph {et~al.}(2011)\citenamefont {Struck},
  \citenamefont {{\"O}lschl{\"a}ger}, \citenamefont {Le~Targat}, \citenamefont
  {Soltan-Panahi}, \citenamefont {Eckardt}, \citenamefont {Lewenstein},
  \citenamefont {Windpassinger},\ and\ \citenamefont {Sengstock}}]{Struck11NP}%
  \BibitemOpen
  \bibfield  {author} {\bibinfo {author} {\bibfnamefont {J.}~\bibnamefont
  {Struck}}, \bibinfo {author} {\bibfnamefont {C.}~\bibnamefont
  {{\"O}lschl{\"a}ger}}, \bibinfo {author} {\bibfnamefont {R.}~\bibnamefont
  {Le~Targat}}, \bibinfo {author} {\bibfnamefont {P.}~\bibnamefont
  {Soltan-Panahi}}, \bibinfo {author} {\bibfnamefont {A.}~\bibnamefont
  {Eckardt}}, \bibinfo {author} {\bibfnamefont {M.}~\bibnamefont {Lewenstein}},
  \bibinfo {author} {\bibfnamefont {P.}~\bibnamefont {Windpassinger}}, \ and\
  \bibinfo {author} {\bibfnamefont {K.}~\bibnamefont {Sengstock}},\ }\href
  {\doibase 10.1126/science.1207239} {\bibfield  {journal} {\bibinfo  {journal}
  {Science}\ }\textbf {\bibinfo {volume} {333}},\ \bibinfo {pages} {996}
  (\bibinfo {year} {2011})},\ \Eprint
  {http://arxiv.org/abs/https://science.sciencemag.org/content/333/6045/996.full.pdf}
  {https://science.sciencemag.org/content/333/6045/996.full.pdf} \BibitemShut
  {NoStop}%
\bibitem [{\citenamefont {Wu}\ \emph {et~al.}(2012)\citenamefont {Wu},
  \citenamefont {Bernevig},\ and\ \citenamefont {Regnault}}]{wu12}%
  \BibitemOpen
  \bibfield  {author} {\bibinfo {author} {\bibfnamefont {Y.-L.}\ \bibnamefont
  {Wu}}, \bibinfo {author} {\bibfnamefont {B.~A.}\ \bibnamefont {Bernevig}}, \
  and\ \bibinfo {author} {\bibfnamefont {N.}~\bibnamefont {Regnault}},\ }\href
  {\doibase 10.1103/PhysRevB.85.075116} {\bibfield  {journal} {\bibinfo
  {journal} {Phys. Rev. B}\ }\textbf {\bibinfo {volume} {85}},\ \bibinfo
  {pages} {075116} (\bibinfo {year} {2012})}\BibitemShut {NoStop}%
\bibitem [{\citenamefont {Lewenstein}\ \emph {et~al.}(2012)\citenamefont
  {Lewenstein}, \citenamefont {Sanpera},\ and\ \citenamefont
  {Ahufinger}}]{Lewenstein2012}%
  \BibitemOpen
  \bibfield  {author} {\bibinfo {author} {\bibfnamefont {M.}~\bibnamefont
  {Lewenstein}}, \bibinfo {author} {\bibfnamefont {A.}~\bibnamefont {Sanpera}},
  \ and\ \bibinfo {author} {\bibfnamefont {V.}~\bibnamefont {Ahufinger}},\
  }\href@noop {} {\emph {\bibinfo {title} {Ultracold atoms in optical lattices:
  simulating quantum many-body systems}}}\ (\bibinfo  {publisher} {Oxford
  University Press},\ \bibinfo {year} {2012})\BibitemShut {NoStop}%
\bibitem [{\citenamefont {Parker}\ \emph {et~al.}(2013)\citenamefont {Parker},
  \citenamefont {Ha},\ and\ \citenamefont {Chin}}]{Chin2013NP}%
  \BibitemOpen
  \bibfield  {author} {\bibinfo {author} {\bibfnamefont {C.~V.}\ \bibnamefont
  {Parker}}, \bibinfo {author} {\bibfnamefont {L.-C.}\ \bibnamefont {Ha}}, \
  and\ \bibinfo {author} {\bibfnamefont {C.}~\bibnamefont {Chin}},\ }\href
  {https://doi.org/10.1038/nphys2789} {\bibfield  {journal} {\bibinfo
  {journal} {Nature Physics}\ }\textbf {\bibinfo {volume} {9}},\ \bibinfo
  {pages} {769 EP } (\bibinfo {year} {2013})}\BibitemShut {NoStop}%
\bibitem [{\citenamefont {Struck}\ \emph {et~al.}(2013)\citenamefont {Struck},
  \citenamefont {Weinberg}, \citenamefont {\"Olschl\"ager}, \citenamefont
  {Windpassinger}, \citenamefont {Simonet}, \citenamefont {Sengstock},
  \citenamefont {H\"oppner}, \citenamefont {Hauke}, \citenamefont {Eckardt},
  \citenamefont {Lewenstein},\ and\ \citenamefont {Mathey}}]{Struck:2013}%
  \BibitemOpen
  \bibfield  {author} {\bibinfo {author} {\bibfnamefont {J.}~\bibnamefont
  {Struck}}, \bibinfo {author} {\bibfnamefont {M.}~\bibnamefont {Weinberg}},
  \bibinfo {author} {\bibfnamefont {C.}~\bibnamefont {\"Olschl\"ager}},
  \bibinfo {author} {\bibfnamefont {P.}~\bibnamefont {Windpassinger}}, \bibinfo
  {author} {\bibfnamefont {J.}~\bibnamefont {Simonet}}, \bibinfo {author}
  {\bibfnamefont {K.}~\bibnamefont {Sengstock}}, \bibinfo {author}
  {\bibfnamefont {R.}~\bibnamefont {H\"oppner}}, \bibinfo {author}
  {\bibfnamefont {P.}~\bibnamefont {Hauke}}, \bibinfo {author} {\bibfnamefont
  {A.}~\bibnamefont {Eckardt}}, \bibinfo {author} {\bibfnamefont
  {M.}~\bibnamefont {Lewenstein}}, \ and\ \bibinfo {author} {\bibfnamefont
  {L.}~\bibnamefont {Mathey}},\ }\href {\doibase 10.1038/nphys2750} {\bibfield
  {journal} {\bibinfo  {journal} {Nat. Phys.}\ }\textbf {\bibinfo {volume}
  {9}},\ \bibinfo {pages} {738} (\bibinfo {year} {2013})}\BibitemShut {NoStop}%
\bibitem [{\citenamefont {Bergholtz}\ and\ \citenamefont
  {Liu}(2013)}]{bergholtz13}%
  \BibitemOpen
  \bibfield  {author} {\bibinfo {author} {\bibfnamefont {E.~J.}\ \bibnamefont
  {Bergholtz}}\ and\ \bibinfo {author} {\bibfnamefont {Z.}~\bibnamefont
  {Liu}},\ }\href {\doibase 10.1142/S021797921330017X} {\bibfield  {journal}
  {\bibinfo  {journal} {Int.\ J.\ Mod.\ Phys.\ B}\ }\textbf {\bibinfo {volume}
  {27}},\ \bibinfo {pages} {1330017} (\bibinfo {year} {2013})}\BibitemShut
  {NoStop}%
\bibitem [{\citenamefont {Parameswaran}\ \emph {et~al.}(2013)\citenamefont
  {Parameswaran}, \citenamefont {Roy},\ and\ \citenamefont
  {Sondhi}}]{parameswaran13}%
  \BibitemOpen
  \bibfield  {author} {\bibinfo {author} {\bibfnamefont {S.~A.}\ \bibnamefont
  {Parameswaran}}, \bibinfo {author} {\bibfnamefont {R.}~\bibnamefont {Roy}}, \
  and\ \bibinfo {author} {\bibfnamefont {S.~L.}\ \bibnamefont {Sondhi}},\
  }\href {\doibase 10.1016/j.crhy.2013.04.003} {\bibfield  {journal} {\bibinfo
  {journal} {C.\ R.\ Phys.}\ }\textbf {\bibinfo {volume} {14}},\ \bibinfo
  {pages} {816} (\bibinfo {year} {2013})}\BibitemShut {NoStop}%
\bibitem [{\citenamefont {Greschner}\ \emph {et~al.}(2014)\citenamefont
  {Greschner}, \citenamefont {Sun}, \citenamefont {Poletti},\ and\
  \citenamefont {Santos}}]{Greschner14}%
  \BibitemOpen
  \bibfield  {author} {\bibinfo {author} {\bibfnamefont {S.}~\bibnamefont
  {Greschner}}, \bibinfo {author} {\bibfnamefont {G.}~\bibnamefont {Sun}},
  \bibinfo {author} {\bibfnamefont {D.}~\bibnamefont {Poletti}}, \ and\
  \bibinfo {author} {\bibfnamefont {L.}~\bibnamefont {Santos}},\ }\href
  {\doibase 10.1103/PhysRevLett.113.215303} {\bibfield  {journal} {\bibinfo
  {journal} {Phys. Phys. Lett.}\ }\textbf {\bibinfo {volume} {113}},\ \bibinfo
  {pages} {215303} (\bibinfo {year} {2014})}\BibitemShut {NoStop}%
\bibitem [{\citenamefont {Anisimovas}\ \emph {et~al.}(2015)\citenamefont
  {Anisimovas}, \citenamefont {{\v Z}labys}, \citenamefont {Anderson},
  \citenamefont {Juzeli{\=u}nas},\ and\ \citenamefont
  {Eckardt}}]{Anisimovas15PRB}%
  \BibitemOpen
  \bibfield  {author} {\bibinfo {author} {\bibfnamefont {E.}~\bibnamefont
  {Anisimovas}}, \bibinfo {author} {\bibfnamefont {G.}~\bibnamefont {{\v
  Z}labys}}, \bibinfo {author} {\bibfnamefont {B.~M.}\ \bibnamefont
  {Anderson}}, \bibinfo {author} {\bibfnamefont {G.}~\bibnamefont
  {Juzeli{\=u}nas}}, \ and\ \bibinfo {author} {\bibfnamefont {A.}~\bibnamefont
  {Eckardt}},\ }\href {\doibase 10.1103/PhysRevB.91.245135} {\bibfield
  {journal} {\bibinfo  {journal} {Phys. Rev. B}\ }\textbf {\bibinfo {volume}
  {91}},\ \bibinfo {pages} {245135} (\bibinfo {year} {2015})}\BibitemShut
  {NoStop}%
\bibitem [{\citenamefont {Meinert}\ \emph {et~al.}(2016)\citenamefont
  {Meinert}, \citenamefont {Mark}, \citenamefont {Lauber}, \citenamefont
  {Daley},\ and\ \citenamefont {N\"agerl}}]{Nagerl2016}%
  \BibitemOpen
  \bibfield  {author} {\bibinfo {author} {\bibfnamefont {F.}~\bibnamefont
  {Meinert}}, \bibinfo {author} {\bibfnamefont {M.~J.}\ \bibnamefont {Mark}},
  \bibinfo {author} {\bibfnamefont {K.}~\bibnamefont {Lauber}}, \bibinfo
  {author} {\bibfnamefont {A.~J.}\ \bibnamefont {Daley}}, \ and\ \bibinfo
  {author} {\bibfnamefont {H.-C.}\ \bibnamefont {N\"agerl}},\ }\href {\doibase
  10.1103/PhysRevLett.116.205301} {\bibfield  {journal} {\bibinfo  {journal}
  {Phys. Rev. Lett.}\ }\textbf {\bibinfo {volume} {116}},\ \bibinfo {pages}
  {205301} (\bibinfo {year} {2016})}\BibitemShut {NoStop}%
\bibitem [{\citenamefont {Desbuquois}\ \emph {et~al.}(2017)\citenamefont
  {Desbuquois}, \citenamefont {Messer}, \citenamefont {G{\"o}rg}, \citenamefont
  {Sandholzer}, \citenamefont {Jotzu},\ and\ \citenamefont
  {Esslinger}}]{Esslinger17PRA}%
  \BibitemOpen
  \bibfield  {author} {\bibinfo {author} {\bibfnamefont {R.}~\bibnamefont
  {Desbuquois}}, \bibinfo {author} {\bibfnamefont {M.}~\bibnamefont {Messer}},
  \bibinfo {author} {\bibfnamefont {F.}~\bibnamefont {G{\"o}rg}}, \bibinfo
  {author} {\bibfnamefont {K.}~\bibnamefont {Sandholzer}}, \bibinfo {author}
  {\bibfnamefont {G.}~\bibnamefont {Jotzu}}, \ and\ \bibinfo {author}
  {\bibfnamefont {T.}~\bibnamefont {Esslinger}},\ }\href@noop {} {\bibfield
  {journal} {\bibinfo  {journal} {Phys. Rev. A}\ }\textbf {\bibinfo {volume}
  {96}},\ \bibinfo {pages} {053602} (\bibinfo {year} {2017})}\BibitemShut
  {NoStop}%
\bibitem [{\citenamefont {Clark}\ \emph {et~al.}(2018)\citenamefont {Clark},
  \citenamefont {Anderson}, \citenamefont {Feng}, \citenamefont {Gaj},
  \citenamefont {Levin},\ and\ \citenamefont {Chin}}]{Chin18PRL}%
  \BibitemOpen
  \bibfield  {author} {\bibinfo {author} {\bibfnamefont {L.~W.}\ \bibnamefont
  {Clark}}, \bibinfo {author} {\bibfnamefont {B.~M.}\ \bibnamefont {Anderson}},
  \bibinfo {author} {\bibfnamefont {L.}~\bibnamefont {Feng}}, \bibinfo {author}
  {\bibfnamefont {A.}~\bibnamefont {Gaj}}, \bibinfo {author} {\bibfnamefont
  {K.}~\bibnamefont {Levin}}, \ and\ \bibinfo {author} {\bibfnamefont
  {C.}~\bibnamefont {Chin}},\ }\href {\doibase 10.1103/PhysRevLett.121.030402}
  {\bibfield  {journal} {\bibinfo  {journal} {Phys. Rev. Lett.}\ }\textbf
  {\bibinfo {volume} {121}},\ \bibinfo {pages} {030402} (\bibinfo {year}
  {2018})}\BibitemShut {NoStop}%
\bibitem [{\citenamefont {Sacha}\ and\ \citenamefont
  {Zakrzewski}(2018)}]{Sacha18RMP}%
  \BibitemOpen
  \bibfield  {author} {\bibinfo {author} {\bibfnamefont {K.}~\bibnamefont
  {Sacha}}\ and\ \bibinfo {author} {\bibfnamefont {J.}~\bibnamefont
  {Zakrzewski}},\ }\href@noop {} {\bibfield  {journal} {\bibinfo  {journal}
  {Rep. Progr. Phys.}\ }\textbf {\bibinfo {volume} {81}},\ \bibinfo {pages}
  {016401} (\bibinfo {year} {2018})}\BibitemShut {NoStop}%
\bibitem [{\citenamefont {Kolovsky}(2011)}]{kolovsky11}%
  \BibitemOpen
  \bibfield  {author} {\bibinfo {author} {\bibfnamefont {A.~R.}\ \bibnamefont
  {Kolovsky}},\ }\href {\doibase 10.1209/0295-5075/93/20003} {\bibfield
  {journal} {\bibinfo  {journal} {Europhys. Lett.}\ }\textbf {\bibinfo {volume}
  {93}},\ \bibinfo {pages} {20003} (\bibinfo {year} {2011})}\BibitemShut
  {NoStop}%
\bibitem [{\citenamefont {Struck}\ \emph {et~al.}(2012)\citenamefont {Struck},
  \citenamefont {\"{O}lschl\"{a}ger}, \citenamefont {Weinberg}, \citenamefont
  {Hauke}, \citenamefont {Simonet}, \citenamefont {Eckardt}, \citenamefont
  {Lewenstein}, \citenamefont {Sengstock},\ and\ \citenamefont
  {Windpassinger}}]{struck12}%
  \BibitemOpen
  \bibfield  {author} {\bibinfo {author} {\bibfnamefont {J.}~\bibnamefont
  {Struck}}, \bibinfo {author} {\bibfnamefont {C.}~\bibnamefont
  {\"{O}lschl\"{a}ger}}, \bibinfo {author} {\bibfnamefont {M.}~\bibnamefont
  {Weinberg}}, \bibinfo {author} {\bibfnamefont {P.}~\bibnamefont {Hauke}},
  \bibinfo {author} {\bibfnamefont {J.}~\bibnamefont {Simonet}}, \bibinfo
  {author} {\bibfnamefont {A.}~\bibnamefont {Eckardt}}, \bibinfo {author}
  {\bibfnamefont {M.}~\bibnamefont {Lewenstein}}, \bibinfo {author}
  {\bibfnamefont {K.}~\bibnamefont {Sengstock}}, \ and\ \bibinfo {author}
  {\bibfnamefont {P.}~\bibnamefont {Windpassinger}},\ }\href {\doibase
  10.1103/PhysRevLett.108.225304} {\bibfield  {journal} {\bibinfo  {journal}
  {Phys. Rev. Lett.}\ }\textbf {\bibinfo {volume} {108}},\ \bibinfo {pages}
  {225304} (\bibinfo {year} {2012})}\BibitemShut {NoStop}%
\bibitem [{\citenamefont {Hauke}\ \emph {et~al.}(2012)\citenamefont {Hauke},
  \citenamefont {Tieleman}, \citenamefont {Celi}, \citenamefont
  {{\"O}lschl{\"a}ger}, \citenamefont {Simonet}, \citenamefont {Struck},
  \citenamefont {Weinberg}, \citenamefont {Windpassinger}, \citenamefont
  {Sengstock}, \citenamefont {Lewenstein},\ and\ \citenamefont
  {Eckardt}}]{Hauke:2012}%
  \BibitemOpen
  \bibfield  {author} {\bibinfo {author} {\bibfnamefont {P.}~\bibnamefont
  {Hauke}}, \bibinfo {author} {\bibfnamefont {O.}~\bibnamefont {Tieleman}},
  \bibinfo {author} {\bibfnamefont {A.}~\bibnamefont {Celi}}, \bibinfo {author}
  {\bibfnamefont {C.}~\bibnamefont {{\"O}lschl{\"a}ger}}, \bibinfo {author}
  {\bibfnamefont {J.}~\bibnamefont {Simonet}}, \bibinfo {author} {\bibfnamefont
  {J.}~\bibnamefont {Struck}}, \bibinfo {author} {\bibfnamefont
  {M.}~\bibnamefont {Weinberg}}, \bibinfo {author} {\bibfnamefont
  {P.}~\bibnamefont {Windpassinger}}, \bibinfo {author} {\bibfnamefont
  {K.}~\bibnamefont {Sengstock}}, \bibinfo {author} {\bibfnamefont
  {M.}~\bibnamefont {Lewenstein}}, \ and\ \bibinfo {author} {\bibfnamefont
  {A.}~\bibnamefont {Eckardt}},\ }\href {\doibase
  10.1103/PhysRevLett.109.145301} {\bibfield  {journal} {\bibinfo  {journal}
  {Phys. Rev. Lett.}\ }\textbf {\bibinfo {volume} {109}},\ \bibinfo {pages}
  {145301} (\bibinfo {year} {2012})}\BibitemShut {NoStop}%
\bibitem [{\citenamefont {Miyake}\ \emph {et~al.}(2013)\citenamefont {Miyake},
  \citenamefont {Siviloglou}, \citenamefont {Kennedy}, \citenamefont {Burton},\
  and\ \citenamefont {Ketterle}}]{Ketterle:2013}%
  \BibitemOpen
  \bibfield  {author} {\bibinfo {author} {\bibfnamefont {H.}~\bibnamefont
  {Miyake}}, \bibinfo {author} {\bibfnamefont {G.~A.}\ \bibnamefont
  {Siviloglou}}, \bibinfo {author} {\bibfnamefont {C.~J.}\ \bibnamefont
  {Kennedy}}, \bibinfo {author} {\bibfnamefont {W.~C.}\ \bibnamefont {Burton}},
  \ and\ \bibinfo {author} {\bibfnamefont {W.}~\bibnamefont {Ketterle}},\
  }\href {\doibase 10.1103/PhysRevLett.111.185302} {\bibfield  {journal}
  {\bibinfo  {journal} {Phys. Phys. Lett.}\ }\textbf {\bibinfo {volume}
  {111}},\ \bibinfo {pages} {185302} (\bibinfo {year} {2013})}\BibitemShut
  {NoStop}%
\bibitem [{\citenamefont {Aidelsburger}\ \emph {et~al.}(2013)\citenamefont
  {Aidelsburger}, \citenamefont {Atala}, \citenamefont {Lohse}, \citenamefont
  {Barreiro}, \citenamefont {Paredes},\ and\ \citenamefont
  {Bloch}}]{Aidelsburger:2013}%
  \BibitemOpen
  \bibfield  {author} {\bibinfo {author} {\bibfnamefont {M.}~\bibnamefont
  {Aidelsburger}}, \bibinfo {author} {\bibfnamefont {M.}~\bibnamefont {Atala}},
  \bibinfo {author} {\bibfnamefont {M.}~\bibnamefont {Lohse}}, \bibinfo
  {author} {\bibfnamefont {J.~T.}\ \bibnamefont {Barreiro}}, \bibinfo {author}
  {\bibfnamefont {B.}~\bibnamefont {Paredes}}, \ and\ \bibinfo {author}
  {\bibfnamefont {I.}~\bibnamefont {Bloch}},\ }\href {\doibase
  10.1103/PhysRevLett.111.185301} {\bibfield  {journal} {\bibinfo  {journal}
  {Phys. Rev. Lett.}\ }\textbf {\bibinfo {volume} {111}},\ \bibinfo {pages}
  {185301} (\bibinfo {year} {2013})}\BibitemShut {NoStop}%
\bibitem [{\citenamefont {Anderson}\ \emph {et~al.}(2013)\citenamefont
  {Anderson}, \citenamefont {Spielman},\ and\ \citenamefont
  {Juzeli{\=u}nas}}]{Anderson2013}%
  \BibitemOpen
  \bibfield  {author} {\bibinfo {author} {\bibfnamefont {B.~M.}\ \bibnamefont
  {Anderson}}, \bibinfo {author} {\bibfnamefont {I.~B.}\ \bibnamefont
  {Spielman}}, \ and\ \bibinfo {author} {\bibfnamefont {G.}~\bibnamefont
  {Juzeli{\=u}nas}},\ }\href {\doibase 10.1103/PhysRevLett.111.125301}
  {\bibfield  {journal} {\bibinfo  {journal} {Phys. Rev. Lett.}\ }\textbf
  {\bibinfo {volume} {111}},\ \bibinfo {pages} {125301} (\bibinfo {year}
  {2013})}\BibitemShut {NoStop}%
\bibitem [{\citenamefont {Xu}\ \emph {et~al.}(2013)\citenamefont {Xu},
  \citenamefont {You},\ and\ \citenamefont {Ueda}}]{Xu2013}%
  \BibitemOpen
  \bibfield  {author} {\bibinfo {author} {\bibfnamefont {Z.-F.}\ \bibnamefont
  {Xu}}, \bibinfo {author} {\bibfnamefont {L.}~\bibnamefont {You}}, \ and\
  \bibinfo {author} {\bibfnamefont {M.}~\bibnamefont {Ueda}},\ }\href {\doibase
  10.1103/PhysRevA.87.063634} {\bibfield  {journal} {\bibinfo  {journal} {Phys.
  Rev. A}\ }\textbf {\bibinfo {volume} {87}},\ \bibinfo {pages} {063634}
  (\bibinfo {year} {2013})}\BibitemShut {NoStop}%
\bibitem [{\citenamefont {Atala}\ \emph {et~al.}(2014)\citenamefont {Atala},
  \citenamefont {Aidelsburger}, \citenamefont {Lohse}, \citenamefont
  {Barreiro}, \citenamefont {Paredes},\ and\ \citenamefont {Bloch}}]{atala14}%
  \BibitemOpen
  \bibfield  {author} {\bibinfo {author} {\bibfnamefont {M.}~\bibnamefont
  {Atala}}, \bibinfo {author} {\bibfnamefont {M.}~\bibnamefont {Aidelsburger}},
  \bibinfo {author} {\bibfnamefont {M.}~\bibnamefont {Lohse}}, \bibinfo
  {author} {\bibfnamefont {J.~T.}\ \bibnamefont {Barreiro}}, \bibinfo {author}
  {\bibfnamefont {B.}~\bibnamefont {Paredes}}, \ and\ \bibinfo {author}
  {\bibfnamefont {I.}~\bibnamefont {Bloch}},\ }\href {\doibase
  10.1038/nphys2998} {\bibfield  {journal} {\bibinfo  {journal} {Nat. Phys.}\
  }\textbf {\bibinfo {volume} {10}},\ \bibinfo {pages} {588} (\bibinfo {year}
  {2014})}\BibitemShut {NoStop}%
\bibitem [{\citenamefont {Goldman}\ \emph {et~al.}(2014)\citenamefont
  {Goldman}, \citenamefont {Juzeli{\=u}nas}, \citenamefont {{\"O}hberg},\ and\
  \citenamefont {Spielman}}]{Goldman2014RPP}%
  \BibitemOpen
  \bibfield  {author} {\bibinfo {author} {\bibfnamefont {N.}~\bibnamefont
  {Goldman}}, \bibinfo {author} {\bibfnamefont {G.}~\bibnamefont
  {Juzeli{\=u}nas}}, \bibinfo {author} {\bibfnamefont {P.}~\bibnamefont
  {{\"O}hberg}}, \ and\ \bibinfo {author} {\bibfnamefont {I.~B.}\ \bibnamefont
  {Spielman}},\ }\href {\doibase 10.1088/0034-4885/77/12/126401} {\bibfield
  {journal} {\bibinfo  {journal} {Rep. Progr. Phys.}\ }\textbf {\bibinfo
  {volume} {77}},\ \bibinfo {pages} {126401} (\bibinfo {year}
  {2014})}\BibitemShut {NoStop}%
\bibitem [{\citenamefont {Fl{\"a}schner}\ \emph {et~al.}(2016)\citenamefont
  {Fl{\"a}schner}, \citenamefont {Rem}, \citenamefont {Tarnowski},
  \citenamefont {Vogel}, \citenamefont {L{\"u}hmann}, \citenamefont
  {Sengstock},\ and\ \citenamefont {Weitenberg}}]{Flaschner16}%
  \BibitemOpen
  \bibfield  {author} {\bibinfo {author} {\bibfnamefont {N.}~\bibnamefont
  {Fl{\"a}schner}}, \bibinfo {author} {\bibfnamefont {B.~S.}\ \bibnamefont
  {Rem}}, \bibinfo {author} {\bibfnamefont {M.}~\bibnamefont {Tarnowski}},
  \bibinfo {author} {\bibfnamefont {D.}~\bibnamefont {Vogel}}, \bibinfo
  {author} {\bibfnamefont {D.-S.}\ \bibnamefont {L{\"u}hmann}}, \bibinfo
  {author} {\bibfnamefont {K.}~\bibnamefont {Sengstock}}, \ and\ \bibinfo
  {author} {\bibfnamefont {C.}~\bibnamefont {Weitenberg}},\ }\href {\doibase
  10.1126/science.aad4568} {\bibfield  {journal} {\bibinfo  {journal}
  {Science}\ }\textbf {\bibinfo {volume} {352}},\ \bibinfo {pages} {1091}
  (\bibinfo {year} {2016})}\BibitemShut {NoStop}%
\bibitem [{\citenamefont {{Luo}}\ \emph {et~al.}(2016)\citenamefont {{Luo}},
  \citenamefont {{Wu}}, \citenamefont {{Chen}}, \citenamefont {{Guan}},
  \citenamefont {{Gao}}, \citenamefont {{Xu}}, \citenamefont {{You}},\ and\
  \citenamefont {{Wang}}}]{Luo16Sci_Rep}%
  \BibitemOpen
  \bibfield  {author} {\bibinfo {author} {\bibfnamefont {X.}~\bibnamefont
  {{Luo}}}, \bibinfo {author} {\bibfnamefont {L.}~\bibnamefont {{Wu}}},
  \bibinfo {author} {\bibfnamefont {J.}~\bibnamefont {{Chen}}}, \bibinfo
  {author} {\bibfnamefont {Q.}~\bibnamefont {{Guan}}}, \bibinfo {author}
  {\bibfnamefont {K.}~\bibnamefont {{Gao}}}, \bibinfo {author} {\bibfnamefont
  {Z.-F.}\ \bibnamefont {{Xu}}}, \bibinfo {author} {\bibfnamefont
  {L.}~\bibnamefont {{You}}}, \ and\ \bibinfo {author} {\bibfnamefont
  {R.}~\bibnamefont {{Wang}}},\ }\href {\doibase 10.1038/srep18983} {\bibfield
  {journal} {\bibinfo  {journal} {Sci. Rep.}\ }\textbf {\bibinfo {volume}
  {6}},\ \bibinfo {pages} {18983} (\bibinfo {year} {2016})}\BibitemShut
  {NoStop}%
\bibitem [{\citenamefont {Shteynas}\ \emph {et~al.}(2019)\citenamefont
  {Shteynas}, \citenamefont {Lee}, \citenamefont {Top}, \citenamefont {Li},
  \citenamefont {Jamison}, \citenamefont {Juzeli\=unas},\ and\ \citenamefont
  {Ketterle}}]{Shteynas19PRL}%
  \BibitemOpen
  \bibfield  {author} {\bibinfo {author} {\bibfnamefont {B.}~\bibnamefont
  {Shteynas}}, \bibinfo {author} {\bibfnamefont {J.}~\bibnamefont {Lee}},
  \bibinfo {author} {\bibfnamefont {F.~C.}\ \bibnamefont {Top}}, \bibinfo
  {author} {\bibfnamefont {J.-R.}\ \bibnamefont {Li}}, \bibinfo {author}
  {\bibfnamefont {A.~O.}\ \bibnamefont {Jamison}}, \bibinfo {author}
  {\bibfnamefont {G.}~\bibnamefont {Juzeli\=unas}}, \ and\ \bibinfo {author}
  {\bibfnamefont {W.}~\bibnamefont {Ketterle}},\ }\href@noop {} {\bibfield
  {journal} {\bibinfo  {journal} {Phys. Phys. Lett.}\ }\textbf {\bibinfo
  {volume} {123}},\ \bibinfo {pages} {033203} (\bibinfo {year} {2019})},\
  \Eprint {http://arxiv.org/abs/arXiv:1807.07041} {arXiv:arXiv:1807.07041}
  \BibitemShut {NoStop}%
\bibitem [{\citenamefont {Galitski}\ \emph {et~al.}(2019)\citenamefont
  {Galitski}, \citenamefont {Juzeli{\=u}nas},\ and\ \citenamefont
  {Spielman}}]{Galitski19PT}%
  \BibitemOpen
  \bibfield  {author} {\bibinfo {author} {\bibfnamefont {V.}~\bibnamefont
  {Galitski}}, \bibinfo {author} {\bibfnamefont {G.}~\bibnamefont
  {Juzeli{\=u}nas}}, \ and\ \bibinfo {author} {\bibfnamefont {I.~B.}\
  \bibnamefont {Spielman}},\ }\href@noop {} {\bibfield  {journal} {\bibinfo
  {journal} {Phys. Today}\ }\textbf {\bibinfo {volume} {72}},\ \bibinfo {pages}
  {38} (\bibinfo {year} {2019})}\BibitemShut {NoStop}%
\bibitem [{\citenamefont {Haldane}\ and\ \citenamefont
  {Raghu}(2008)}]{Haldane:2008cc}%
  \BibitemOpen
  \bibfield  {author} {\bibinfo {author} {\bibfnamefont {F.~D.~M.}\
  \bibnamefont {Haldane}}\ and\ \bibinfo {author} {\bibfnamefont
  {S.}~\bibnamefont {Raghu}},\ }\href {\doibase 10.1103/PhysRevLett.100.013904}
  {\bibfield  {journal} {\bibinfo  {journal} {Phys. Rev. Lett.}\ }\textbf
  {\bibinfo {volume} {100}},\ \bibinfo {pages} {013904} (\bibinfo {year}
  {2008})}\BibitemShut {NoStop}%
\bibitem [{\citenamefont {Rechtsman}\ \emph {et~al.}(2013)\citenamefont
  {Rechtsman}, \citenamefont {Zeuner}, \citenamefont {Plotnik}, \citenamefont
  {Lumer}, \citenamefont {Podolsky}, \citenamefont {Dreisow}, \citenamefont
  {Nolte}, \citenamefont {Segev},\ and\ \citenamefont
  {Szameit}}]{Rechtsman:2013fe}%
  \BibitemOpen
  \bibfield  {author} {\bibinfo {author} {\bibfnamefont {M.~C.}\ \bibnamefont
  {Rechtsman}}, \bibinfo {author} {\bibfnamefont {J.~M.}\ \bibnamefont
  {Zeuner}}, \bibinfo {author} {\bibfnamefont {Y.}~\bibnamefont {Plotnik}},
  \bibinfo {author} {\bibfnamefont {Y.}~\bibnamefont {Lumer}}, \bibinfo
  {author} {\bibfnamefont {D.}~\bibnamefont {Podolsky}}, \bibinfo {author}
  {\bibfnamefont {F.}~\bibnamefont {Dreisow}}, \bibinfo {author} {\bibfnamefont
  {S.}~\bibnamefont {Nolte}}, \bibinfo {author} {\bibfnamefont
  {M.}~\bibnamefont {Segev}}, \ and\ \bibinfo {author} {\bibfnamefont
  {A.}~\bibnamefont {Szameit}},\ }\href {\doibase 10.1038/nature12066}
  {\bibfield  {journal} {\bibinfo  {journal} {Nature}\ }\textbf {\bibinfo
  {volume} {496}},\ \bibinfo {pages} {196} (\bibinfo {year}
  {2013})}\BibitemShut {NoStop}%
\bibitem [{\citenamefont {Mukherjee}\ \emph {et~al.}(2017)\citenamefont
  {Mukherjee}, \citenamefont {Spracklen}, \citenamefont {Valiente},
  \citenamefont {Andersson}, \citenamefont {{\"O}hberg}, \citenamefont
  {Goldman},\ and\ \citenamefont {Thomson}}]{Mukherjee17Ncommun}%
  \BibitemOpen
  \bibfield  {author} {\bibinfo {author} {\bibfnamefont {S.}~\bibnamefont
  {Mukherjee}}, \bibinfo {author} {\bibfnamefont {A.}~\bibnamefont
  {Spracklen}}, \bibinfo {author} {\bibfnamefont {M.}~\bibnamefont {Valiente}},
  \bibinfo {author} {\bibfnamefont {E.}~\bibnamefont {Andersson}}, \bibinfo
  {author} {\bibfnamefont {P.}~\bibnamefont {{\"O}hberg}}, \bibinfo {author}
  {\bibfnamefont {N.}~\bibnamefont {Goldman}}, \ and\ \bibinfo {author}
  {\bibfnamefont {R.~R.}\ \bibnamefont {Thomson}},\ }\href@noop {} {\bibfield
  {journal} {\bibinfo  {journal} {Nat. Commun.}\ }\textbf {\bibinfo {volume}
  {8}} (\bibinfo {year} {2017})}\BibitemShut {NoStop}%
\bibitem [{\citenamefont {J.~Noh}\ and\ \citenamefont
  {Rechtsman}(2018)}]{Rechtsman18PRL}%
  \BibitemOpen
  \bibfield  {author} {\bibinfo {author} {\bibfnamefont {K.~P.~C.}\
  \bibnamefont {J.~Noh}, \bibfnamefont {S.~Huang}}\ and\ \bibinfo {author}
  {\bibfnamefont {M.~C.}\ \bibnamefont {Rechtsman}},\ }\href@noop {} {\bibfield
   {journal} {\bibinfo  {journal} {Phys. Phys. Lett.}\ }\textbf {\bibinfo
  {volume} {120}},\ \bibinfo {pages} {063902} (\bibinfo {year}
  {2018})}\BibitemShut {NoStop}%
\bibitem [{\citenamefont {Cardano}\ \emph {et~al.}(2016)\citenamefont
  {Cardano}, \citenamefont {Maffei}, \citenamefont {Massa}, \citenamefont
  {Piccirillo}, \citenamefont {de~Lisio}, \citenamefont {De~Filippis},
  \citenamefont {Cataudella}, \citenamefont {Santamato},\ and\ \citenamefont
  {Marrucci}}]{Cardano16NPhoton}%
  \BibitemOpen
  \bibfield  {author} {\bibinfo {author} {\bibfnamefont {F.}~\bibnamefont
  {Cardano}}, \bibinfo {author} {\bibfnamefont {M.}~\bibnamefont {Maffei}},
  \bibinfo {author} {\bibfnamefont {F.}~\bibnamefont {Massa}}, \bibinfo
  {author} {\bibfnamefont {B.}~\bibnamefont {Piccirillo}}, \bibinfo {author}
  {\bibfnamefont {C.}~\bibnamefont {de~Lisio}}, \bibinfo {author}
  {\bibfnamefont {G.}~\bibnamefont {De~Filippis}}, \bibinfo {author}
  {\bibfnamefont {V.}~\bibnamefont {Cataudella}}, \bibinfo {author}
  {\bibfnamefont {E.}~\bibnamefont {Santamato}}, \ and\ \bibinfo {author}
  {\bibfnamefont {L.}~\bibnamefont {Marrucci}},\ }\href
  {https://doi.org/10.1038/ncomms11439} {\bibfield  {journal} {\bibinfo
  {journal} {Nature Communications}\ }\textbf {\bibinfo {volume} {7}},\
  \bibinfo {pages} {11439 EP } (\bibinfo {year} {2016})}\BibitemShut {NoStop}%
\bibitem [{\citenamefont {Ozawa}\ \emph {et~al.}(2019)\citenamefont {Ozawa},
  \citenamefont {Price}, \citenamefont {Amo}, \citenamefont {Goldman},
  \citenamefont {Hafezi}, \citenamefont {Lu}, \citenamefont {Rechtsman},
  \citenamefont {Schuster}, \citenamefont {Simon}, \citenamefont {Zilberberg},\
  and\ \citenamefont {Carusotto}}]{Ozawa19RMP}%
  \BibitemOpen
  \bibfield  {author} {\bibinfo {author} {\bibfnamefont {T.}~\bibnamefont
  {Ozawa}}, \bibinfo {author} {\bibfnamefont {H.~M.}\ \bibnamefont {Price}},
  \bibinfo {author} {\bibfnamefont {A.}~\bibnamefont {Amo}}, \bibinfo {author}
  {\bibfnamefont {N.}~\bibnamefont {Goldman}}, \bibinfo {author} {\bibfnamefont
  {M.}~\bibnamefont {Hafezi}}, \bibinfo {author} {\bibfnamefont
  {L.}~\bibnamefont {Lu}}, \bibinfo {author} {\bibfnamefont {M.~C.}\
  \bibnamefont {Rechtsman}}, \bibinfo {author} {\bibfnamefont {D.}~\bibnamefont
  {Schuster}}, \bibinfo {author} {\bibfnamefont {J.}~\bibnamefont {Simon}},
  \bibinfo {author} {\bibfnamefont {O.}~\bibnamefont {Zilberberg}}, \ and\
  \bibinfo {author} {\bibfnamefont {I.}~\bibnamefont {Carusotto}},\ }\href
  {\doibase 10.1103/RevModPhys.91.015006} {\bibfield  {journal} {\bibinfo
  {journal} {Rev. Mod. Phys.}\ }\textbf {\bibinfo {volume} {91}},\ \bibinfo
  {pages} {015006} (\bibinfo {year} {2019})}\BibitemShut {NoStop}%
\bibitem [{\citenamefont {Kitagawa}\ \emph {et~al.}(2011)\citenamefont
  {Kitagawa}, \citenamefont {Oka}, \citenamefont {Brataas}, \citenamefont
  {Fu},\ and\ \citenamefont {Demler}}]{Kitagawa2011}%
  \BibitemOpen
  \bibfield  {author} {\bibinfo {author} {\bibfnamefont {T.}~\bibnamefont
  {Kitagawa}}, \bibinfo {author} {\bibfnamefont {T.}~\bibnamefont {Oka}},
  \bibinfo {author} {\bibfnamefont {A.}~\bibnamefont {Brataas}}, \bibinfo
  {author} {\bibfnamefont {L.}~\bibnamefont {Fu}}, \ and\ \bibinfo {author}
  {\bibfnamefont {E.}~\bibnamefont {Demler}},\ }\href {\doibase
  10.1103/PhysRevB.84.235108} {\bibfield  {journal} {\bibinfo  {journal} {Phys.
  Rev. B}\ }\textbf {\bibinfo {volume} {84}},\ \bibinfo {pages} {235108}
  (\bibinfo {year} {2011})}\BibitemShut {NoStop}%
\bibitem [{\citenamefont {Fregoso}\ \emph {et~al.}(2013)\citenamefont
  {Fregoso}, \citenamefont {Wang}, \citenamefont {Gedik},\ and\ \citenamefont
  {Galitski}}]{Galitski13PRB}%
  \BibitemOpen
  \bibfield  {author} {\bibinfo {author} {\bibfnamefont {B.~M.}\ \bibnamefont
  {Fregoso}}, \bibinfo {author} {\bibfnamefont {Y.~H.}\ \bibnamefont {Wang}},
  \bibinfo {author} {\bibfnamefont {N.}~\bibnamefont {Gedik}}, \ and\ \bibinfo
  {author} {\bibfnamefont {V.}~\bibnamefont {Galitski}},\ }\href {\doibase
  10.1103/PhysRevB.88.155129} {\bibfield  {journal} {\bibinfo  {journal} {Phys.
  Rev. B}\ }\textbf {\bibinfo {volume} {88}},\ \bibinfo {pages} {155129}
  (\bibinfo {year} {2013})}\BibitemShut {NoStop}%
\bibitem [{\citenamefont {{Tong}}\ \emph {et~al.}(2013)\citenamefont {{Tong}},
  \citenamefont {{An}}, \citenamefont {{Gong}}, \citenamefont {{Luo}},\ and\
  \citenamefont {{Oh}}}]{tong13majorana}%
  \BibitemOpen
  \bibfield  {author} {\bibinfo {author} {\bibfnamefont {Q.-J.}\ \bibnamefont
  {{Tong}}}, \bibinfo {author} {\bibfnamefont {J.-H.}\ \bibnamefont {{An}}},
  \bibinfo {author} {\bibfnamefont {J.}~\bibnamefont {{Gong}}}, \bibinfo
  {author} {\bibfnamefont {H.-G.}\ \bibnamefont {{Luo}}}, \ and\ \bibinfo
  {author} {\bibfnamefont {C.~H.}\ \bibnamefont {{Oh}}},\ }\href {\doibase
  10.1103/PhysRevB.87.201109} {\bibfield  {journal} {\bibinfo  {journal} {Phys.
  Rev. B}\ }\textbf {\bibinfo {volume} {87}},\ \bibinfo {pages} {201109(R)}
  (\bibinfo {year} {2013})}\BibitemShut {NoStop}%
\bibitem [{\citenamefont {Grushin}\ \emph {et~al.}(2014)\citenamefont
  {Grushin}, \citenamefont {G\'{o}mez-Le\'{o}n},\ and\ \citenamefont
  {Neupert}}]{grushin14}%
  \BibitemOpen
  \bibfield  {author} {\bibinfo {author} {\bibfnamefont {A.~G.}\ \bibnamefont
  {Grushin}}, \bibinfo {author} {\bibfnamefont {A.}~\bibnamefont
  {G\'{o}mez-Le\'{o}n}}, \ and\ \bibinfo {author} {\bibfnamefont
  {T.}~\bibnamefont {Neupert}},\ }\href {\doibase
  10.1103/PhysRevLett.112.156801} {\bibfield  {journal} {\bibinfo  {journal}
  {Phys. Rev. Lett.}\ }\textbf {\bibinfo {volume} {112}},\ \bibinfo {pages}
  {156801} (\bibinfo {year} {2014})}\BibitemShut {NoStop}%
\bibitem [{\citenamefont {{Usaj}}\ \emph {et~al.}(2014)\citenamefont {{Usaj}},
  \citenamefont {{Perez-Piskunow}}, \citenamefont {{Foa Torres}},\ and\
  \citenamefont {{Balseiro}}}]{usaj14}%
  \BibitemOpen
  \bibfield  {author} {\bibinfo {author} {\bibfnamefont {G.}~\bibnamefont
  {{Usaj}}}, \bibinfo {author} {\bibfnamefont {P.~M.}\ \bibnamefont
  {{Perez-Piskunow}}}, \bibinfo {author} {\bibfnamefont {L.~E.~F.}\
  \bibnamefont {{Foa Torres}}}, \ and\ \bibinfo {author} {\bibfnamefont
  {C.~A.}\ \bibnamefont {{Balseiro}}},\ }\href {\doibase
  10.1103/PhysRevB.90.115423} {\bibfield  {journal} {\bibinfo  {journal} {Phys.
  Rev. B}\ }\textbf {\bibinfo {volume} {90}},\ \bibinfo {pages} {115423}
  (\bibinfo {year} {2014})}\BibitemShut {NoStop}%
\bibitem [{\citenamefont {{Quelle}}\ \emph {et~al.}(2015)\citenamefont
  {{Quelle}}, \citenamefont {{Beugeling}},\ and\ \citenamefont {{Morais
  Smith}}}]{quelle15}%
  \BibitemOpen
  \bibfield  {author} {\bibinfo {author} {\bibfnamefont {A.}~\bibnamefont
  {{Quelle}}}, \bibinfo {author} {\bibfnamefont {W.}~\bibnamefont
  {{Beugeling}}}, \ and\ \bibinfo {author} {\bibfnamefont {C.}~\bibnamefont
  {{Morais Smith}}},\ }\href {\doibase 10.1016/j.ssc.2014.10.024} {\bibfield
  {journal} {\bibinfo  {journal} {Solid St. Commun.}\ }\textbf {\bibinfo
  {volume} {215}},\ \bibinfo {pages} {27} (\bibinfo {year} {2015})}\BibitemShut
  {NoStop}%
\bibitem [{\citenamefont {Peralta~Gavensky}\ \emph {et~al.}(2018)\citenamefont
  {Peralta~Gavensky}, \citenamefont {Usaj},\ and\ \citenamefont
  {Balseiro}}]{Gavensky18PRB}%
  \BibitemOpen
  \bibfield  {author} {\bibinfo {author} {\bibfnamefont {L.}~\bibnamefont
  {Peralta~Gavensky}}, \bibinfo {author} {\bibfnamefont {G.}~\bibnamefont
  {Usaj}}, \ and\ \bibinfo {author} {\bibfnamefont {C.~A.}\ \bibnamefont
  {Balseiro}},\ }\href {\doibase 10.1103/PhysRevB.98.165414} {\bibfield
  {journal} {\bibinfo  {journal} {Phys. Rev. B}\ }\textbf {\bibinfo {volume}
  {98}},\ \bibinfo {pages} {165414} (\bibinfo {year} {2018})}\BibitemShut
  {NoStop}%
\bibitem [{\citenamefont {Novi\v{c}enko}\ \emph {et~al.}(2017)\citenamefont
  {Novi\v{c}enko}, \citenamefont {Anisimovas},\ and\ \citenamefont
  {Juzeli\={u}nas}}]{Novicenko2017}%
  \BibitemOpen
  \bibfield  {author} {\bibinfo {author} {\bibfnamefont {V.}~\bibnamefont
  {Novi\v{c}enko}}, \bibinfo {author} {\bibfnamefont {E.}~\bibnamefont
  {Anisimovas}}, \ and\ \bibinfo {author} {\bibfnamefont {G.}~\bibnamefont
  {Juzeli\={u}nas}},\ }\href {\doibase 10.1103/PhysRevA.95.023615} {\bibfield
  {journal} {\bibinfo  {journal} {Phys. Rev. A}\ }\textbf {\bibinfo {volume}
  {95}},\ \bibinfo {pages} {023615} (\bibinfo {year} {2017})}\BibitemShut
  {NoStop}%
\bibitem [{\citenamefont {Novi\v{c}enko}\ and\ \citenamefont
  {G.~Juzeli\=unas}(2019)}]{Novicenko19PRA}%
  \BibitemOpen
  \bibfield  {author} {\bibinfo {author} {\bibfnamefont {V.}~\bibnamefont
  {Novi\v{c}enko}}\ and\ \bibinfo {author} {\bibfnamefont {G.}~\bibnamefont
  {G.~Juzeli\=unas}},\ }\href {\doibase 10.1103/PhysRevA.100.012127} {\bibfield
   {journal} {\bibinfo  {journal} {Phys. Rev. A}\ }\textbf {\bibinfo {volume}
  {100}},\ \bibinfo {pages} {012127} (\bibinfo {year} {2019})}\BibitemShut
  {NoStop}%
\bibitem [{\citenamefont {{Chen}}\ \emph {et~al.}(2019)\citenamefont {{Chen}},
  \citenamefont {{Murphree}},\ and\ \citenamefont
  {{Bigelow}}}]{Bigelow19Arxiv}%
  \BibitemOpen
  \bibfield  {author} {\bibinfo {author} {\bibfnamefont {Z.}~\bibnamefont
  {{Chen}}}, \bibinfo {author} {\bibfnamefont {J.~D.}\ \bibnamefont
  {{Murphree}}}, \ and\ \bibinfo {author} {\bibfnamefont {N.~P.}\ \bibnamefont
  {{Bigelow}}},\ }\href@noop {} {\bibfield  {journal} {\bibinfo  {journal}
  {arXiv e-prints}\ ,\ \bibinfo {eid} {arXiv:1907.10721}} (\bibinfo {year}
  {2019})},\ \Eprint {http://arxiv.org/abs/1907.10721} {arXiv:1907.10721
  [quant-ph]} \BibitemShut {NoStop}%
\bibitem [{\citenamefont {Ruseckas}\ \emph {et~al.}(2005)\citenamefont
  {Ruseckas}, \citenamefont {Juzeli\=unas}, \citenamefont {\"Ohberg},\ and\
  \citenamefont {Fleischhauer}}]{Ruseckas2005}%
  \BibitemOpen
  \bibfield  {author} {\bibinfo {author} {\bibfnamefont {J.}~\bibnamefont
  {Ruseckas}}, \bibinfo {author} {\bibfnamefont {G.}~\bibnamefont
  {Juzeli\=unas}}, \bibinfo {author} {\bibfnamefont {P.}~\bibnamefont
  {\"Ohberg}}, \ and\ \bibinfo {author} {\bibfnamefont {M.}~\bibnamefont
  {Fleischhauer}},\ }\href@noop {} {\bibfield  {journal} {\bibinfo  {journal}
  {Phys. Rev. Lett.}\ }\textbf {\bibinfo {volume} {95}},\ \bibinfo {pages}
  {010404} (\bibinfo {year} {2005})}\BibitemShut {NoStop}%
\bibitem [{\citenamefont {Stanescu}\ \emph {et~al.}(2007)\citenamefont
  {Stanescu}, \citenamefont {Zhang},\ and\ \citenamefont
  {Galitski}}]{Stanescu2007}%
  \BibitemOpen
  \bibfield  {author} {\bibinfo {author} {\bibfnamefont {T.~D.}\ \bibnamefont
  {Stanescu}}, \bibinfo {author} {\bibfnamefont {C.}~\bibnamefont {Zhang}}, \
  and\ \bibinfo {author} {\bibfnamefont {V.}~\bibnamefont {Galitski}},\ }\href
  {\doibase 10.1103/PhysRevLett.99.110403} {\bibfield  {journal} {\bibinfo
  {journal} {Phys. Rev. Lett.}\ }\textbf {\bibinfo {volume} {99}},\ \bibinfo
  {pages} {110403} (\bibinfo {year} {2007})}\BibitemShut {NoStop}%
\bibitem [{\citenamefont {Jacob}\ \emph {et~al.}(2007)\citenamefont {Jacob},
  \citenamefont {{\"O}hberg}, \citenamefont {Juzeli{\=u}nas},\ and\
  \citenamefont {Santos}}]{Jacob2007}%
  \BibitemOpen
  \bibfield  {author} {\bibinfo {author} {\bibfnamefont {A.}~\bibnamefont
  {Jacob}}, \bibinfo {author} {\bibfnamefont {P.}~\bibnamefont {{\"O}hberg}},
  \bibinfo {author} {\bibfnamefont {G.}~\bibnamefont {Juzeli{\=u}nas}}, \ and\
  \bibinfo {author} {\bibfnamefont {L.}~\bibnamefont {Santos}},\ }\href@noop {}
  {\bibfield  {journal} {\bibinfo  {journal} {Appl. Phys. B}\ }\textbf
  {\bibinfo {volume} {89}},\ \bibinfo {pages} {439} (\bibinfo {year}
  {2007})}\BibitemShut {NoStop}%
\bibitem [{\citenamefont {Juzeli\=unas}\ \emph {et~al.}(2008)\citenamefont
  {Juzeli\=unas}, \citenamefont {Ruseckas}, \citenamefont {Lindberg},
  \citenamefont {Santos},\ and\ \citenamefont
  {{\"O}hberg}}]{Juzeliunas2008PRA}%
  \BibitemOpen
  \bibfield  {author} {\bibinfo {author} {\bibfnamefont {G.}~\bibnamefont
  {Juzeli\=unas}}, \bibinfo {author} {\bibfnamefont {J.}~\bibnamefont
  {Ruseckas}}, \bibinfo {author} {\bibfnamefont {M.}~\bibnamefont {Lindberg}},
  \bibinfo {author} {\bibfnamefont {L.}~\bibnamefont {Santos}}, \ and\ \bibinfo
  {author} {\bibfnamefont {P.}~\bibnamefont {{\"O}hberg}},\ }\href@noop {}
  {\bibfield  {journal} {\bibinfo  {journal} {Phys. Rev. A}\ }\textbf {\bibinfo
  {volume} {77}},\ \bibinfo {pages} {011802(R)} (\bibinfo {year}
  {2008})}\BibitemShut {NoStop}%
\bibitem [{\citenamefont {Stanescu}\ \emph {et~al.}(2008)\citenamefont
  {Stanescu}, \citenamefont {Anderson},\ and\ \citenamefont
  {Galitski}}]{Stanescu2008}%
  \BibitemOpen
  \bibfield  {author} {\bibinfo {author} {\bibfnamefont {T.~D.}\ \bibnamefont
  {Stanescu}}, \bibinfo {author} {\bibfnamefont {B.}~\bibnamefont {Anderson}},
  \ and\ \bibinfo {author} {\bibfnamefont {V.}~\bibnamefont {Galitski}},\
  }\href@noop {} {\bibfield  {journal} {\bibinfo  {journal} {Phys. Rev. A}\
  }\textbf {\bibinfo {volume} {78}},\ \bibinfo {pages} {023616} (\bibinfo
  {year} {2008})}\BibitemShut {NoStop}%
\bibitem [{\citenamefont {Campbell}\ \emph {et~al.}(2011)\citenamefont
  {Campbell}, \citenamefont {Juzeli\={u}nas},\ and\ \citenamefont
  {Spielman}}]{Campbell2011}%
  \BibitemOpen
  \bibfield  {author} {\bibinfo {author} {\bibfnamefont {D.~L.}\ \bibnamefont
  {Campbell}}, \bibinfo {author} {\bibfnamefont {G.}~\bibnamefont
  {Juzeli\={u}nas}}, \ and\ \bibinfo {author} {\bibfnamefont {I.~B.}\
  \bibnamefont {Spielman}},\ }\href {\doibase 10.1103/PhysRevA.84.025602}
  {\bibfield  {journal} {\bibinfo  {journal} {Phys. Rev. A}\ }\textbf {\bibinfo
  {volume} {84}},\ \bibinfo {pages} {025602} (\bibinfo {year}
  {2011})}\BibitemShut {NoStop}%
\bibitem [{\citenamefont {Anderson}\ \emph {et~al.}(2012)\citenamefont
  {Anderson}, \citenamefont {Juzeli\=unas}, \citenamefont {Galitski},\ and\
  \citenamefont {Spielman}}]{Anderson2012PRL}%
  \BibitemOpen
  \bibfield  {author} {\bibinfo {author} {\bibfnamefont {B.~M.}\ \bibnamefont
  {Anderson}}, \bibinfo {author} {\bibfnamefont {G.}~\bibnamefont
  {Juzeli\=unas}}, \bibinfo {author} {\bibfnamefont {V.~M.}\ \bibnamefont
  {Galitski}}, \ and\ \bibinfo {author} {\bibfnamefont {I.~B.}\ \bibnamefont
  {Spielman}},\ }\href@noop {} {\bibfield  {journal} {\bibinfo  {journal}
  {Phys. Rev. Lett.}\ }\textbf {\bibinfo {volume} {108}},\ \bibinfo {pages}
  {235301} (\bibinfo {year} {2012})}\BibitemShut {NoStop}%
\bibitem [{\citenamefont {Huang}\ \emph {et~al.}(2016)\citenamefont {Huang},
  \citenamefont {Meng}, \citenamefont {Wang}, \citenamefont {Peng},
  \citenamefont {Zhang}, \citenamefont {Chen}, \citenamefont {Li},
  \citenamefont {Zhou},\ and\ \citenamefont {Zhang}}]{Huang16NP}%
  \BibitemOpen
  \bibfield  {author} {\bibinfo {author} {\bibfnamefont {L.}~\bibnamefont
  {Huang}}, \bibinfo {author} {\bibfnamefont {Z.}~\bibnamefont {Meng}},
  \bibinfo {author} {\bibfnamefont {P.}~\bibnamefont {Wang}}, \bibinfo {author}
  {\bibfnamefont {P.}~\bibnamefont {Peng}}, \bibinfo {author} {\bibfnamefont
  {S.-L.}\ \bibnamefont {Zhang}}, \bibinfo {author} {\bibfnamefont
  {L.}~\bibnamefont {Chen}}, \bibinfo {author} {\bibfnamefont {D.}~\bibnamefont
  {Li}}, \bibinfo {author} {\bibfnamefont {Q.}~\bibnamefont {Zhou}}, \ and\
  \bibinfo {author} {\bibfnamefont {J.}~\bibnamefont {Zhang}},\ }\href
  {\doibase doi:10.1038/nphys3672} {\bibfield  {journal} {\bibinfo  {journal}
  {Nature Phys.}\ }\textbf {\bibinfo {volume} {12}},\ \bibinfo {pages} {540}
  (\bibinfo {year} {2016})}\BibitemShut {NoStop}%
\bibitem [{\citenamefont {Meng}\ \emph {et~al.}(2016)\citenamefont {Meng},
  \citenamefont {Huang}, \citenamefont {Peng}, \citenamefont {Li},
  \citenamefont {Chen}, \citenamefont {Xu}, \citenamefont {Zhang},
  \citenamefont {Wang},\ and\ \citenamefont {Zhang}}]{Meng16PRL}%
  \BibitemOpen
  \bibfield  {author} {\bibinfo {author} {\bibfnamefont {Z.}~\bibnamefont
  {Meng}}, \bibinfo {author} {\bibfnamefont {L.}~\bibnamefont {Huang}},
  \bibinfo {author} {\bibfnamefont {P.}~\bibnamefont {Peng}}, \bibinfo {author}
  {\bibfnamefont {D.}~\bibnamefont {Li}}, \bibinfo {author} {\bibfnamefont
  {L.}~\bibnamefont {Chen}}, \bibinfo {author} {\bibfnamefont {Y.}~\bibnamefont
  {Xu}}, \bibinfo {author} {\bibfnamefont {C.}~\bibnamefont {Zhang}}, \bibinfo
  {author} {\bibfnamefont {P.}~\bibnamefont {Wang}}, \ and\ \bibinfo {author}
  {\bibfnamefont {J.}~\bibnamefont {Zhang}},\ }\href {\doibase
  10.1103/PhysRevLett.117.235304} {\bibfield  {journal} {\bibinfo  {journal}
  {Phys. Rev. Lett.}\ }\textbf {\bibinfo {volume} {117}},\ \bibinfo {pages}
  {235304} (\bibinfo {year} {2016})}\BibitemShut {NoStop}%
\bibitem [{\citenamefont {Vald{\'e}s-Curiel}\ \emph {et~al.}()\citenamefont
  {Vald{\'e}s-Curiel}, \citenamefont {Trypogeorgos}, \citenamefont {Liang},
  \citenamefont {Anderson},\ and\ \citenamefont
  {Spielman}}]{Spielman19_2D_SOC}%
  \BibitemOpen
  \bibfield  {author} {\bibinfo {author} {\bibfnamefont {A.}~\bibnamefont
  {Vald{\'e}s-Curiel}}, \bibinfo {author} {\bibfnamefont {D.}~\bibnamefont
  {Trypogeorgos}}, \bibinfo {author} {\bibfnamefont {Q.-Y.}\ \bibnamefont
  {Liang}}, \bibinfo {author} {\bibfnamefont {R.~P.}\ \bibnamefont {Anderson}},
  \ and\ \bibinfo {author} {\bibfnamefont {I.~B.}\ \bibnamefont {Spielman}},\
  }\href@noop {} {\enquote {\bibinfo {title} {Unconventional topology with a
  rashba spin-orbit coupled quantum gas},}\ }\Eprint
  {http://arxiv.org/abs/1907.08637} {arXiv:1907.08637 [cond-mat.quant-gas]}
  \BibitemShut {NoStop}%
\bibitem [{\citenamefont {Landau}\ and\ \citenamefont
  {Lifshitz}(1987)}]{Landau:1987}%
  \BibitemOpen
  \bibfield  {author} {\bibinfo {author} {\bibfnamefont {L.~D.}\ \bibnamefont
  {Landau}}\ and\ \bibinfo {author} {\bibfnamefont {E.~M.}\ \bibnamefont
  {Lifshitz}},\ }\href@noop {} {\emph {\bibinfo {title} {Quantum mechanics}}}\
  (\bibinfo  {publisher} {Pergamon Press},\ \bibinfo {address} {New York},\
  \bibinfo {year} {1987})\BibitemShut {NoStop}%
\bibitem [{\citenamefont {Zhou}\ \emph {et~al.}(2018)\citenamefont {Zhou},
  \citenamefont {Wu}, \citenamefont {Guo}, \citenamefont {Wang}, \citenamefont
  {Pu},\ and\ \citenamefont {Zhou}}]{Pu18PRL}%
  \BibitemOpen
  \bibfield  {author} {\bibinfo {author} {\bibfnamefont {X.-F.}\ \bibnamefont
  {Zhou}}, \bibinfo {author} {\bibfnamefont {C.}~\bibnamefont {Wu}}, \bibinfo
  {author} {\bibfnamefont {G.-C.}\ \bibnamefont {Guo}}, \bibinfo {author}
  {\bibfnamefont {R.}~\bibnamefont {Wang}}, \bibinfo {author} {\bibfnamefont
  {H.}~\bibnamefont {Pu}}, \ and\ \bibinfo {author} {\bibfnamefont {Z.-W.}\
  \bibnamefont {Zhou}},\ }\href {\doibase 10.1103/PhysRevLett.120.130402}
  {\bibfield  {journal} {\bibinfo  {journal} {Phys. Rev. Lett.}\ }\textbf
  {\bibinfo {volume} {120}},\ \bibinfo {pages} {130402} (\bibinfo {year}
  {2018})}\BibitemShut {NoStop}%
\bibitem [{\citenamefont {Goldman}\ and\ \citenamefont
  {Dalibard}(2014)}]{Goldman2014}%
  \BibitemOpen
  \bibfield  {author} {\bibinfo {author} {\bibfnamefont {N.}~\bibnamefont
  {Goldman}}\ and\ \bibinfo {author} {\bibfnamefont {J.}~\bibnamefont
  {Dalibard}},\ }\href {\doibase 10.1103/PhysRevX.4.031027} {\bibfield
  {journal} {\bibinfo  {journal} {Phys. Rev. X}\ }\textbf {\bibinfo {volume}
  {4}},\ \bibinfo {pages} {031027} (\bibinfo {year} {2014})}\BibitemShut
  {NoStop}%
\bibitem [{\citenamefont {Eckardt}\ and\ \citenamefont
  {Anisimovas}(2015)}]{Eckardt2015}%
  \BibitemOpen
  \bibfield  {author} {\bibinfo {author} {\bibfnamefont {A.}~\bibnamefont
  {Eckardt}}\ and\ \bibinfo {author} {\bibfnamefont {E.}~\bibnamefont
  {Anisimovas}},\ }\href {\doibase 10.1088/1367-2630/17/9/093039} {\bibfield
  {journal} {\bibinfo  {journal} {New J. Phys.}\ }\textbf {\bibinfo {volume}
  {17}},\ \bibinfo {pages} {093039} (\bibinfo {year} {2015})}\BibitemShut
  {NoStop}%
\bibitem [{\citenamefont {Bukov}\ \emph {et~al.}(2015)\citenamefont {Bukov},
  \citenamefont {D'Alessio},\ and\ \citenamefont {Polkovnikov}}]{Bukov2015}%
  \BibitemOpen
  \bibfield  {author} {\bibinfo {author} {\bibfnamefont {M.}~\bibnamefont
  {Bukov}}, \bibinfo {author} {\bibfnamefont {L.}~\bibnamefont {D'Alessio}}, \
  and\ \bibinfo {author} {\bibfnamefont {A.}~\bibnamefont {Polkovnikov}},\
  }\href {\doibase 10.1080/00018732.2015.1055918} {\bibfield  {journal}
  {\bibinfo  {journal} {Adv.\ Phys.}\ }\textbf {\bibinfo {volume} {64}},\
  \bibinfo {pages} {139} (\bibinfo {year} {2015})}\BibitemShut {NoStop}%
\bibitem [{\citenamefont {{Cheng}}\ \emph {et~al.}(2019)\citenamefont
  {{Cheng}}, \citenamefont {{Gong}}, \citenamefont {{Guo}}, \citenamefont
  {{Zhou}},\ and\ \citenamefont {{Zhou}}}]{Cheng19arXiv}%
  \BibitemOpen
  \bibfield  {author} {\bibinfo {author} {\bibfnamefont {J.-M.}\ \bibnamefont
  {{Cheng}}}, \bibinfo {author} {\bibfnamefont {M.}~\bibnamefont {{Gong}}},
  \bibinfo {author} {\bibfnamefont {G.-C.}\ \bibnamefont {{Guo}}}, \bibinfo
  {author} {\bibfnamefont {Z.-W.}\ \bibnamefont {{Zhou}}}, \ and\ \bibinfo
  {author} {\bibfnamefont {X.-F.}\ \bibnamefont {{Zhou}}},\ }\href@noop {}
  {\bibfield  {journal} {\bibinfo  {journal} {arXiv e-prints}\ ,\ \bibinfo
  {eid} {arXiv:1907.02216}} (\bibinfo {year} {2019})},\ \Eprint
  {http://arxiv.org/abs/1907.02216} {arXiv:1907.02216 [cond-mat.quant-gas]}
  \BibitemShut {NoStop}%
\bibitem [{\citenamefont {Suchet}\ \emph {et~al.}(2016)\citenamefont {Suchet},
  \citenamefont {Rabinovic}, \citenamefont {Reimann}, \citenamefont
  {Kretschmar}, \citenamefont {Sievers}, \citenamefont {Salomon}, \citenamefont
  {Lau}, \citenamefont {Goulko}, \citenamefont {Lobo},\ and\ \citenamefont
  {Chevy}}]{Suchet_2016}%
  \BibitemOpen
  \bibfield  {author} {\bibinfo {author} {\bibfnamefont {D.}~\bibnamefont
  {Suchet}}, \bibinfo {author} {\bibfnamefont {M.}~\bibnamefont {Rabinovic}},
  \bibinfo {author} {\bibfnamefont {T.}~\bibnamefont {Reimann}}, \bibinfo
  {author} {\bibfnamefont {N.}~\bibnamefont {Kretschmar}}, \bibinfo {author}
  {\bibfnamefont {F.}~\bibnamefont {Sievers}}, \bibinfo {author} {\bibfnamefont
  {C.}~\bibnamefont {Salomon}}, \bibinfo {author} {\bibfnamefont
  {J.}~\bibnamefont {Lau}}, \bibinfo {author} {\bibfnamefont {O.}~\bibnamefont
  {Goulko}}, \bibinfo {author} {\bibfnamefont {C.}~\bibnamefont {Lobo}}, \ and\
  \bibinfo {author} {\bibfnamefont {F.}~\bibnamefont {Chevy}},\ }\href
  {\doibase 10.1209/0295-5075/114/26005} {\bibfield  {journal} {\bibinfo
  {journal} {{EPL} (Europhysics Letters)}\ }\textbf {\bibinfo {volume} {114}},\
  \bibinfo {pages} {26005} (\bibinfo {year} {2016})}\BibitemShut {NoStop}%
\bibitem [{\citenamefont {Marzlin}\ \emph {et~al.}(1997)\citenamefont
  {Marzlin}, \citenamefont {Zhang},\ and\ \citenamefont
  {Wright}}]{Marzlin97PRL}%
  \BibitemOpen
  \bibfield  {author} {\bibinfo {author} {\bibfnamefont {K.-P.}\ \bibnamefont
  {Marzlin}}, \bibinfo {author} {\bibfnamefont {W.}~\bibnamefont {Zhang}}, \
  and\ \bibinfo {author} {\bibfnamefont {E.~M.}\ \bibnamefont {Wright}},\
  }\href {\doibase 10.1103/PhysRevLett.79.4728} {\bibfield  {journal} {\bibinfo
   {journal} {Phys. Rev. Lett.}\ }\textbf {\bibinfo {volume} {79}},\ \bibinfo
  {pages} {4728} (\bibinfo {year} {1997})}\BibitemShut {NoStop}%
\bibitem [{\citenamefont {Dutton}\ and\ \citenamefont
  {Ruostekoski}(2004)}]{Ruostekoski04PRL}%
  \BibitemOpen
  \bibfield  {author} {\bibinfo {author} {\bibfnamefont {Z.}~\bibnamefont
  {Dutton}}\ and\ \bibinfo {author} {\bibfnamefont {J.}~\bibnamefont
  {Ruostekoski}},\ }\href {\doibase 10.1103/PhysRevLett.93.193602} {\bibfield
  {journal} {\bibinfo  {journal} {Phys. Rev. Lett.}\ }\textbf {\bibinfo
  {volume} {93}},\ \bibinfo {pages} {193602} (\bibinfo {year}
  {2004})}\BibitemShut {NoStop}%
\bibitem [{\citenamefont {Nandi}\ \emph {et~al.}(2004)\citenamefont {Nandi},
  \citenamefont {Walser},\ and\ \citenamefont {Schleich}}]{Nandi04PRA}%
  \BibitemOpen
  \bibfield  {author} {\bibinfo {author} {\bibfnamefont {G.}~\bibnamefont
  {Nandi}}, \bibinfo {author} {\bibfnamefont {R.}~\bibnamefont {Walser}}, \
  and\ \bibinfo {author} {\bibfnamefont {W.~P.}\ \bibnamefont {Schleich}},\
  }\href {\doibase 10.1103/PhysRevA.69.063606} {\bibfield  {journal} {\bibinfo
  {journal} {Phys. Rev. A}\ }\textbf {\bibinfo {volume} {69}},\ \bibinfo
  {pages} {063606} (\bibinfo {year} {2004})}\BibitemShut {NoStop}%
\bibitem [{\citenamefont {Juzeli\=unas}\ and\ \citenamefont
  {\"Ohberg}(2004)}]{Juzeliunas2004}%
  \BibitemOpen
  \bibfield  {author} {\bibinfo {author} {\bibfnamefont {G.}~\bibnamefont
  {Juzeli\=unas}}\ and\ \bibinfo {author} {\bibfnamefont {P.}~\bibnamefont
  {\"Ohberg}},\ }\href@noop {} {\bibfield  {journal} {\bibinfo  {journal}
  {Phys. Rev. Lett.}\ }\textbf {\bibinfo {volume} {93}},\ \bibinfo {pages}
  {033602} (\bibinfo {year} {2004})}\BibitemShut {NoStop}%
\bibitem [{\citenamefont {Juzeli\=unas}\ \emph {et~al.}(2005)\citenamefont
  {Juzeli\=unas}, \citenamefont {\"Ohberg}, \citenamefont {Ruseckas},\ and\
  \citenamefont {Klein}}]{Juzeliunas2005}%
  \BibitemOpen
  \bibfield  {author} {\bibinfo {author} {\bibfnamefont {G.}~\bibnamefont
  {Juzeli\=unas}}, \bibinfo {author} {\bibfnamefont {P.}~\bibnamefont
  {\"Ohberg}}, \bibinfo {author} {\bibfnamefont {J.}~\bibnamefont {Ruseckas}},
  \ and\ \bibinfo {author} {\bibfnamefont {A.}~\bibnamefont {Klein}},\
  }\href@noop {} {\bibfield  {journal} {\bibinfo  {journal} {Phys. Rev. A}\
  }\textbf {\bibinfo {volume} {71}},\ \bibinfo {pages} {053614} (\bibinfo
  {year} {2005})}\BibitemShut {NoStop}%
\bibitem [{\citenamefont {Wright}\ \emph {et~al.}(2008)\citenamefont {Wright},
  \citenamefont {Leslie},\ and\ \citenamefont {Bigelow}}]{Bigelow08PRA}%
  \BibitemOpen
  \bibfield  {author} {\bibinfo {author} {\bibfnamefont {K.~C.}\ \bibnamefont
  {Wright}}, \bibinfo {author} {\bibfnamefont {L.~S.}\ \bibnamefont {Leslie}},
  \ and\ \bibinfo {author} {\bibfnamefont {N.~P.}\ \bibnamefont {Bigelow}},\
  }\href {\doibase 10.1103/PhysRevA.78.053412} {\bibfield  {journal} {\bibinfo
  {journal} {Phys. Rev. A}\ }\textbf {\bibinfo {volume} {78}},\ \bibinfo
  {pages} {053412} (\bibinfo {year} {2008})}\BibitemShut {NoStop}%
\bibitem [{\citenamefont {Wright}\ \emph {et~al.}(2009)\citenamefont {Wright},
  \citenamefont {Leslie}, \citenamefont {Hansen},\ and\ \citenamefont
  {Bigelow}}]{Bigelow09PRL}%
  \BibitemOpen
  \bibfield  {author} {\bibinfo {author} {\bibfnamefont {K.~C.}\ \bibnamefont
  {Wright}}, \bibinfo {author} {\bibfnamefont {L.~S.}\ \bibnamefont {Leslie}},
  \bibinfo {author} {\bibfnamefont {A.}~\bibnamefont {Hansen}}, \ and\ \bibinfo
  {author} {\bibfnamefont {N.~P.}\ \bibnamefont {Bigelow}},\ }\href {\doibase
  10.1103/PhysRevLett.102.030405} {\bibfield  {journal} {\bibinfo  {journal}
  {Phys. Rev. Lett.}\ }\textbf {\bibinfo {volume} {102}},\ \bibinfo {pages}
  {030405} (\bibinfo {year} {2009})}\BibitemShut {NoStop}%
\bibitem [{\citenamefont {DeMarco}\ and\ \citenamefont {Pu}(2015)}]{Pu15PRA}%
  \BibitemOpen
  \bibfield  {author} {\bibinfo {author} {\bibfnamefont {M.}~\bibnamefont
  {DeMarco}}\ and\ \bibinfo {author} {\bibfnamefont {H.}~\bibnamefont {Pu}},\
  }\href {\doibase 10.1103/PhysRevA.91.033630} {\bibfield  {journal} {\bibinfo
  {journal} {Phys. Rev. A}\ }\textbf {\bibinfo {volume} {91}},\ \bibinfo
  {pages} {033630} (\bibinfo {year} {2015})}\BibitemShut {NoStop}%
\bibitem [{\citenamefont {Qu}\ \emph {et~al.}(2015)\citenamefont {Qu},
  \citenamefont {Sun},\ and\ \citenamefont {Zhang}}]{Qu15PRA}%
  \BibitemOpen
  \bibfield  {author} {\bibinfo {author} {\bibfnamefont {C.}~\bibnamefont
  {Qu}}, \bibinfo {author} {\bibfnamefont {K.}~\bibnamefont {Sun}}, \ and\
  \bibinfo {author} {\bibfnamefont {C.}~\bibnamefont {Zhang}},\ }\href
  {\doibase 10.1103/PhysRevA.91.053630} {\bibfield  {journal} {\bibinfo
  {journal} {Phys. Rev. A}\ }\textbf {\bibinfo {volume} {91}},\ \bibinfo
  {pages} {053630} (\bibinfo {year} {2015})}\BibitemShut {NoStop}%
\bibitem [{\citenamefont {Chen}\ \emph
  {et~al.}(2018{\natexlab{a}})\citenamefont {Chen}, \citenamefont {Lin},
  \citenamefont {Chen}, \citenamefont {Chiu}, \citenamefont {Wang},
  \citenamefont {Chen}, \citenamefont {Huang}, \citenamefont {Yip},
  \citenamefont {Kawaguchi},\ and\ \citenamefont {Lin}}]{Lin18PRL-a}%
  \BibitemOpen
  \bibfield  {author} {\bibinfo {author} {\bibfnamefont {H.-R.}\ \bibnamefont
  {Chen}}, \bibinfo {author} {\bibfnamefont {K.-Y.}\ \bibnamefont {Lin}},
  \bibinfo {author} {\bibfnamefont {P.-K.}\ \bibnamefont {Chen}}, \bibinfo
  {author} {\bibfnamefont {N.-C.}\ \bibnamefont {Chiu}}, \bibinfo {author}
  {\bibfnamefont {J.-B.}\ \bibnamefont {Wang}}, \bibinfo {author}
  {\bibfnamefont {C.-A.}\ \bibnamefont {Chen}}, \bibinfo {author}
  {\bibfnamefont {P.-P.}\ \bibnamefont {Huang}}, \bibinfo {author}
  {\bibfnamefont {S.-K.}\ \bibnamefont {Yip}}, \bibinfo {author} {\bibfnamefont
  {Y.}~\bibnamefont {Kawaguchi}}, \ and\ \bibinfo {author} {\bibfnamefont
  {Y.-J.}\ \bibnamefont {Lin}},\ }\href {\doibase
  10.1103/PhysRevLett.121.113204} {\bibfield  {journal} {\bibinfo  {journal}
  {Phys. Rev. Lett.}\ }\textbf {\bibinfo {volume} {121}},\ \bibinfo {pages}
  {113204} (\bibinfo {year} {2018}{\natexlab{a}})}\BibitemShut {NoStop}%
\bibitem [{\citenamefont {Chen}\ \emph
  {et~al.}(2018{\natexlab{b}})\citenamefont {Chen}, \citenamefont {Liu},
  \citenamefont {Tsai}, \citenamefont {Chiu}, \citenamefont {Kawaguchi},
  \citenamefont {Yip}, \citenamefont {Chang},\ and\ \citenamefont
  {Lin}}]{Lin18PRL-b}%
  \BibitemOpen
  \bibfield  {author} {\bibinfo {author} {\bibfnamefont {P.-K.}\ \bibnamefont
  {Chen}}, \bibinfo {author} {\bibfnamefont {L.-R.}\ \bibnamefont {Liu}},
  \bibinfo {author} {\bibfnamefont {M.-J.}\ \bibnamefont {Tsai}}, \bibinfo
  {author} {\bibfnamefont {N.-C.}\ \bibnamefont {Chiu}}, \bibinfo {author}
  {\bibfnamefont {Y.}~\bibnamefont {Kawaguchi}}, \bibinfo {author}
  {\bibfnamefont {S.-K.}\ \bibnamefont {Yip}}, \bibinfo {author} {\bibfnamefont
  {M.-S.}\ \bibnamefont {Chang}}, \ and\ \bibinfo {author} {\bibfnamefont
  {Y.-J.}\ \bibnamefont {Lin}},\ }\href {\doibase
  10.1103/PhysRevLett.121.250401} {\bibfield  {journal} {\bibinfo  {journal}
  {Phys. Rev. Lett.}\ }\textbf {\bibinfo {volume} {121}},\ \bibinfo {pages}
  {250401} (\bibinfo {year} {2018}{\natexlab{b}})}\BibitemShut {NoStop}%
\bibitem [{\citenamefont {Zhang}\ \emph {et~al.}(2019)\citenamefont {Zhang},
  \citenamefont {Gao}, \citenamefont {Zou}, \citenamefont {Kong}, \citenamefont
  {Li}, \citenamefont {Shen}, \citenamefont {Chen}, \citenamefont {Peng},
  \citenamefont {Zhan}, \citenamefont {Pu},\ and\ \citenamefont
  {Jiang}}]{Zhang19PRL}%
  \BibitemOpen
  \bibfield  {author} {\bibinfo {author} {\bibfnamefont {D.}~\bibnamefont
  {Zhang}}, \bibinfo {author} {\bibfnamefont {T.}~\bibnamefont {Gao}}, \bibinfo
  {author} {\bibfnamefont {P.}~\bibnamefont {Zou}}, \bibinfo {author}
  {\bibfnamefont {L.}~\bibnamefont {Kong}}, \bibinfo {author} {\bibfnamefont
  {R.}~\bibnamefont {Li}}, \bibinfo {author} {\bibfnamefont {X.}~\bibnamefont
  {Shen}}, \bibinfo {author} {\bibfnamefont {X.-L.}\ \bibnamefont {Chen}},
  \bibinfo {author} {\bibfnamefont {S.-G.}\ \bibnamefont {Peng}}, \bibinfo
  {author} {\bibfnamefont {M.}~\bibnamefont {Zhan}}, \bibinfo {author}
  {\bibfnamefont {H.}~\bibnamefont {Pu}}, \ and\ \bibinfo {author}
  {\bibfnamefont {K.}~\bibnamefont {Jiang}},\ }\href {\doibase
  10.1103/PhysRevLett.122.110402} {\bibfield  {journal} {\bibinfo  {journal}
  {Phys. Rev. Lett.}\ }\textbf {\bibinfo {volume} {122}},\ \bibinfo {pages}
  {110402} (\bibinfo {year} {2019})}\BibitemShut {NoStop}%
\bibitem [{\citenamefont {Luo}\ \emph {et~al.}(2017)\citenamefont {Luo},
  \citenamefont {Sun},\ and\ \citenamefont {Zhang}}]{Zhang17PRL}%
  \BibitemOpen
  \bibfield  {author} {\bibinfo {author} {\bibfnamefont {X.-W.}\ \bibnamefont
  {Luo}}, \bibinfo {author} {\bibfnamefont {K.}~\bibnamefont {Sun}}, \ and\
  \bibinfo {author} {\bibfnamefont {C.}~\bibnamefont {Zhang}},\ }\href
  {\doibase 10.1103/PhysRevLett.119.193001} {\bibfield  {journal} {\bibinfo
  {journal} {Phys. Rev. Lett.}\ }\textbf {\bibinfo {volume} {119}},\ \bibinfo
  {pages} {193001} (\bibinfo {year} {2017})}\BibitemShut {NoStop}%
\bibitem [{\citenamefont {Juzeli\=unas}\ \emph {et~al.}(2006)\citenamefont
  {Juzeli\=unas}, \citenamefont {Ruseckas}, \citenamefont {{\"O}hberg},\ and\
  \citenamefont {Fleischhauer}}]{Juzeliunas2006}%
  \BibitemOpen
  \bibfield  {author} {\bibinfo {author} {\bibfnamefont {G.}~\bibnamefont
  {Juzeli\=unas}}, \bibinfo {author} {\bibfnamefont {J.}~\bibnamefont
  {Ruseckas}}, \bibinfo {author} {\bibfnamefont {P.}~\bibnamefont
  {{\"O}hberg}}, \ and\ \bibinfo {author} {\bibfnamefont {M.}~\bibnamefont
  {Fleischhauer}},\ }\href@noop {} {\bibfield  {journal} {\bibinfo  {journal}
  {Phys. Rev. A}\ }\textbf {\bibinfo {volume} {73}},\ \bibinfo {pages} {025602}
  (\bibinfo {year} {2006})}\BibitemShut {NoStop}%
\bibitem [{\citenamefont {Li}\ and\ \citenamefont
  {Wu}(2013)}]{Congjun-Wu13PRL}%
  \BibitemOpen
  \bibfield  {author} {\bibinfo {author} {\bibfnamefont {Y.}~\bibnamefont
  {Li}}\ and\ \bibinfo {author} {\bibfnamefont {C.}~\bibnamefont {Wu}},\ }\href
  {\doibase 10.1103/PhysRevLett.110.216802} {\bibfield  {journal} {\bibinfo
  {journal} {Phys. Rev. Lett.}\ }\textbf {\bibinfo {volume} {110}},\ \bibinfo
  {pages} {216802} (\bibinfo {year} {2013})}\BibitemShut {NoStop}%
\bibitem [{\citenamefont {Ui}\ and\ \citenamefont
  {Takeda}(1984)}]{Ui-Takeda84PTP}%
  \BibitemOpen
  \bibfield  {author} {\bibinfo {author} {\bibfnamefont {H.}~\bibnamefont
  {Ui}}\ and\ \bibinfo {author} {\bibfnamefont {G.}~\bibnamefont {Takeda}},\
  }\href {\doibase 10.1143/PTP.72.266} {\bibfield  {journal} {\bibinfo
  {journal} {Progress of Theoretical Physics}\ }\textbf {\bibinfo {volume}
  {72}},\ \bibinfo {pages} {266} (\bibinfo {year} {1984})},\ \Eprint
  {http://arxiv.org/abs/http://oup.prod.sis.lan/ptp/article-pdf/72/2/266/5178964/72-2-266.pdf}
  {http://oup.prod.sis.lan/ptp/article-pdf/72/2/266/5178964/72-2-266.pdf}
  \BibitemShut {NoStop}%
\bibitem [{\citenamefont {Bagchi}(2001)}]{Bagchi2001}%
  \BibitemOpen
  \bibfield  {author} {\bibinfo {author} {\bibfnamefont {B.~K.}\ \bibnamefont
  {Bagchi}},\ }\href@noop {} {\emph {\bibinfo {title} {Supersymmetry In Quantum
  and Classical Mechanics}}}\ (\bibinfo  {publisher} {Chapman and Hall/CRC},\
  \bibinfo {year} {2001})\BibitemShut {NoStop}%
\bibitem [{\citenamefont {Dunlap}\ and\ \citenamefont
  {Kenkre}(1986)}]{Kenkre1986PRB}%
  \BibitemOpen
  \bibfield  {author} {\bibinfo {author} {\bibfnamefont {D.~H.}\ \bibnamefont
  {Dunlap}}\ and\ \bibinfo {author} {\bibfnamefont {V.~M.}\ \bibnamefont
  {Kenkre}},\ }\href {\doibase 10.1103/PhysRevB.34.3625} {\bibfield  {journal}
  {\bibinfo  {journal} {Phys. Rev. B}\ }\textbf {\bibinfo {volume} {34}},\
  \bibinfo {pages} {3625} (\bibinfo {year} {1986})}\BibitemShut {NoStop}%
\bibitem [{\citenamefont {Lignier}\ \emph {et~al.}(2007)\citenamefont
  {Lignier}, \citenamefont {Sias}, \citenamefont {Ciampini}, \citenamefont
  {Singh}, \citenamefont {Zenesini}, \citenamefont {Morsch},\ and\
  \citenamefont {Arimondo}}]{Arimondo07PRL}%
  \BibitemOpen
  \bibfield  {author} {\bibinfo {author} {\bibfnamefont {H.}~\bibnamefont
  {Lignier}}, \bibinfo {author} {\bibfnamefont {C.}~\bibnamefont {Sias}},
  \bibinfo {author} {\bibfnamefont {D.}~\bibnamefont {Ciampini}}, \bibinfo
  {author} {\bibfnamefont {Y.}~\bibnamefont {Singh}}, \bibinfo {author}
  {\bibfnamefont {A.}~\bibnamefont {Zenesini}}, \bibinfo {author}
  {\bibfnamefont {O.}~\bibnamefont {Morsch}}, \ and\ \bibinfo {author}
  {\bibfnamefont {E.}~\bibnamefont {Arimondo}},\ }\href {\doibase
  10.1103/PhysRevLett.99.220403} {\bibfield  {journal} {\bibinfo  {journal}
  {Phys. Rev. Lett.}\ }\textbf {\bibinfo {volume} {99}},\ \bibinfo {pages}
  {220403} (\bibinfo {year} {2007})}\BibitemShut {NoStop}%
\bibitem [{\citenamefont {Arimondo}\ \emph {et~al.}(2012)\citenamefont
  {Arimondo}, \citenamefont {Ciampini}, \citenamefont {Eckardt}, \citenamefont
  {Holthaus},\ and\ \citenamefont {Morsch}}]{Arimondo2012}%
  \BibitemOpen
  \bibfield  {author} {\bibinfo {author} {\bibfnamefont {E.}~\bibnamefont
  {Arimondo}}, \bibinfo {author} {\bibfnamefont {D.}~\bibnamefont {Ciampini}},
  \bibinfo {author} {\bibfnamefont {A.}~\bibnamefont {Eckardt}}, \bibinfo
  {author} {\bibfnamefont {M.}~\bibnamefont {Holthaus}}, \ and\ \bibinfo
  {author} {\bibfnamefont {O.}~\bibnamefont {Morsch}},\ }\href {\doibase
  10.1016/B978-0-12-396482-3.00010-7} {\bibfield  {journal} {\bibinfo
  {journal} {Adv. At. Molec. Opt. Phys.}\ }\textbf {\bibinfo {volume} {61}},\
  \bibinfo {pages} {515} (\bibinfo {year} {2012})}\BibitemShut {NoStop}%
\bibitem [{\citenamefont {Li}\ \emph {et~al.}()\citenamefont {Li},
  \citenamefont {Shteynas},\ and\ \citenamefont
  {Ketterle}}]{Ketterle19Floquet-heating}%
  \BibitemOpen
  \bibfield  {author} {\bibinfo {author} {\bibfnamefont {J.-R.}\ \bibnamefont
  {Li}}, \bibinfo {author} {\bibfnamefont {B.}~\bibnamefont {Shteynas}}, \ and\
  \bibinfo {author} {\bibfnamefont {W.}~\bibnamefont {Ketterle}},\ }\href@noop
  {} {\enquote {\bibinfo {title} {Floquet heating in interacting atomic gases
  with an oscillating force},}\ }\Eprint {http://arxiv.org/abs/1906.08747}
  {arXiv:1906.08747 [physics.atom-ph]} \BibitemShut {NoStop}%
\bibitem [{\citenamefont {Metcalf}\ and\ \citenamefont {van~der
  Straten}(1999)}]{Metcalf1999}%
  \BibitemOpen
  \bibfield  {author} {\bibinfo {author} {\bibfnamefont {H.~J.}\ \bibnamefont
  {Metcalf}}\ and\ \bibinfo {author} {\bibfnamefont {P.}~\bibnamefont {van~der
  Straten}},\ }\href@noop {} {\emph {\bibinfo {title} {Laser Cooling and
  Trapping}}}\ (\bibinfo  {publisher} {Springer-Verlag, New York},\ \bibinfo
  {year} {1999})\BibitemShut {NoStop}%
\bibitem [{\citenamefont {Lin}\ \emph {et~al.}(2011)\citenamefont {Lin},
  \citenamefont {Jim{\'e}nez-Garc{\'\i}a},\ and\ \citenamefont
  {Spielman}}]{Lin2011}%
  \BibitemOpen
  \bibfield  {author} {\bibinfo {author} {\bibfnamefont {Y.-J.}\ \bibnamefont
  {Lin}}, \bibinfo {author} {\bibfnamefont {K.}~\bibnamefont
  {Jim{\'e}nez-Garc{\'\i}a}}, \ and\ \bibinfo {author} {\bibfnamefont {I.~B.}\
  \bibnamefont {Spielman}},\ }\href@noop {} {\bibfield  {journal} {\bibinfo
  {journal} {Nature}\ }\textbf {\bibinfo {volume} {471}},\ \bibinfo {pages}
  {83} (\bibinfo {year} {2011})}\BibitemShut {NoStop}%
\bibitem [{\citenamefont {Han}\ \emph {et~al.}(2015)\citenamefont {Han},
  \citenamefont {Juzeli\={u}nas}, \citenamefont {Zhang},\ and\ \citenamefont
  {Liu}}]{Han15PRA}%
  \BibitemOpen
  \bibfield  {author} {\bibinfo {author} {\bibfnamefont {W.}~\bibnamefont
  {Han}}, \bibinfo {author} {\bibfnamefont {G.}~\bibnamefont {Juzeli\={u}nas}},
  \bibinfo {author} {\bibfnamefont {W.}~\bibnamefont {Zhang}}, \ and\ \bibinfo
  {author} {\bibfnamefont {W.-M.}\ \bibnamefont {Liu}},\ }\href {\doibase
  10.1103/PhysRevA.91.013607} {\bibfield  {journal} {\bibinfo  {journal} {Phys.
  Rev. A}\ }\textbf {\bibinfo {volume} {91}},\ \bibinfo {pages} {013607}
  (\bibinfo {year} {2015})}\BibitemShut {NoStop}%
\end{thebibliography}%

\end{document}